\documentclass[a4paper,11pt]{article}
\usepackage{caption}
\usepackage{graphicx}
\usepackage{subcaption}
\usepackage{hyperref}
\usepackage{amssymb}
\usepackage{color}
\usepackage{tikz}
\usepackage{amsmath}
\usepackage{cite}
\usetikzlibrary{decorations.markings}
\usepackage{xparse}
\usetikzlibrary{arrows}
\usetikzlibrary{calc,decorations.pathreplacing}
\usetikzlibrary{arrows,shapes,positioning}
\numberwithin{equation}{section}
\newcommand{\ii}{\text{i}}

\newcommand{\Tr}{\text{Tr} \,}
\newcommand{\be}{\begin{equation}}
\newcommand{\ee}{\end{equation}}
\newcommand{\p}{\partial}

\newcommand{\C}{{\mathbb C}}

\newcommand{\one}{{\rm 1\kern -.9mm l}}

\newdimen\tableauside\tableauside=1.0ex
\newdimen\tableaurule\tableaurule=0.4pt
\newdimen\tableaustep
\def\phantomhrule#1{\hbox{\vbox to0pt{\hrule height\tableaurule
width#1\vss}}}
\def\phantomvrule#1{\vbox{\hbox to0pt{\vrule width\tableaurule
height#1\hss}}}
\def\sqr{\vbox{%
  \phantomhrule\tableaustep
\hbox{\phantomvrule\tableaustep\kern\tableaustep\phantomvrule\tableaustep}%
  \hbox{\vbox{\phantomhrule\tableauside}\kern-\tableaurule}}}
\def\squares#1{\hbox{\count0=#1\noindent\loop\sqr
  \advance\count0 by-1 \ifnum\count0>0\repeat}}
\def\tableau#1{\vcenter{\offinterlineskip
  \tableaustep=\tableauside\advance\tableaustep by-\tableaurule
  \kern\normallineskip\hbox
    {\kern\normallineskip\vbox
      {\gettableau#1 0 }%
     \kern\normallineskip\kern\tableaurule}%
  \kern\normallineskip\kern\tableaurule}}
\def\gettableau#1 {\ifnum#1=0\let\next=\null\else
  \squares{#1}\let\next=\gettableau\fi\next}
\newcommand{\eu}{\epsilon_1}
\newcommand{\ed}{\epsilon_2}

\def\XXint#1#2#3{{\setbox0=\hbox{$#1{#2#3}{\int}$}
     \vcenter{\hbox{$#2#3$}}\kern-.5\wd0}}

\usetikzlibrary{decorations.markings}
\usepackage{xparse}
\usetikzlibrary{calc,decorations.pathreplacing}
\usetikzlibrary{arrows,shapes,positioning}
\usetikzlibrary{shapes.misc}
\tikzset{cross/.style={cross out, draw=black, minimum size=2*(#1-\pgflinewidth), inner sep=0pt, outer sep=0pt},
cross/.default={1pt}}

\begin{document}

\begin{titlepage}

\vskip 1.5in
\begin{center}
{\bf\Large{Exact WKB Analysis of ${\cal N}=2$
	 Gauge Theories} }\vskip 1.9cm
\large{Sujay K. Ashok$^{a,}$\footnote{\href{mailto:sashok@imsc.res.in}{sashok@imsc.res.in}}, Dileep P. Jatkar$^{b,}$\footnote{\href{mailto:dileep@hri.res.in}{dileep@hri.res.in}}, Renjan R. John$^{a,}$\footnote{\href{mailto:renjan@imsc.res.in}{renjan@imsc.res.in}},\\ 
Madhusudhan Raman$^{a,}$\footnote{\href{mailto:madhur@imsc.res.in}{madhur@imsc.res.in}}, Jan Troost$^{c,}$\footnote{\href{mailto:troost@lpt.ens.fr}{troost@lpt.ens.fr}}}
\vskip0.3in

 \emph{$^{a}$}
\emph{ Institute of Mathematical Sciences \\
   C.~I.~T.~Campus, Taramani\\
   Chennai, India 600113\\ 
\vspace{.5cm}}

 \emph{$^{b}$}
\emph{ Harish-Chandra Research Institute \\
Chhatnag Road, Jhusi,\\ 
Allahabad, India 211019\\
\vspace{.2cm}}
  \emph{\\${}^{c}$ Laboratoire de Physique Th\'eorique \\ de l'\'Ecole Normale Sup\'erieure  \\ CNRS \\ PSL Research University \\  Sorbonne Universit\'es \\
75005 Paris, France}

\end{center}


\baselineskip 16pt
%



\begin{abstract}
  We study $\mathcal{N}=2$ supersymmetric gauge theories with gauge
  group SU$(2)$ coupled to fundamental flavours, covering all asymptotically free
  and conformal cases. We re-derive, from the conformal field theory
  perspective, the differential equations satisfied by $\epsilon_1$-
  and $\epsilon_2$-deformed instanton partition functions. We confirm
  their validity at leading order in $\epsilon_2$ via a saddle-point
  analysis of the partition function. In the semi-classical
  limit we show that these differential equations take a form amenable
  to exact WKB analysis. We compute the monodromy group associated to
  the differential equations in terms of $\epsilon_1$-deformed and
  Borel resummed Seiberg-Witten data. For each case, we study pairs of
  Stokes graphs that are related by flips and pops, and show that the
  monodromy groups allow one to confirm the Stokes automorphisms that
  arise as the phase of $\epsilon_1$ is varied. Finally, we relate
  the Borel resummed monodromies with the traditional Seiberg-Witten variables
in the semi-classical limit.
\end{abstract}

\end{titlepage}
\vfill\eject

\tableofcontents

\section{Introduction}

For some time now, we have been able to compute the low-energy
effective action of ${\cal N}=2$ supersymmetric gauge theories in four
dimensions. In \cite{Seiberg:1994rs,Seiberg:1994aj}, the solution for
the low-energy theory was given in terms of an algebraic curve and an
associated differential. Subsequent works have simplified and
clarified many aspects of the Seiberg-Witten solution. The
Seiberg-Witten curves may be intuitively pictured in terms of M-theory
five-branes \cite{Witten:1997sc}, and this geometric picture has inspired
a description of class $\mathcal{S}$ theories in terms of punctured
Riemann surfaces \cite{Gaiotto:2009we}. In a parallel development, it
has also become possible to compute instanton contributions by
invoking the powerful machinery of equivariant localization
\cite{Nekrasov:2002qd}. Of particular note, the calculation of the
gauge theory partition function on $S^4$ via localization
naturally incorporates these instanton sums \cite{Pestun:2007rz}. All
these developments were key to writing a dictionary between
observables in four-dimensional gauge theories and those in two-dimensional
conformal field theories: the 2d/4d correspondence
\cite{Alday:2009aq}.

The 2d/4d correspondence makes it possible to use the technology of
conformal field theory to gain deeper insights into the behavior of
${\cal N}=2$ gauge theories. For instance, the $\Omega$-deformed gauge
theory partition function with a surface operator insertion maps to
the meromorphic solution of a null vector decoupling equation
\cite{Drukker:2009id,Alday:2009fs}. Thus, an analysis of conformal
blocks in two-dimensional conformal field theory yields information
about surface operators in gauge theories. These conformal blocks can
be viewed as solutions to Riemann-Hilbert problems specified by a
differential equation with singularities and associated monodromies \cite{Vartanov:2013ima}. We
expect this exact picture to be valid in gauge theory (and the field
theory limit of topological string theory) \cite{Awata:2010bz}.

In this paper, we study quantum chromodynamics with ${\cal N}=2$ supersymmetry and gauge group SU$(2)$, and the
corresponding Virasoro conformal blocks. 
In particular, we study the differential equation
that the instanton partition function with surface operator insertion
satisfies. This corresponds to an analysis of null vector decoupling equations in the presence
of irregular blocks. The differential equations satisfied by
correlators involving irregular blocks were described in \cite{Awata:2010bz,AwataYamada,Gaiotto:2012sf}. The 
equations are exact in the $\Omega$-deformation parameters
$\left(\epsilon_{1},\epsilon_2\right)$, and provide for a map to standard gauge theory
expressions for the Seiberg-Witten curve, including $\epsilon_i$
corrections.

We then concentrate on the limit $\epsilon_2/\epsilon_1 \rightarrow 0$ \cite{NSlimit}, which is a large central charge limit in the conformal field
theory. It has been shown in e.g. \cite{MironovMorozov,Fateev:2009aw,TroostToroidal} that a WKB analysis of the null vector decoupling equations in this semi-classical limit reproduces the non-convergent $\epsilon_1$-expansion of the
instanton partition function of the gauge theory.\footnote{For non-perturbative results in the context of topological strings we refer to \cite{Marino,Marino1308}.} There is  a rich 
 literature 
\cite{Voros,Delabaere1,Jidoumou,Delabaere2,Delabaere3,ZinnJustin,ZinnJustin2,book,Costin,NakanishiandIwaki} on methods which may be used to enhance these results non-perturbatively.
Using the exact WKB analysis, we study the resulting differential equations satisfied by
the (ir)regular conformal blocks (equivalently, the
$\epsilon_1$-deformed surface operator partition function). This allows us to compute the monodromy group of
each of the differential equations as a function of (i) the parameters
of the differential equations, and (ii) the Borel resummed monodromies
that are properties of individual solutions. The monodromy group
contains information about the instanton partition function with
surface operator insertion, which is non-perturbative in $\epsilon_1$.
In doing so, we provide the underlying exact picture
\cite{Vartanov:2013ima} with a detailed description of how these
beautiful and abstract mathematical constructs reduce to the more
hands-on limiting analysis of ${\cal N}=2$ gauge theories to which we have become accustomed.

In this physical set-up, we apply the theorems of
\cite{NakanishiandIwaki}, thereby drawing on intuition from both gauge
theory and the mathematical study of singular perturbation theory
\cite{book}. As a by-product, we add details to the WKB analysis and
provide a calculation of the monodromy group of the differential
equation in terms of deformed gauge theory data. For instance, we
analyze the occurrence of a double flip, consisting of simultaneous
single flips. Two different ways of splitting the double flip into two
single flips give the same monodromy group and Stokes automorphism.
Although we demonstrate this result in the context of $N_f=4$ theory,
this is a new result in the exact WKB method and we believe it is
valid in a more general context.

In \cite{GMN}, a WKB analysis of the Hitchin systems corresponding to
circle compactifications of undeformed SU$(2)$ gauge theories was
undertaken. Our work may be viewed as an alternative route to the WKB analysis, which is closely
related to \cite{GMN} at zeroth order in $\epsilon_1$.

Our broader goal is to communicate the extreme generality of the correspondence between $\epsilon_1$-deformed ${\cal N}=2$ 
gauge theories --- specifically, their instanton partition functions with surface operator insertions ---
and certain Schr\"odinger equations amenable to exact WKB analysis. As a first step, we show the extent
to which the program applied to pure ${\cal N}=2$ super Yang-Mills in \cite{Kashani-Poor:2015pca} generalizes to theories with matter.

We will now briefly present the structure of our paper. In section
\ref{CFT}, we present a derivation of the null vector decoupling
equation satisfied by the five-point conformal block with a light
degenerate insertion, which has a null vector at level two. We apply
the collision procedure of \cite{Gaiotto:2009ma} to produce
irregular conformal blocks and derive the null vector decoupling
equations satisfied by the limit blocks. We then consider the
semi-classical limit (of infinite central charge) of these
differential equations. These equations will be the starting point for
the exact WKB analysis of section \ref{exactWKB}. In this section, we
briefly review the exact WKB approach, and in section
\ref{monodromygroup} apply it to the calculation of the monodromy
groups of our differential equations. 
We make contact with the standard undeformed Seiberg-Witten perspective in
section \ref{gaugetheory} and end with comments and future directions
for work in section \ref{conclusions}. The appendices collect details
regarding the derivation of the $\epsilon_2$-exact differential
equations for the asymptotically free theories, and an independent
check of the semi-classical differential equations via the
saddle-point analyses of Nekrasov partition functions
\cite{NPS13}.

\section{The Conformal Field Theory Perspective}
\label{CFT}

In this section, we present the null vector decoupling equation
satisfied by the five-point conformal block with one degenerate
operator insertion. We then list the corresponding equations satisfied
by irregular blocks that arise when punctures collide \cite{Gaiotto:2009ma}. We  study these equations within the framework of conformal field theory, and finally, exploit the fact that these conformal
blocks also capture the $\epsilon_i$-deformed instanton partition function of ${\cal N}=2$ supersymmetric gauge theories in four dimensions with SU$(2)$ gauge group and a varying number of flavours \cite{Nekrasov:2002qd}. We thus lay the groundwork for further analysis of these partition functions, which will be non-perturbative in the deformation parameter $\epsilon_1$.
For completeness, we provide the details of the derivation of all these equations in appendix \ref{diffeqns}.

We start our analysis by considering regular conformal blocks with four
ordinary primary operator insertions on the sphere and one degenerate operator insertion with a null vector at level two, which remains light in the limit of large central charge. On the gauge theory side of the $2$d/$4$d correspondence, this set-up corresponds to the conformal $N_f=4$ case.
To get asymptotically free (lower $N_f$) theories, we sequentially collide primary operators on the sphere in such a way that they
generate irregular conformal blocks \cite{Gaiotto:2009ma}. The case of three flavours will correspond to one irregular block, the case
of two flavours can correspond to either one or two irregular blocks, while a lower number of flavours corresponds to 
two irregular blocks in the conformal field theory. For all these collision limits, we give the corresponding
null vector decoupling equations. 

\subsection{The Five-Point Block}

We study a conformal field theory with central charge
\begin{equation}
c = 1 + 6 Q^2 \, , \qquad \mbox{where} \qquad Q = b + b^{-1} \qquad \mbox{and} \qquad b=\sqrt{\frac{\epsilon_2}{\epsilon_1}} \, .
\label{Q}
\end{equation} 
We consider a five-point chiral conformal block $\Psi$ with four primary operator insertions $V_{\alpha_i}$ 
and an insertion of a degenerate field $\Phi_{2,1}(z)$ of the Virasoro algebra \cite{Alday:2009fs}:
\be
\Psi(z_i, z) = \Big\langle 
\Phi_{2,1}(z)\!:\prod_{i=1}^{4}V_{\alpha_i}(z_i) \!: \Big\rangle \, .
\label{psi}
\ee
The degenerate field $\Phi_{2,1}$ has conformal dimension $\Delta_{2,1}$
\be
\Delta_{2,1}=-\frac{1}{2} - \frac{3}{4}\, \frac{\ed}{\eu} \, ,
\label{del}
\ee
while the conformal dimensions of the generic primaries are denoted $\Delta_{\alpha_i}$. We have chosen the degenerate insertion 
such that it remains light in the limit of large central charge $\epsilon_2/\epsilon_1 \rightarrow 0$.
The degenerate field $\Phi_{2,1}$ has a null vector at level two, and consequently satisfies the null vector condition
\be
\frac{\eu}{\ed}\, \partial^2 
\Phi_{2,1}(z)
+\, :\! T(z) \Phi_{2,1}(z)\!:\,\ = 0~ ,
\label{nuv}
\ee
where the operator $T(z)$ is the holomorphic stress tensor of the conformal field theory. Using the operator
product expansion between the stress tensor and the primary fields, the second term can be written as:
\begin{equation}
\begin{aligned}
\Big\langle\!\! :\!T(z) \Phi_{2,1}(z)\!:\prod_{i=1}^{4}V_{\alpha_i}(z_i)\Big\rangle &=
\sum_{i=1}^{4}\left(\frac{\Delta_{\alpha_i}}{(z-z_i)^2}+\frac{1}{z-z_i}
\frac{\p}{\p z_i} \right)\,\Big\langle \Phi_{2,1}(z)\prod_{i=1}^{4}V_{\alpha_i}(z_i)\Big\rangle~.
\end{aligned}
\label{tpsi}
\end{equation}
Imposing global conformal invariance allows us to express the 
derivatives with respect to $z_1, z_3$ and $z_4$ in terms of the derivatives at $z_2$ and $z$. 
Then, setting the insertions to be at $(z,0,q,1,\infty)$, the null vector decoupling equation takes the form
\begin{align}
\label{tpsi2}
\nonumber \left[\frac{\eu}{\ed}\,\frac{\p^2}{\p z^2}\, + \left(\frac{\Delta_{\alpha_2}}{(z-q)^2}+\frac{q(q-1)}{z(z-1)(z-q)}\frac{\p}{\p q}\right) - \frac{2z-1}{z(z-1)}\frac{\p}{\p z} +\frac{\Delta_{\alpha_1}}{z^2}+ \frac{\Delta_{\alpha_{3}}}{(z-1)^2} \right. &\\
\left. -\frac{\Delta_{2,1}+\Delta_{\alpha_1} +\Delta_{\alpha_2} + \Delta_{\alpha_{3}} -\Delta_{\alpha_{4}} }{z(z-1)} \right] & \Psi(z, q) = 0
\end{align}
The null vector decoupling on the five point conformal block was also studied in \cite{Alday:2009fs,Kashani-Poor:2013oza}.
The conformal dimensions $\Delta_i$ of the primary fields
$V_{\alpha_i}$ can be written in terms of the momenta $\alpha_i$ as
 \begin{equation}
 \Delta_{\alpha_i}=\alpha_i(Q-\alpha_i)~.
 \label{deltaalpha}
 \end{equation}
We further parameterize the  momenta $\alpha_i$ in terms of the four  masses $m_i$:
\begin{equation}
\begin{aligned}
\alpha_1&=\frac{Q}{2}+\frac{m_1-m_2}{2\sqrt{\eu\ed}}~,\qquad
\alpha_2=\frac{Q}{2}+\frac{m_1+m_2}{2\sqrt{\eu\ed}}~,\\
\alpha_{3}&=\frac{Q}{2}-\frac{m_3+m_4}{2\sqrt{\eu\ed}}~,\qquad
\alpha_{4}=\frac{Q}{2}-\frac{m_3-m_4}{2\sqrt{\eu\ed}}~.
\end{aligned}
\label{alphas}
\end{equation}
As a function of the masses, the conformal dimensions are
 \begin{equation}
 \begin{aligned}
 \Delta_{\alpha_1}&=\frac{(\eu+\ed)^2-(m_1-m_2)^2}{4\eu\ed}~,\qquad
 \Delta_{\alpha_2}=\frac{(\eu+\ed)^2-(m_1+m_2)^2}{4\eu\ed}~,\\
 \Delta_{\alpha_{3}}&=\frac{(\eu+\ed)^2-(m_3+m_4)^2}{4\eu\ed}~,\qquad
 \Delta_{\alpha_{4}}=\frac{(\eu+\ed)^2-(m_3-m_4)^2}{4\eu\ed}~.
 \end{aligned}
 \label{Deltas}
 \end{equation}
In terms of these variables that are appropriate for comparison to the four dimensional gauge theory, the null vector decoupling equation for the $N_f=4$ theory takes the following form:

\begin{align}
\nonumber &\left[ -\epsilon_1^2\frac{\p^2}{\p z^2} +\frac{(m_1-m_2)^2}{4z^2} + \frac{(m_1+m_2)^2}{4(z-q)^2} +\frac{(m_3+m_4)^2}{4(z-1)^2} + \frac{m_1^2+m_2^2+2m_3m_4}{2z(1-z)} \right. \\ 
\nonumber &\quad -\epsilon_1^2 \left( \frac{q^2-2 q z+z^2 \left(z^2-2 z+2\right)}{4 (z-1)^2 z^2 (q-z)^2} \right) + \epsilon_1\epsilon_2\left(\frac{q(1-q)}{z(z-1)(z-q)}\frac{\p}{\p q} +\frac{2z-1}{z(z-1)}\frac{\p}{\p z}\right.\\
\nonumber &\quad \quad \left. +\frac{ q^2 \left(-z^2+z-1\right)+2 q z \left(z^2-z+1\right)+z^2 \left(-2 z^2+3 z-2\right)}{2 (z-1)^2 z^2 (q-z)^2}\right) \\
&\quad \left. +\epsilon_2^2 \left(\frac{q^2 \left(-3 z^2+3 z-1\right)+2 q z \left(3 z^2-3 z+1\right)+z^2 \left(-4 z^2+5 z-2\right)}{4 (z-1)^2 z^2 (q-z)^2}\right) \right] \Psi(z,q) = 0\,.
\end{align}

\subsection{The Null Vector Decoupling Equations for Irregular Blocks}

We now take limits of the five-point null vector decoupling equation
(\ref{tpsi2}) in which various primary operators $V_{\alpha_i}$
collide to form irregular conformal blocks of order one
\cite{Gaiotto:2009ma}. These limiting configurations are in direct
correspondence with the $\epsilon_i$-deformed SU$(2)$ gauge theories
with $N_f <4$. We list below the null vector decoupling equations for each
of these cases and refer to appendix \ref{diffeqns} for a
detailed derivation. A summary of these equations can also be found
in \cite{Awata:2010bz}.

\paragraph{$N_f=3:$} In this case, we have one irregular block of order
one with a fourth order pole at $z=0$. In the gauge theory variables,
we take $q\rightarrow 0$ and $m_2\rightarrow\infty$, keeping the dynamical scale
$\Lambda_3 = q \, m_2$ finite. The resulting differential equation is:

\begin{align}
\nonumber &\left[-\epsilon_1^2 \frac{\p^2}{\p z^2} +\frac{(m_3+m_4)^2}{4(z-1)^2}+ \frac{m_3 m_4}{z(1-z)} +\frac{m_1\Lambda_3}{z^3}+\frac{\Lambda_3^2}{4z^4} +\epsilon_1\epsilon_2\left(\frac{1-2z}{z\left(1-z\right)}\frac{\p}{\p z} + \frac{1-2z}{2z(z-1)^2} \right)\right. \\
\nonumber &\left. \ +\frac{1}{z^2\left(1-z\right)}\left(-\epsilon_1\epsilon_2\Lambda_3\frac{\p}{\p\Lambda_3}+ m_1^2+m_1(\epsilon_1+\epsilon_2) \right) -\frac{\epsilon_1^2}{4(z-1)^2} +\epsilon_2^2\frac{(3-4z)}{4z(z-1)^2} \right]\Psi_3(z,\Lambda_3) = 0 \, .
\end{align}

\paragraph{$N_f=2:$} There are two ways to reach the case
with two flavours from the case with three flavours. One could
decouple either the flavour with mass $m_1$ or one of those with
masses $m_{3,4}$. As shown in \cite{GMN}, these lead to inequivalent
Hitchin systems and give rise to distinct differential equations.

Let us first consider  the irregular block of order one with a third
order pole at $z=0$. This corresponds to decoupling $m_1$. We refer to this as the
asymmetric configuration and the associated null vector decoupling equation becomes:
\begin{multline}
\left[-\epsilon_1^2 \frac{\p^2}{\p z^2} +\frac{(m_3+m_4)^2}{4(z-1)^2}+ \frac{m_3 m_4}{z(1-z)} +\frac{\Lambda_2^2}{z^3} -\frac{\epsilon_1\epsilon_2}{2z^2(1-z)}\Lambda_2\frac{\p}{\p\Lambda_2}
\right.\cr
\left.+\epsilon_1\epsilon_2\left(\frac{1-2z}{z(1-z)}\frac{\p}{\p z} + \frac{1-2z}{2z(z-1)^2} \right)-\frac{\epsilon_1^2}{4(z-1)^2} +\epsilon_2^2\frac{(3-4z)}{4z(z-1)^2} \right]\Psi_{2,A}(z,\Lambda_2) = 0
\end{multline}

Alternatively, one can consider two irregular blocks of order one,
with equal fourth order poles. This corresponds to decoupling $m_3$
while keeping $m_1$ and $m_4$ finite. We refer to this as the
symmetric configuration and the associated null vector decoupling
equation reads:

\begin{align}
\nonumber &\left[-\epsilon_1^2 \frac{\p^2}{\p z^2} +\frac{\Lambda_2^2}{4z^4}  + \frac{\Lambda_2m_1}{z^3(z-1)^2}-\frac{\Lambda_2 m_4}{z(z-1)^3} +\frac{\Lambda_2^2}{4(z-1)^4} +\frac{2-3z}{4z(z-1)^2}(2\epsilon_1\epsilon_2+3\epsilon_2^2)\right. \\
\nonumber & \left. \ +\frac{1}{z^2(z-1)^2}\left(-\epsilon_1\epsilon_2\Lambda_2\frac{\p}{\p\Lambda_2} -2\Lambda_2m_1+ m_1^2+m_1(\epsilon_1+\epsilon_2)\right) +\epsilon_1\epsilon_2 \frac{3z-1}{z(z-1)}\frac{\p}{\p z} \right]\Psi_{2,S}(z, \Lambda_2) = 0\,.  
\end{align}

\paragraph{$N_f=1:$} We consider two irregular blocks of order one with one fourth order pole and one third order pole. This corresponds to decoupling $m_4$ and the null vector decoupling equation takes the form

\begin{align}
\nonumber \Bigg[-\epsilon_1^2 \frac{\p^2}{\p z^2} &+\frac{\Lambda_1^2}{4z^4}  + \frac{\Lambda_1m_1}{z^3(z-1)^2}-\frac{\Lambda_1^2}{4z(z-1)^3} 
+\frac{2-3z}{4z(z-1)^2}(2\epsilon_1\epsilon_2+3\epsilon_2^2) +\epsilon_1\epsilon_2 \frac{3z-1}{z(z-1)}\frac{\p}{\p z}
\\
& \quad +\frac{1}{z^2(z-1)^2}(-\epsilon_1\epsilon_2\Lambda_1\frac{\p}{\p\Lambda_1} -2\Lambda_1m_1+ m_1^2+m_1(\epsilon_1+\epsilon_2))  
\Bigg]\Psi_1(z, \Lambda_1) = 0\,.
\end{align}

\paragraph{$N_f=0:$} Finally, we consider the case with two irregular blocks of order one with equal third order poles. All masses have been decoupled and the null vector decoupling equation becomes
\begin{multline}
\left[-\epsilon_1^2 \frac{\p^2}{\p z^2}  + \frac{\Lambda_0^2}{z^3(z-1)^2} +\frac{1}{z^2(z-1)^2} \left(-\frac{1}{2}\epsilon_1\epsilon_2\Lambda_0\frac{\p}{\p\Lambda_0 }- 2\Lambda_0^2\right)
+ \frac{\Lambda_0^2}{z(z-1)^3}
\right.\cr
\left. +\epsilon_1\epsilon_2 \frac{3z-1}{z(z-1)}\frac{\p}{\p z} +\frac{2-3z}{4z(z-1)^2}(2\epsilon_1\epsilon_2+3\epsilon_2^2)
\right]\Psi_0(z, \Lambda_0) = 0\,.
\end{multline}
This completes the list of six differential equations that we refer to throughout.


\subsection{The Semi-Classical Limit}
\label{semiclassical:subsec}
In the rest of our paper, we will concentrate on the limit $\epsilon_2/\epsilon_1 \rightarrow 0$, which is a large central charge
limit. We  keep the ratio of the mass parameters $m_i$ and the deformation parameter $\epsilon_1$
fixed. In this limit, the primary insertions $V_{\alpha_i}$ are heavy, while the degenerate insertion $\Phi_{2,1}$ is light. Thus, in this limit,
the differential equation (\ref{tpsi2}) simplifies, and we can drop the term proportional to $\partial_z$, while the terms proportional
to the conformal dimensions $\Delta_{\alpha_i}$ grow large. 
To simplify the equation further, we must specify the leading
dependence of the $q$-derivative of the five-point block on $\epsilon_2$. To that end, we make the semi-classical $\epsilon_2\rightarrow 0$ ansatz
\be\label{semiclassicalF}
\Psi(z,q) = 
 \text{exp}\left(-\frac{\widetilde{F}(q, m_i,\epsilon_i)}{\epsilon_1\epsilon_2}\right)  \, \psi(z,q) \, . 
\ee
We suppose that the $q$-derivative of the remaining function $\psi(z,q)$ is sub-dominant in the small $\epsilon_2/\epsilon_1$ limit,
and observe that the leading dependence in $\epsilon_2$ is only on the cross-ratio $q$ of the heavy operators.
We then define the quantity
\begin{eqnarray}
\widetilde{u} &=& q (1-q) \partial_q \widetilde{F} \, .
\end{eqnarray}
The parameter $\widetilde{u}$ is identified with the Coulomb modulus of the gauge
theory up to shifts that depend on the masses. Substituting this parameterization
into the null vector decoupling equation and taking the semi-classical
limit $\epsilon_2 \rightarrow 0$ leads to the Schr\"odinger
equation
\begin{equation}
\label{diffeqngeneral1}
\left( -\epsilon_1^2 \frac{\text{d}^2}{\text{d}z^2} + Q(z,\epsilon_1) \right) \psi(z,q) = 0 \, ,
\end{equation}
where the potential function $Q$ has an $\epsilon_1$ expansion which
terminates at second order
\begin{equation}
\label{diffeqngeneral2}
Q(z) = Q_0 (z) + \epsilon_1 \ Q_1 (z) + \epsilon_1^2 \ Q_2 (z) \, .
\end{equation}
The coefficient functions are
\begin{align}\label{nullNf4Qfunctions}
Q_0(z) &= -\frac{\widetilde{u}}{z(z-1)(z-q)} + \frac{(m_1-m_2)^2}{4z^2} + \frac{(m_1+m_2)^2}{4(z-q)^2} +\frac{(m_3+m_4)^2}{4(z-1)^2}+ \frac{m_1^2 + m_2^2 + 2 m_3 m_4}{2 z (1- z)} \, ,
\cr
Q_1(z)&= 0 \, , \cr  
Q_2(z) &= -\frac{1}{4z^2}-\frac{1}{4(z-1)^2}-\frac{1}{4(z-q)^2}+\frac{1}{2z(z-1)} \, . 
\end{align}
%

\subsection{The Semi-Classical Irregular Blocks}
\label{semiclassicalIrregular:subsec}

The same type of  ansatz \eqref{semiclassicalF} can be used in order to obtain the differential equations for the irregular blocks in the
semi-classical  $\epsilon_2\rightarrow 0$ limit. The variable parameterizing the Coulomb modulus is now defined as
\be
\widetilde{u} = \Lambda_{N_f}\frac{\p \widetilde{F}}{\p \Lambda_{N_f}} \,,
\ee
where $\Lambda_{N_f}$ is the corresponding strong coupling scale of the $N_f < 4$ gauge theory. As in the conformal case, 
the prepotential of the gauge theory will differ mildly from $\widetilde{F}$. 
However, what is of importance to us is the pole structure of the functions $Q_k(z)$, and we choose a parameterization
that descends naturally from the conformal theory and that allows for a simple presentation of the differential equations.
In the following, we present all the asymptotically free cases:
\begin{itemize}

\item {$N_f=3$}: The Schr\"odinger equation which governs the $\epsilon_1$-deformed gauge theory is given by 
\begin{align}
\label{Nf=3 semiclassical}
\left[-\epsilon_1^2 \frac{\p^2}{\p z^2} +\frac{(m_3+m_4)^2}{4(z-1)^2}+ \frac{m_3 m_4}{z(1-z)} +\frac{m_1\Lambda_3}{z^3}+\frac{\Lambda_3^2}{4z^4} +\frac{\widetilde{u}}{z^2(1-z)}
-\frac{\epsilon_1^2}{4(z-1)^2} 
\right]\psi_3(z,\Lambda_3) 
= 0 \,.
\end{align}

\item {$N_f=2$ (asymmetric  realization)}: The differential equation in the semi-classical limit takes the form
\be
\label{Nf=2 asymmetric semiclassical}
\left[-\epsilon_1^2 \frac{\p^2}{\p z^2} +\frac{(m_3+m_4)^2}{4(z-1)^2}+ \frac{m_3 m_4}{z(1-z)} +\frac{\Lambda_2^2}{z^3} +\frac{\widetilde{u}}{z^2(1-z)}
-\frac{\epsilon_1^2}{4(z-1)^2} 
\right]\psi_{2,A}(z,\Lambda_2) = 0
\ee
%

\item {$N_f=2$ (symmetric  realization)}:
\be
\label{Nf=2 symmetric semiclassical}
\left[-\epsilon_1^2 \frac{\p^2}{\p z^2} +\frac{\Lambda_2^2}{4z^4}  + \frac{\Lambda_2m_1}{z^3(z-1)^2}-\frac{\Lambda_2 m_4}{z(z-1)^3} +\frac{\Lambda_2^2}{4(z-1)^4}
+\frac{\widetilde{u}}{z^2(z-1)^2}
\right]\psi_{2,S}(z, \Lambda_2) = 0\,.
\ee
\item {$N_f=1$}:
\be
\label{Nf=1 semiclassical}
\left[-\epsilon_1^2 \frac{\p^2}{\p z^2} +\frac{\Lambda_1^2}{4z^4}  + \frac{\Lambda_1m_1}{z^3(z-1)^2}-\frac{\Lambda_1^2}{4z(z-1)^3} 
+\frac{\widetilde{u}}{z^2(z-1)^2}
\right]\psi_1(z, \Lambda_1) = 0\,.
\ee

\item {$N_f=0$}: Finally, for the pure super Yang-Mills theory, the  equation reads 
\be
\label{PureNVD}
\left[-\epsilon_1^2 \frac{\p^2}{\p z^2}  
+\frac{\Lambda_0^2}{z^3(z-1)^2} +\frac{\widetilde{u}}{z^2(z-1)^2}  
+ \frac{\Lambda_0^2}{z(z-1)^3}
\right]\psi_0(z, \Lambda_0) = 0\,.
\ee
\end{itemize}
We have thus obtained the  differential equations which we analyze in detail in section \ref{monodromygroup}.

\section{The Exact WKB Analysis of Differential Equations}
\label{sec:IntroExactWKB}
\label{exactWKB}

In this section, we review the exact WKB approach to the analysis of
differential equations and apply it to the
null vector decoupling equations in the semi-classical
limit. We will carry out the  exact WKB analysis
with respect to the  small parameter $\epsilon_1$. Our analysis
will therefore be valid to zeroth order in $\epsilon_2$ and non-perturbatively
in $\epsilon_1$. Below, we briefly review the salient features of the exact WKB analysis and refer the reader to \cite{book, NakanishiandIwaki} for a more comprehensive treatment
of the same.

\subsection{The Exact WKB Method}

The differential equations that we study can be written in the form of
a Schr\"odinger equation: 
\begin{equation}
\left( -\epsilon_1^2 \frac{\text{d}^2}{\text{d}x^2} + Q(x) \right)
\psi(x,\epsilon_1) = 0 \, .
\end{equation}
We allow the function $Q$ to have an expansion of the form
\begin{equation}
Q(x) = Q_0 (x) + \epsilon_1 \ Q_1 (x) + \epsilon_1^2 \ Q_2 (x) + \cdots \, .
\end{equation}
For the null vector
decoupling equations that we study, the only non-zero coefficient functions are
$Q_0,Q_1$ and $Q_2$. We choose a WKB ansatz for the solution to this
differential equation, which takes the form
\begin{equation}
\psi (x,\epsilon_1) = \text{exp} \left( \int_{x_0}^{x} \text{d}x' \ S(x',\epsilon_1) \right) \, ,
\end{equation}
with $S(x,\epsilon_1)$ expanded as a formal power series in
$\epsilon_1$ as
\begin{equation}
S(x,\epsilon_1) = \frac{1}{\epsilon_1} \ S_{-1} (x) + S_0 (x) + \epsilon_1 \ S_1 (x) + \cdots \, .
\end{equation}
Substituting this ansatz into the differential equation, we get
recursion relations governing the coefficients $S_k$
\begin{align}
S_{-1}^2 &= Q_0 \, , \\
2 S_{-1} S_{n+1} + \sum_{k+l = n} S_k S_l +
  \frac{\text{d}S_n}{\text{d}x} &= Q_{n+2} \quad \text{for} \quad n
                                  \geq -1 \, .
\end{align}
We see that the initial conditions governing the system of recursion
relations allow for two possible sets of solutions to these recursion
relations, as $S_{-1} = \pm \sqrt{ Q_0} $. We also note the crucial
feature that the zeroes of $Q_0$, which we call turning points,
introduce branch cuts on the Riemann surface $\Sigma$ on which our
differential equation and its exact solutions live. Thus, in our exact
WKB treatment, we introduce a new manifold $\hat{\Sigma}$, which is a
double cover of the Riemann surface, and we move between sheets as we
pass branch cuts that emanate from turning points, or odd order poles.
From hereon, we will distinguish the choice of WKB solution by
attaching to it the subscript ($\pm$). We also observe that in the $\epsilon_1$-expansion of
$S(x,\epsilon_1)$, the sets of odd and even coefficients are
dependent. If we define
\begin{equation}
S_{\text{odd}} = \sum_{j \geq 0} S_{2j-1} \; \epsilon_1^{2j-1} \quad
\text{and} \quad S_{\text{even}} = \sum_{j \geq 0} S_{2j} \;
\epsilon_1^{2j} \, ,
\end{equation}
we have the relation
\begin{equation}
S_{\text{even}} = -\frac{1}{2} \frac{\text{d}}{\text{d}x} \log S_{\text{odd}} \, .
\end{equation}
Putting all this together, we can write down a formal expression for
the two linearly independent solutions to our differential equation:
\begin{equation}
\label{eq:WKBSoln}
\psi_{\pm} = \frac{1}{\sqrt{S_{\text{odd}}}} \ \text{exp} \left\lbrace
  \pm \int_{x_0}^{x} \text{d}x' \ S_{\text{odd}} \right\rbrace \, .
\end{equation}
This formal expression should be understood as an analytic function of
$x$ multiplying an asymptotic series in $\epsilon_1$\ :
\begin{equation}
\psi_{\pm} = \text{exp} \left\lbrace \pm \frac{1}{\epsilon_1}
  \int_{x_0}^{x} \text{d}x' \ \sqrt{Q_0 (x')} \right\rbrace
\epsilon_1^{1/2} \sum_{k=0}^{\infty} \epsilon_1^k \ \psi_{\pm,k} (x)
\, .
\end{equation}

\subsubsection*{Borel Resummation}

In the exact WKB approach, it is convenient to normalize
wave-functions at distinguished points of the differential
equation. As mentioned earlier, in addition to the
singularities of the coefficient functions of the differential
equations, their zeros (turning points) also play an important role. We will normalize our solutions with respect to
the turning points, i.e. choose the starting point $x_0$ of the
integration path to be a turning point $t$,
\begin{equation}
\psi_{\pm} = \frac{1}{\sqrt{S_{\text{odd}}}} \ \text{exp} \left\lbrace
  \pm \int_{t}^{x} \text{d}x' \ S_{\text{odd}} \right\rbrace \, .
\end{equation}
Formal WKB solutions are generically
divergent. To remedy this, we invoke Borel resummation: a technique
that constructs an analytic function whose asymptotic expansion
matches the formal WKB series. The Borel transformed series is defined as
\begin{equation}
\psi (\epsilon_1) = \sum_{k=0}^{\infty} \psi_k \ \epsilon_1^k \quad
\xrightarrow{\text{Borel transform}} \quad  \widetilde{\psi}(y) =
\sum_{k=1}^{\infty} \psi_k \frac{y^{k-1}}{(k-1)!} \, .
\end{equation}
Next, define the function \cite{NakanishiandIwaki}
\begin{equation}
\Psi(\epsilon_1) = \psi_0 + \int_{\ell_\theta}
\text{d}y \ e^{-y/\epsilon_1} \widetilde{\psi}(y) \, ,
\end{equation}
where $\ell_\theta$ is the line connecting a point at which the series
$\widetilde{\psi}(y)$ converges\footnote{To be precise, this is true
  for Gevrey-1 series, which in our context corresponds to the
  following statement. If $\psi_k$ is the $k$th coefficient of the
  asymptotic series then the series is Gevrey-1 type if growth of
  $\psi_k$ is bounded by $\psi_k \leq A B^k k!$ for some constants $A$
  and $B$. If $\psi_k$ is a function of a continuous variable, say,
  $x\in \C$ then
  this condition applies to the supremum of $\psi_k(x)$ in a compact
  subset of $\C$.} --- typically, a turning point --- to the point at infinity at an angle $\theta$. If this integral exists, $\Psi(\epsilon_1)$ is the requisite analytic function, called the Borel sum. 

Notice that the Borel sum contains an angular dependence. In order to
understand this better, one must appreciate that Borel sums are
typically defined only in regions of the complex $\epsilon_1$-plane,
and not throughout. These regions are bounded by Stokes
lines, defined by the condition
\begin{equation}
\text{Im} \left[ \int_{x_0}^{x} \text{d}x' \sqrt{Q_0 (x')} \right] = 0 \, ,
\end{equation}
and different Stokes regions are assigned different linear
combinations of a given basis of analytic solutions to the
differential equation, arrived at via Borel resummation. One of the
key components of the exact WKB analysis is understanding how
solutions in different Stokes regions are related by analytic continuation; these often go by
the name of ``connection formulae''. However, before we address this
transition behaviour, we will find it necessary to endow Stokes lines
with an orientation. To this end, we adopt the convention that Stokes
lines are oriented \emph{away} (i.e. the arrow on the Stokes
line is pointing away from a turning point) if
\begin{equation}
\text{Re} \left[ \int_{x_0}^{x} \text{d}x' \sqrt{Q_0 (x')} \right] > 0 \, 
\end{equation}
along the Stokes line. Else, the arrow points towards the turning point.

Three Stokes lines emanate from a first order zero of $Q_0(x)$, which
is also referred to as a simple turning point. Thus one end of any
Stokes line is at a turning point. The other end can either be at a
singularity or at a turning point. When both the end points of a
Stokes line in a given Stokes graph terminate at turning points then
the corresponding Stokes graph is called ``critical''.\footnote{The general behaviour of Stokes lines is discussed in \cite{NakanishiandIwaki}. We restrict ourselves to situations that are relevant
  in this work.}

\subsubsection*{Connection Formulae}

We are now in a position to state the connection formulae. For a Stokes graph which is not critical, consider two
regions $U_{1}$ and $U_{2}$ separated by a Stokes curve $\Gamma$, and
consider $\Psi_{\pm}^{j}$ to be the Borel sums of WKB solutions in
each of the regions $U_j$. The connection formulae for the Borel sums
in different Stokes domains are given by:
\begin{align}
\text{if} \ \text{Re} \left[ \int_{x_0}^{x} \text{d}x' \sqrt{Q_0 (x')} \right] < 0 \ \text{on} \ \Gamma \quad : \quad
\begin{cases}
\Psi_+^1 &= \Psi_+^2 \, , \\
\Psi_-^1 &= \Psi_-^2 \pm \text{i} \Psi_+^2 \, ,
\end{cases}
\end{align}
\begin{align}
\text{if} \ \text{Re} \left[ \int_{x_0}^{x} \text{d}x' \sqrt{Q_0 (x')} \right] > 0 \ \text{on} \ \Gamma \quad : \quad
\begin{cases}
\Psi_+^1 &= \Psi_+^2 \pm \text{i} \Psi_-^2 \, , \\
\Psi_-^1 &= \Psi_-^2 \, .
\end{cases}
\end{align}
In the above connection formulae, there is an ambiguity ($\pm$) that
is fixed by noting that the turning point that $\Gamma$ originates
from serves as a point of reference. If the path of analytic
continuation crosses $\Gamma$ counter-clockwise as seen from the
turning point, we pick the ($+$) sign, and if this path crosses
$\Gamma$ clockwise, we pick the ($-$) sign. Later in this
section, we will write down the Stokes matrices that multiply
wave-functions; these are equivalent to the above result.

The global properties of solutions to the differential equations we consider are governed by the monodromy group and the
Stokes phenomena around singular points. The monodromy group of these
differential equations can be expressed entirely in terms
of two sets of quantities: (a) the characteristic exponents at each
singular point $s_k$, and (b) the contour integrals of
$S_{\text{odd}}$ around branch cuts. We now parameterize the
characteristic exponents conveniently.

\begin{figure}[ht!]
\centering
\includegraphics[width=100mm]{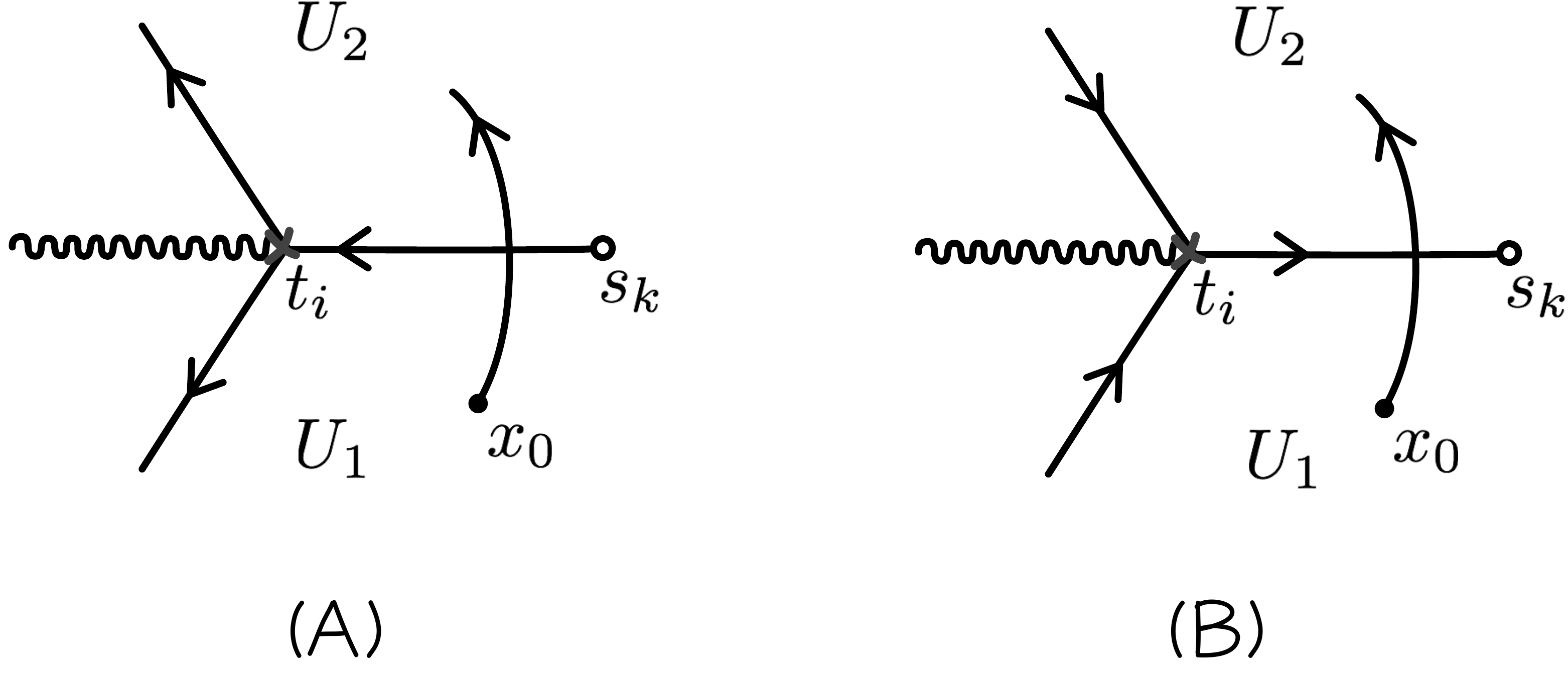}
\caption{Analytic continuation of wave-functions from $U_1$ to $U_2$}
\label{analytic_continuation}
\end{figure}

\begin{figure}[ht!]
\centering
\includegraphics[width=85mm]{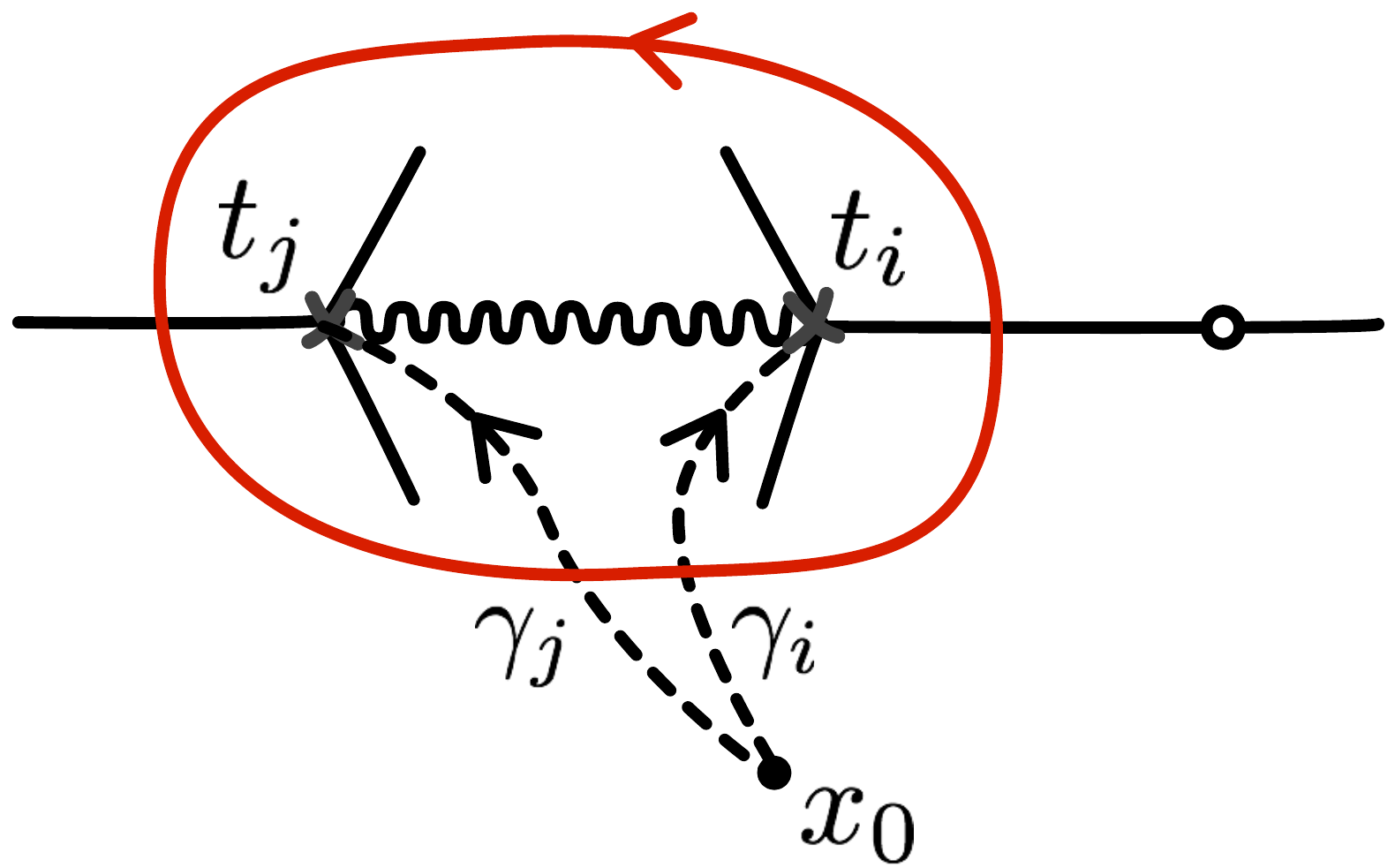}
\caption{Encircling branch cuts}
\label{branchcutcontour}
\end{figure}

As a system of solutions to our differential equation, we consider the
WKB solutions \eqref{eq:WKBSoln}, and define the characteristic
exponents as residues of the differential:
\begin{equation}
M_k = \text{Res} \ \sqrt{Q_0(x)} \Big\vert_{x=s_k} \, .
\end{equation}
From the null vector decoupling equations we derived in the
previous section, one can check that the residues $M_k$ are linear
combinations of the mass parameters of the gauge theory. As the
monodromy group computations will use WKB wave-functions \eqref{eq:WKBSoln}, we relate the
residues of $S_{\text{odd}}$ to our characteristic exponents
as\footnote{This is true under the assumptions that $\text{Re} \ M_k
  \neq 0$.}
\begin{equation}
\text{Res} \ S_{\text{odd}} (x,\epsilon) \Big\vert_{x=s_k} = \frac{M_k}{\epsilon_1} \sqrt{1+\frac{\epsilon_1^2}{4M_k^2}} \, . 
\end{equation}
Finally, upon exponentiating this contribution, we get the multiplier that affects WKB wave-functions:
\begin{equation}
\nu_k^\pm = \text{exp} \left[ \text{i}\pi \left( 1 \pm \sqrt{\frac{4M_k^2}{\epsilon_1^2}+1}\right)\right] \, .
\end{equation}
Notice that $\nu^+_k = 1/\nu_k^-$, a fact that we will use repeatedly.
Since the base point $x_0$ will not always be a turning point, the modified connection formulae can be obtained by a
composition of the contour integrals. We find it convenient to use a
matrix notation to exhibit the connection formulae. As an example, let us consider
analytically continuing the Borel resummed wave-functions from Stokes
region $U_1$ to Stokes region $U_2$. As shown in figure \ref{analytic_continuation}, there are two distinct possibilities.
If the contour crosses a Stokes line that is directed inwards to a turning point as in figure \ref{analytic_continuation}\ (A), we find the  connection formula:
\be
\begin{pmatrix}
\Psi_+^1\ , & \Psi_-^1
\end{pmatrix} \Longrightarrow
\begin{pmatrix}
\Psi_+^2\ , & \Psi_-^2
\end{pmatrix} 
\begin{pmatrix}
1 &\pm \ii u_i^{-1} \\
0& 1
\end{pmatrix} \,.
\ee
In the above equation, we use the notation,
\begin{equation}
u_j=\mbox{exp}\left(2\int_{\gamma_j} \text{d}x \ S_{\text{odd}}\right)
\, ,
\end{equation}
where $\gamma_j$ is an oriented curve from the base point to the
turning point $t_j$. Along a contour that crosses a Stokes line which
is directed outwards from a turning point as in figure \ref{analytic_continuation}\ (B), we have the
connection formula:
\be
\begin{pmatrix}
\Psi_+^1\ , & \Psi_-^1
\end{pmatrix} \Longrightarrow
\begin{pmatrix}
\Psi_+^2\ , & \Psi_-^2
\end{pmatrix} 
\begin{pmatrix}
1 &0 \\
\pm \ii u_i & 1
\end{pmatrix} \, .
\ee
In the above, the $+(-)$ sign is chosen for counter-clockwise (clockwise)
crossing of the contour from one Stokes region to the other, with respect to the turning point.
For more complicated contours, it is important to take into account
contributions from any branch cuts and/or singularities enclosed along
the closed contour from the base point to the intersection point,
the turning point and then back to the base point. As a simple example of this phenomenon, let us suppose the
contour chosen happens to encircle a branch cut --- say between $t_j$
and $t_i$ as in figure \ref{branchcutcontour} -- counter-clockwise.
Here, the curves $\gamma_i$ are those that define the parameter $u_i$.
%
%
The closed contour $\gamma_{ji}$ that encircles the branch cut has a contribution of the form
\begin{equation}
u_{ji} = \text{exp} \left( \int_{\gamma_{ji}} \text{d}x' \ S_{\text{odd}} \right) \, ,
\end{equation}
where from the figure it is clear that
\begin{equation}
u_{ji} = u_j^{-1} u_i \, .
\end{equation}
One can see that although the $u_i$ by itself is dependent on the base point, the contour integral is independent of this choice.

\subsubsection*{Contour Encircling a Turning Point}

Let us make another 
important preliminary point regarding the choice of cycles. In order to define the
monodromy group, we first choose a base point and define a basis of
closed loops that encircle just the singularities. In some of the
cases we encounter, there are branch cuts between turning
points and singularities. In such cases, we choose the contours to
also include these turning points.

\begin{figure}[ht!]
\centering
\includegraphics[width=50mm]{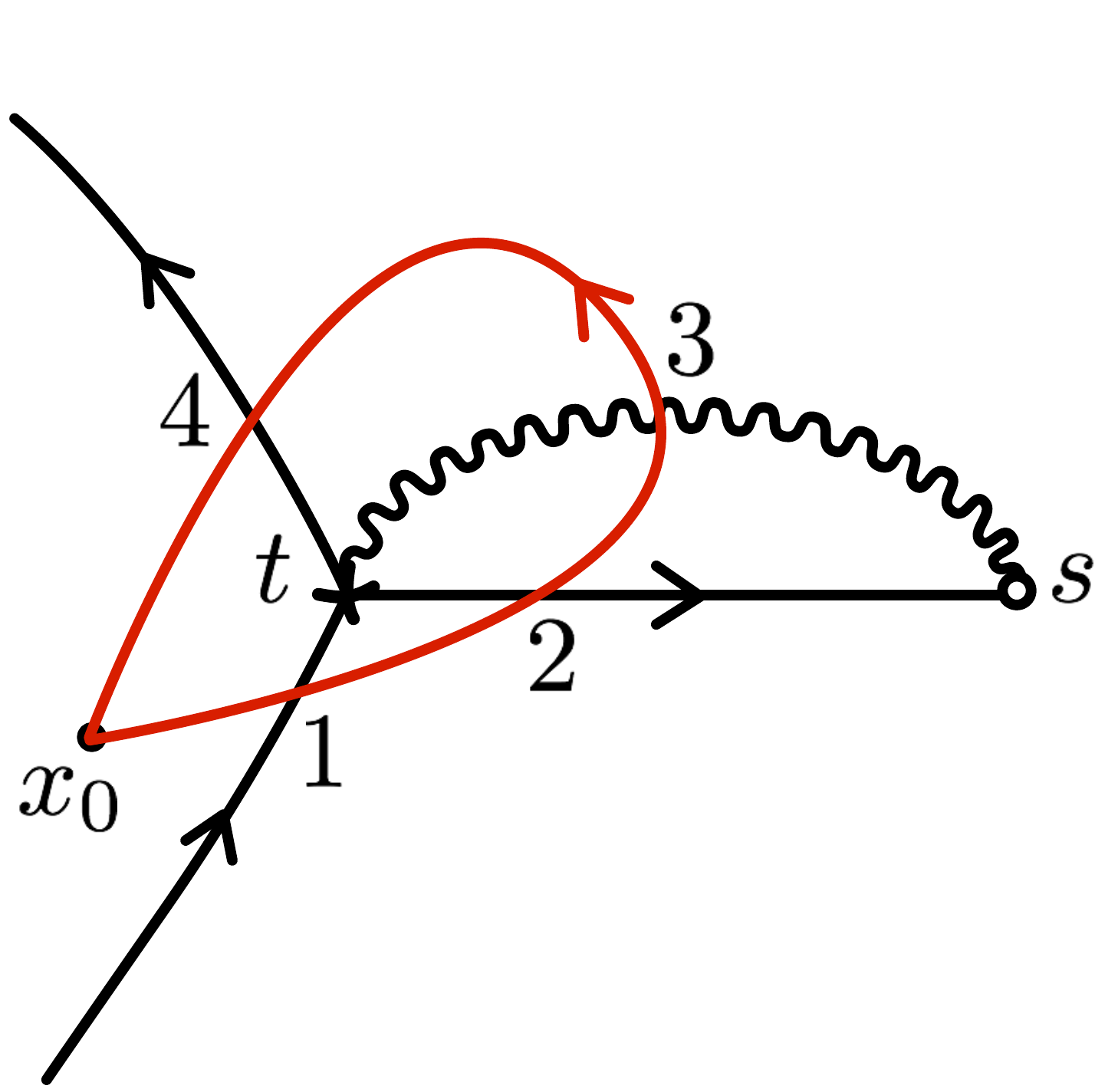}
\caption{Contour with base point $x_0$ encircling a turning point $t$}
\label{tpath}
\end{figure}

In order to prove that this is consistent with the usual definition of the monodromy group, let us consider a contour that only encircles the turning point, as shown in figure \ref{tpath}. 

If we choose to normalize the wave-functions at $x_0$, the wave-functions undergo the following transformation as we travel along the path:
\begin{align}
M_{x_0,\text{path}} &= 
\begin{pmatrix}
1 & 0 \\
\frac{\ii}{u_1} &1
\end{pmatrix}
\begin{pmatrix}
0 & -\ii \\
-\ii &0
\end{pmatrix}
\begin{pmatrix}
1 & 0 \\
\ii u_1 &1
\end{pmatrix}
\begin{pmatrix}
1 & \frac{\ii}{u_1} \\
0 &1
\end{pmatrix}\\
&=
\begin{pmatrix}
u_1 & 0 \\
0 &\frac{1}{u_1}
\end{pmatrix}
\end{align}
%
Here we have associated the matrix $-\ii\sigma_1$ to the branch-cut crossing, which ensures that we remain on the same
sheet of the Riemann surface. 
It can be easily shown that for any base point that one may choose, the answer
is trivial as above. If we
chose the turning point itself to be the base-point, $u_1=1$ and the
matrix reduces to the identity matrix. Since the net result is simply
the identity matrix, in order to calculate the monodromy matrix for
the contour that encircles the singularity $s$, one may just as well
compute the monodromy of the wave-functions around the cycle that
encircles both the turning point $t$ and the singularity $s$. 
We will make use of this
repeatedly in those cases in which the branch cut extends between a
turning point and a singularity.

It is instructive to square this situation with the solution of a
differential equation near an ordinary point. It is known that any
solution of a differential equation can be written as a Taylor series
in the neighbourhood of an ordinary point. The radius of convergence
of this solution is at least as much as the distance from the chosen point
to the nearest singularity. The Taylor series solution will clearly have trivial
monodromy property. Although the WKB analysis assigns a special status to
turning points, from the differential equation point of view the turning
point is an ordinary point. Clearly, the branch cut and the Stokes
lines emanating from a turning point are artefacts of the WKB
approximation and the insertion of the matrix $-\ii\sigma_1$ restores the
fact that the turning point is an ordinary point of the differential
equation.  

\subsubsection*{Contours Encircling a Singular Point}
\begin{figure}[ht!]
\centering
\includegraphics[width=75mm]{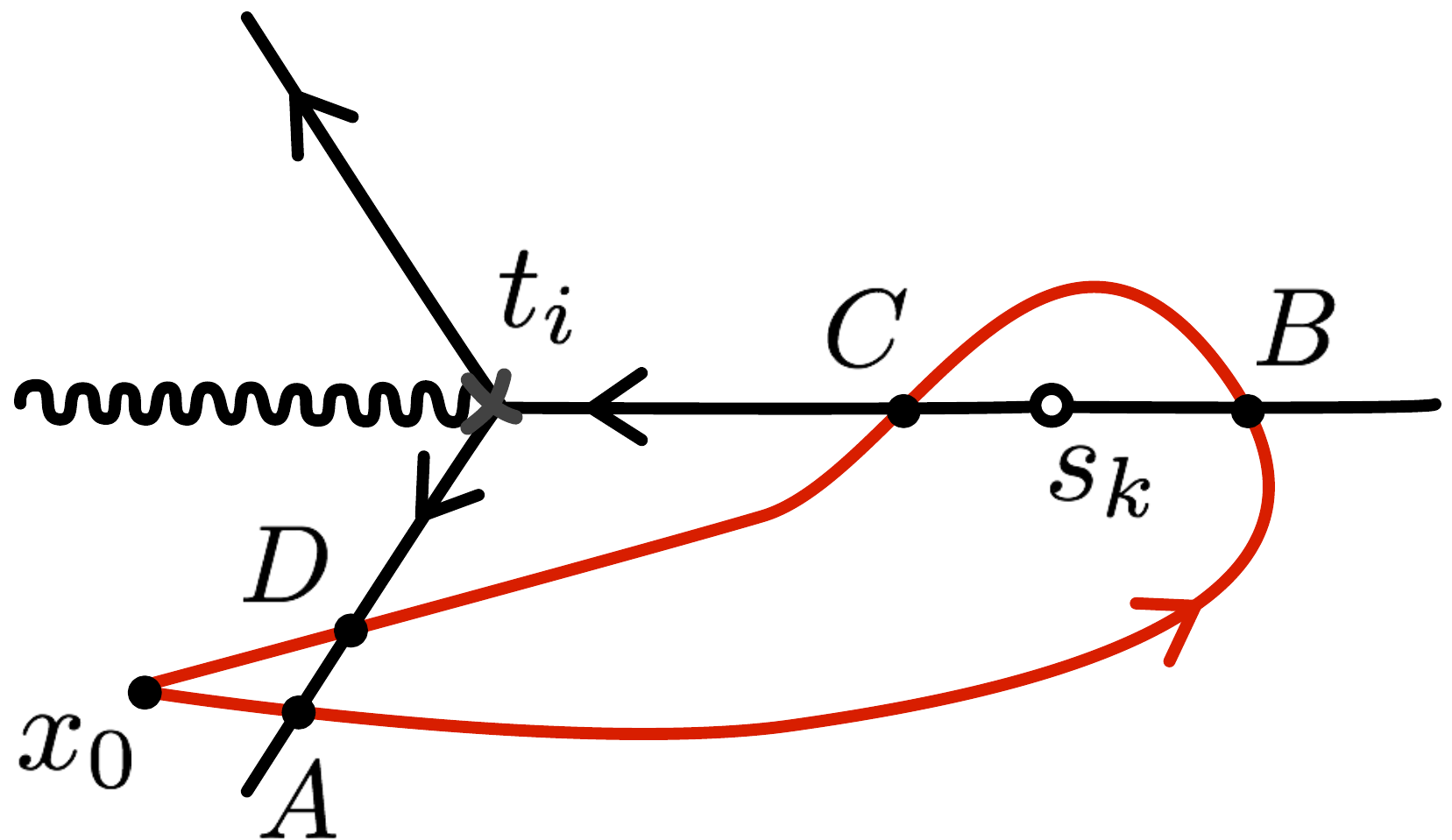}
\caption{Evaluation of Stokes matrices: effect of singularities}
\label{Stokes_example}
\end{figure}

Let us now consider the toy example, as shown in figure \ref{Stokes_example}, where the contour encloses 
a singularity.\footnote{This example will illustrate the manner in which the Stokes matrices at each intersection are written down. The Stokes lines here don't end at turning points or singularities; the reader is encouraged to think of the figure as a part of a complete Stokes graph that has been zoomed into.}

In figure \ref{Stokes_example}, at the first intersection point $A$, the contour crosses counter-clockwise a Stokes line emanating from $t_i$. 
Thus, the Stokes matrix is
\begin{equation}
\begin{pmatrix}
1 & 0 \\
+ \text{i} u_i & 0
\end{pmatrix} \, .
\end{equation} 
In order to determine the Stokes matrix at $B$, we need to know to which turning point the Stokes line is connected. 
Since this is irrelevant to the present discussion, we move on to consider the third intersection point $C$. 
This time the contour crosses a Stokes line going into $t_i$, and the crossing is clockwise as seen from $t_i$. 
Further, when this contour is completed using $\gamma_j$, we see that a singularity is encircled counter-clockwise. 
Taking this into account, the Stokes matrix is

\begin{equation}
\begin{pmatrix}
1 & -\text{i} u_i^{-1} \nu_k^{-2} \\
0 & 1
\end{pmatrix} \, .
\end{equation}
Finally, at the fourth intersection point $D$, the contour crosses the Stokes line clockwise. In fact, it is very similar to the first intersection, except that now there is a singularity encircled. Consequently, the Stokes matrix is
\begin{equation}
\begin{pmatrix}
1 & 0 \\
- \text{i} u_j \nu_k^2 & 0
\end{pmatrix} \, .
\end{equation} 
This concludes our brief review of the exact WKB analysis. We refer the reader to \cite{book, NakanishiandIwaki} for a more detailed discussion and further references.

\subsection{The Applicability of the Exact WKB Analysis}
The application of the exact WKB techniques depends on the precise
differential equation under consideration. 
Before we apply the exact WKB method to the equations derived in the
previous section, it is important to point out the subtleties in the
applicability of this analysis.
In the Schr\"odinger type differential equations listed in sections \ref{semiclassical:subsec} and \ref{semiclassicalIrregular:subsec}, the parameter $\epsilon_1$ functions as the
Planck's constant $\hbar$ in the WKB approximation scheme. For our
null vector decoupling equations, the potential has
zeroth, first and second order terms in $\epsilon_1$. In order to apply the exact WKB
techniques to the solution of the differential equation, 
the $\epsilon_1$-deformed potential must satisfy certain 
conditions. These consistency conditions not only ensure normalizability of the
wave-functions at singularities but also are useful in proving Borel
summability of the WKB wave-functions. 


The necessary conditions (eq.~(2.8) and (2.9) in  \cite{NakanishiandIwaki}) are:
\begin{itemize}
\item  If the leading coefficient $Q_0$ has a pole of order $m \ge 3$, then the order of $Q_{n \ge1}$ at 
that pole should be smaller than $1+m/2$. 
\item If the pole of $Q_0$ (at, say $z=z_0$) is of order $m=2$, then $Q_{n \neq 2}$ may have at most
a simple pole there and $Q_2$ should have a double pole\ : 
\begin{eqnarray}
Q_2 &=& - \frac{1}{4 (z-z_0)^2} (1+ O(z-z_0)) \quad \mbox{as} \quad z \rightarrow z_.
\label{condition2}
\end{eqnarray}
\end{itemize}
It is easily checked that the potentials that appear in the various
Schr\"odinger type differential equations in sections
\ref{semiclassical:subsec} and \ref{semiclassicalIrregular:subsec} satisfy these conditions.

\subsection{Theorems on Stokes Automorphisms}

Since all the equations listed in sections \ref{semiclassical:subsec}
and \ref{semiclassicalIrregular:subsec} satisfy the necessary
conditions, the theorems proved in \cite{NakanishiandIwaki} using
these conditions can be directly applied to our equations. There is
however, an interesting exception and we will comment on it
momentarily. In particular, the results of \cite{NakanishiandIwaki} include theorems
on the Stokes automorphisms that relate WKB resummed monodromies with
a given Borel resummation angle, to monodromies with another Borel
resummation angle.

We will now list the relevant results from these theorems. Consider a
closed curve $\gamma$ on the double cover $\hat{\Sigma}$ of the
Riemann surface $\Sigma$ encircling either a singularity or a turning
point. We then define the Voros symbol $ e^{V_\gamma}$ as a formal
power series using the integral
\begin{equation}
  \label{eq:voros}
  V_\gamma(\epsilon_1)=\oint_\gamma dz\ S_{\text{odd}}(z, \epsilon_1) \ .
\end{equation}
The Borel sums of the Voros symbol are then defined as
$S_\pm [ e^{V_\gamma}]$. They satisfy the Stokes automorphism
formula
\begin{eqnarray}
S_-[e^{V_\gamma}] &=& S_+ [ e^{V_\gamma}] ( 1 + S_+ [e^{V_{\gamma_0}}])^{-(\gamma_0,\gamma)}
\end{eqnarray}
whereby we suppose a simple flip, with the critical Stokes cycle being denoted by
$\gamma_0$, and $(\gamma_0,\gamma)$ is the intersection
number of the critical cycle with the cycle $\gamma$ defining the
Voros symbol. The resummations $S_\pm$ are the Borel resummations of
the Voros symbol on either side of (and close enough to) the critical
graph. The intersection numbers are defined using the 
convention that, if the cycle $\gamma_1$ has the arrow pointing outwards in
the positive $x$ direction and the cycle $\gamma_2$, which crosses
$\gamma_1$, with the arrow pointing towards the upper half-plane, then
$(\gamma_1,\gamma_2)= +1$. When we have Borel sums on either side of a
pop rather than a flip, the Voros symbols (importantly, associated to closed cycles) are
trivially related
\begin{eqnarray}
\label{stokes pops}
S_-[e^{V_\gamma}] &=& S_+ [ e^{V_\gamma}] \, .
\end{eqnarray}
These two theorems govern the transformation of Voros symbols
associated to closed cycles. In the next section we will frequently
use results of these theorems to study global properties of our
differential equations.

In the case of the conformal SU$(2)$ gauge theory (with $N_f=4$
flavours) however, the extra assumptions of \cite{NakanishiandIwaki} are
not always fully satisfied. In particular, in this case we find that pairs of Stokes
graphs that are related by a simultaneous or double flip, excluded in
\cite{NakanishiandIwaki}. When such a double flip occurs, we show
that the formulae for the Stokes automorphisms derived 
for single flips compose without change to give 
the Stokes automorphism for the double flip. This is an extension of
the results of \cite{NakanishiandIwaki}. We will discuss this case
in detail in the next section.

\section{The Monodromy Group}
\label{monodromygroup}
In this section, we study global properties of the  differential equations derived in section \ref{CFT}. The differential
equations are second order and hence have two linearly independent
global solutions. The solutions undergo a monodromy as we analytically continue them around a singular point. The monodromies,
defined up to a change of basis, form a group called the monodromy group. The monodromy
group of the differential equations we consider can be expressed
entirely in terms of two sets of quantities: (i) the characteristic
exponents $\nu_k$ at the singular points $s_k$
and (ii) the Borel resummed contour integrals of the WKB differential $S_{\text{odd}}$ around branch cuts,
which we denote by $u_{ij}$.

The connection formulae which relate the Borel
resummed wave functions in the various Stokes regions are sufficient
to completely determine the monodromy group associated to the relevant
null vector decoupling equation. The Borel resummed exact WKB contour integrals depend on the Borel resummation angle, (equivalently, 
on the phase of $\epsilon_1$) and undergo Stokes automorphisms as a function of these parameters. Thus,
the expression of the monodromy group in terms of the resummed integrals varies, and we determine the explicit
transformation rules as we pass through a critical graph. 
In this section, we calculate the monodromy groups, starting with the
simplest case of zero flavours, with no regular singular points
in the differential equation, and we end with the
conformal case ($N_f=4$) which has four regular singular points.

We stress the fact that  there is a dictionary between the Borel resummation
angle $\theta$, and the phase of the zeroth order differential which is determined by the phase of $\epsilon_1$ in
our set-up. (See e.g. \cite{NakanishiandIwaki} for the details, which follow from the definition of the Borel
sum.)
 We see that this dictionary is given a natural home in $\epsilon_1$-deformed ${\cal N}=2$ gauge theories.
The formal dependence on the Borel resummation angle that induces the Stokes automorphism, has a physical counterpart in the dependence of all non-perturbatively resummed monodromies
on the phase of the deformation parameter $\epsilon_1$.

\subsubsection*{A Brief Summary of our Analysis}

Throughout this section, we perform the calculation of the monodromy group in a strong coupling regime. In all the examples, we will plot the Stokes graphs emphasizing
the connectivity of the graphs and the choice of branch
cuts; we refer to \cite{GMN} for various possible sequences of
Stokes graphs. To illustrate the detailed coding of the monodromy group in terms of the 
characteristic exponents and the resummed monodromies, as well as the ambiguity of their formal
expression in terms of the monodromies, we calculate the
monodromy groups associated to two distinct Stokes graphs. Equating the invariants constructed
from the monodromy groups of the two graphs gives us the Stokes
automorphism relating the variables in each description. We will thus find concrete descriptions of the monodromy group, 
as well as the Stokes automorphisms that the exact WKB parameters undergo. The Stokes automorphisms must satisfy the theorems
of \cite{NakanishiandIwaki} and this fact serves as a consistency check of our analysis.

\subsection{Pure Super Yang-Mills}
\label{pureSYM}
The semi-classical null vector decoupling equation corresponding to the case  of pure super Yang-Mills theory has been discussed in detail in  \cite{Kashani-Poor:2015pca}.
The description was mostly in terms of variables that resulted after mapping the sphere onto a cylinder,
such that the differential equation became the Mathieu equation, and the monodromy group was coded
in the Floquet exponent. Below, we perform an equivalent analysis on the sphere, which will prepare
us to include flavours. A WKB analysis of the Mathieu equation can be found in \cite{HeMiao1,HeMiao2}
and further in \cite{Dunne} in the context of exact WKB and the 2d/4d correspondence.

The Stokes graph only depends on the leading potential term $Q_0(z)$. For the pure ${\cal N}=2$
super Yang-Mills theory, the zeroth order term is given by \eqref{PureNVD}
\be
Q_0(z) = 
\frac{\Lambda_0^2}{z^3(z-1)^2} +\frac{\widetilde{u}}{z^2(z-1)^2}  
+ \frac{\Lambda_0^2}{z(z-1)^3} \, .
\ee
\begin{figure}[ht!]
\centering
\includegraphics[width=170mm]{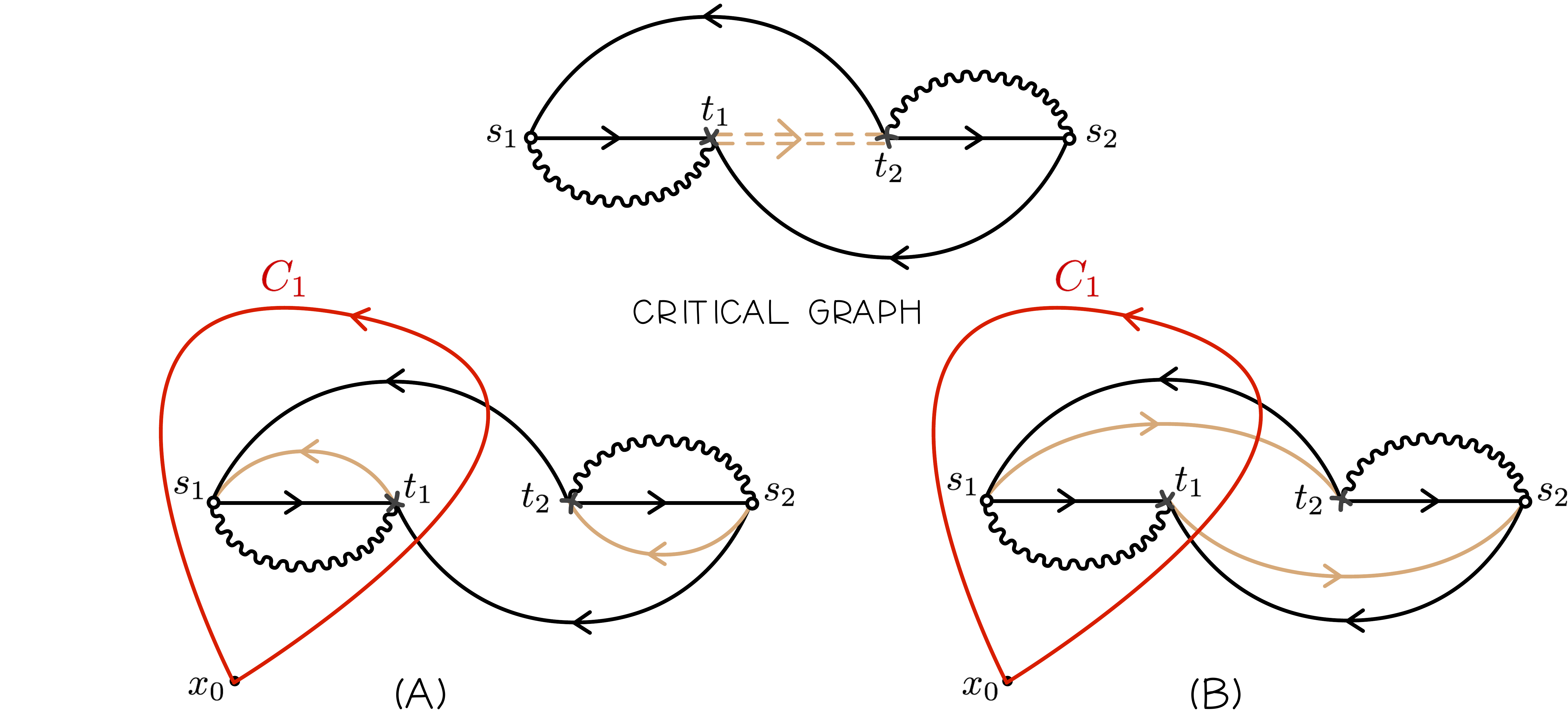}
\caption{The two Stokes graphs of the $N_f=0$ case that are related by a simple flip. We also exhibit the contour used to calculate the monodromy matrix. These graphs were obtained with the parameters $\Lambda_0 = e^{-\ii\frac{\pi}{4}}$ and $\widetilde{u} = -1+\ii$, with the critical graph observed at $\theta = \pi$.}
\label{Nf=0graphs}
\end{figure}
In figure \ref{Nf=0graphs}, we exhibit various Stokes graphs in the
strong coupling region of the pure super Yang-Mills
theory.\footnote{Using the form of the $N_f = 0$ differential as in
  \cite{GMN}, we are within the strong coupling region if we make the
  choice $\Lambda = 1$ and $u = 1/2$. A series of conformal
  transformations and rescalings relate the differential presented
  here and the one presented in \cite{GMN}. At the end of this series of transformations, we are led to the choice of parameters presented in the caption of figure \ref{Nf=0graphs}.} We
first draw the critical graph \ref{Nf=0graphs} that has a finite
WKB line connecting the turning points $t_1$ and $t_2$. The Stokes
graphs we work with are related by a flip \cite{GMN} about this finite
WKB line.  
\subsubsection*{The Monodromy Group}
In this case, there is a single independent generator
of the monodromy group and we choose the contour enclosing the
singularity $s_1$ and the turning point $t_1$ to be it. 
Consider first the Stokes graph \ref{Nf=0graphs}(A). The
contour intersects two Stokes lines; the monodromy matrix is given by
\begin{align}
\begin{split}
M_{A, s_1} &= \left( \begin{array}{cc}
x &\  0\\
0 &\  x^{-1} \end{array} \right)
\left( \begin{array}{cc}
1 &\  0\\
-\ii u_2 &\  1 \end{array} \right)
\left( \begin{array}{cc}
1 &\  +\ii u_1^{-1}\\
0 &\  1 \end{array} \right) \, . \\
\end{split}
\end{align}
Note that the matrices are written from right to left as we go around the branch cut. The final matrix encodes the overall normalization factor as we return to the base point $x_0$. The variable $x$ which appears there is identified with the overall monodromy around the branch cut connecting $s_1$ and $t_1$. 

We now turn to the second Stokes graph \ref{Nf=0graphs}(B). We see that the contour intersects four Stokes lines, including two lines arising from the flip. The monodromy matrix is given by 
\begin{align}
\begin{split}
M_{B, s_1} &= 
 \left( \begin{array}{cc}
\tilde x &\  0\\
0 &\  \tilde{x}^{-1} \end{array} \right)
\left( \begin{array}{cc}
1 &\  0\\
-\ii\tilde u_2 &\  1 \end{array} \right)
\left( \begin{array}{cc}
1 &\  -\ii\tilde u_2^{-1}\\
0 &\  1 \end{array} \right)
\left( \begin{array}{cc}
1 &\  0\\
+\ii\tilde u_1 &\  1 \end{array} \right)
\left( \begin{array}{cc}
1 &\  +\ii\tilde u_1^{-1}\\
0 &\  1 \end{array} \right)\\
&=\left( \begin{array}{cc}
\frac{\tilde{x}(\tilde u_1+\tilde u_2)}{\tilde u_2} &\  \frac{\ii \tilde{x}}{\tilde u_1}\\
-i\frac{\tilde u_2}{\tilde{x}} &\  \frac{\tilde u_2}{\tilde{x}\tilde
                                 u_1} \end{array} \right)\,.   
\end{split}
\end{align}
We have denoted the variables in Stokes graph \ref{Nf=0graphs}(B) by variables with tildes since they correspond to a different
Borel resummation. The  monodromy matrix $M_{A,s_1}$ must be equivalent to the monodromy matrix calculated on the basis of
graph (B), since the monodromy (equivalence class) is a property of the exact solutions on the Riemann surface $\Sigma$.

\subsubsection*{The Stokes Automorphism}

Above, we have the explicit expressions for the monodromy matrices for the two Stokes graphs. 
The independent Stokes variables are given by $x$ and $u_{21}$ in graph \ref{Nf=0graphs}(A) and the tilde-variables in graph \ref{Nf=0graphs}(B).
Using this notation, we calculate the conjugation invariant traces of the two monodromy matrices:
\begin{align}
\Tr\left(M_{A, s_1} \right)&=x+\frac{1}{x}+\frac{1}{u_{21}x}\\
\Tr\left(M_{B, s_1} \right)&=\tilde{x}+\tilde{u}_{21}\tilde{x}+\frac{1}{\tilde{u}_{21}\tilde{x}}\, .
\end{align}
Requiring that the traces of the two monodromy matrices match leads to the map between the parameters appearing in the two graphs:
\begin{align}
u_{21}&=\tilde{u}_{21}\nonumber\\
x&=\tilde{x}(1+\tilde{u}_{21}) \, .
\end{align}
This agrees with the Stokes automorphisms derived in \cite{Kashani-Poor:2015pca}. 
This is also consistent with the general analysis in \cite{NakanishiandIwaki}. Let us expand on this briefly: 
the two Stokes graphs lie on either side of the $t_1-t_2$ flip in the critical graph \ref{Nf=0graphs}. Since the $t_1-t_2$ cycle
corresponding to  $u_{12}$ has zero intersection number with itself, the variable $u_{12}$ is unaffected by the flip. However, the $x$ variable changes because the contour around the branch cut has intersection number $1$ with  the $t_1-t_2$ cycle. 
\subsection{One Flavour}

The Stokes graphs corresponding to the differential in the case of one flavour are determined by the corresponding zeroth order differential \eqref{Nf=1 semiclassical}:
\be
Q_0(z) =\frac{\Lambda_1^2}{4z^4}  + \frac{\Lambda_1m_1}{z^3(z-1)^2}-\frac{\Lambda_1^2}{4z(z-1)^3} 
+\frac{\widetilde{u}}{z^2(z-1)^2} \, .
\ee
{From} the form of the differential,
one can see that in the $z$-plane, there are three turning points and
two irregular singularities (at $z=0$ and $z=1$). The WKB triangulations are
given in figure 61 of \cite{GMN}. We consider a particular pair that
are separated by a flip\footnote{These are the first and the third
out of the six triangulations given in figure 61 of \cite{GMN}.} and
draw only the corresponding Stokes graphs. The two Stokes graphs
correspond to a flip about the $t_2-t_3$ finite Stokes line in the
critical graph (see figure \ref{Nf=1graphs}).

\subsubsection*{The Monodromy Group}
\begin{figure}[ht!]
\centering
\includegraphics[width=170mm]{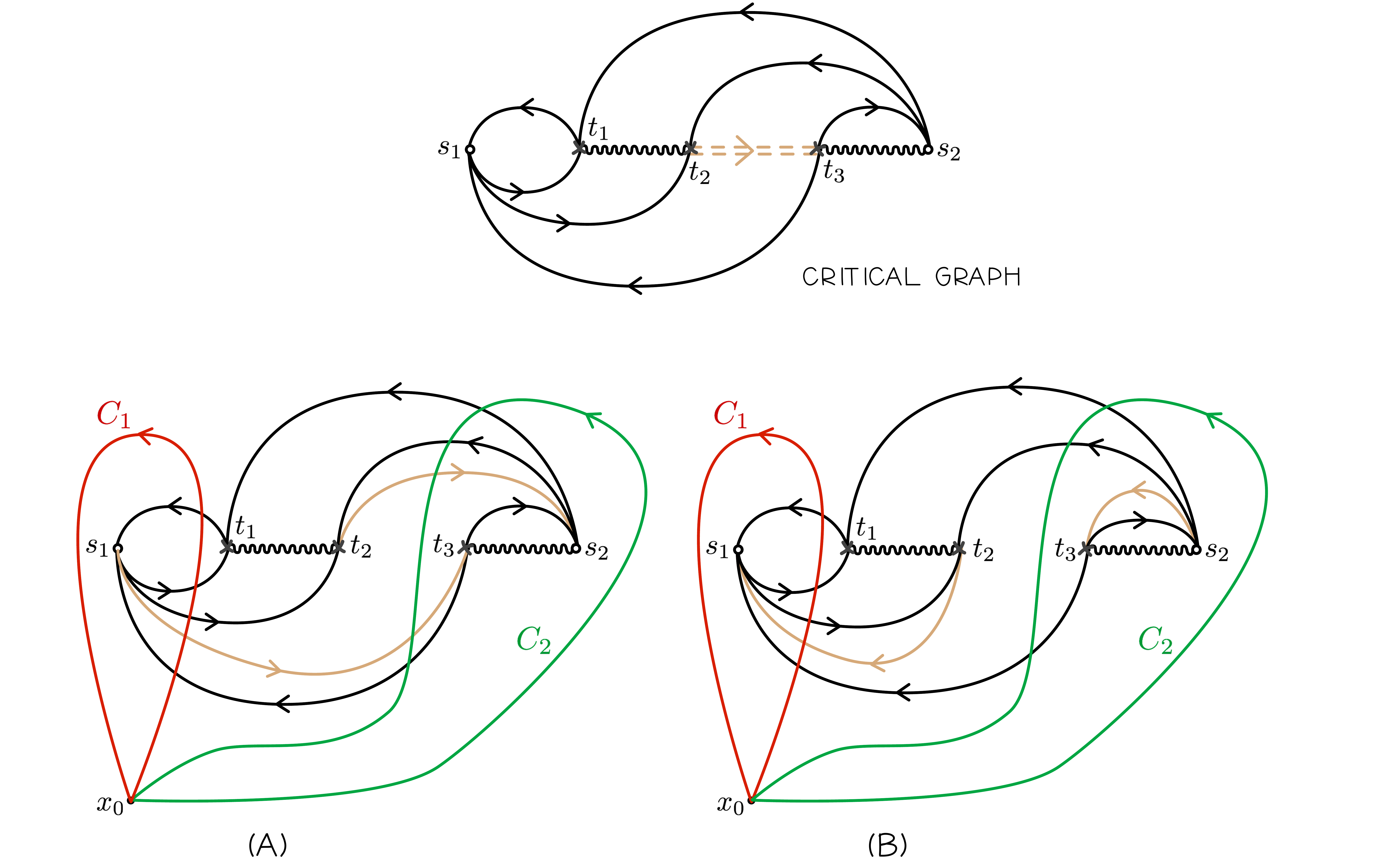}
\caption{The critical graph, the pair of Stokes graphs related by a flip and the contours that define the monodromy group for the $N_f=1$ case. The parameters chosen were $\Lambda_1 = 2$, $\widetilde{u} = -1/2$, and $m_1 = 1$, and the critical graph was observed at $\theta = \pi$.}
\label{Nf=1graphs}
\end{figure}

We proceed to calculate the monodromy group for both the Stokes
graphs. There are two irregular singularities in the graphs and one
expects two independent generators of the monodromy group. We choose
the two corresponding generators of the monodromy group as shown in
figure \ref{Nf=1graphs}. The contour around the singularity $s_2$ is
treated in much the same way as the irregular singular point in the
$N_f=0$ case, while the singularity $s_1$ behaves slightly differently.

Let us first consider Stokes graph \ref{Nf=1graphs}(A) and calculate the monodromy matrices; we find 
%
%
\begin{align}
\begin{split}
M_{A, s_1} &= 
\begin{pmatrix} 
\nu_1 &0\\
0&\frac{1}{\nu_1}
\end{pmatrix}
\begin{pmatrix} 
1 &0\\
-\ii u_1&1
\end{pmatrix}
\begin{pmatrix} 
1 &-\frac{\ii}{u_1}\\
0&1
\end{pmatrix}
\begin{pmatrix} 
1 &-\frac{\ii}{u_2}\\
0&1
\end{pmatrix}
\begin{pmatrix} 
1 &-\frac{\ii}{u_3}\\
0&1
\end{pmatrix}
\begin{pmatrix} 
1 &0\\
-\ii u_3&1
\end{pmatrix} \, ,\\
%
%
M_{A, s_2}&=\left( \begin{array}{cc}
x &\  0\\
0 &\frac{1}{x} \end{array} \right) 
\left( \begin{array}{cc}
1 &\ 0\\
\ii u_3x^2 &1 \end{array} \right) 
\left( \begin{array}{cc}
1 &\  \frac{\ii}{u_3x^2}\\
0 &1 \end{array} \right) 
\left( \begin{array}{cc}
1 &\  0\\
-\ii u_2x^2 &1 \end{array} \right) 
\left( \begin{array}{cc}
1 &\  -\frac{\ii}{u_2x^2}\\
0 &\ 1 \end{array} \right) 
\left( \begin{array}{cc}
1 &\  -\frac{\ii}{u_1u_{12}^2x^2}\\
0 &1 \end{array} \right)  \, .
\end{split}
\end{align}
The matrix element on the extreme left in the second monodromy matrix is the naive WKB monodromy  around the branch cut connecting $t_3$ and $s_2$. This contribution $x$  satisfies the relation, 
\be
x\, u_{12}\nu_1=1\,.
\ee
Let us now turn to the Stokes graph \ref{Nf=1graphs}(B). 
The monodromy matrices are given by
\begin{align}
\begin{split}
M_{B, s_1} &= 
\begin{pmatrix} 
\nu_1 &0\\
0&\frac{1}{\nu_1}
\end{pmatrix}
\begin{pmatrix} 
1 &0\\
-\ii \tilde{u}_1&1
\end{pmatrix}
\begin{pmatrix} 
1 &-\frac{\ii}{\tilde{u}_1}\\
0&1
\end{pmatrix}
\begin{pmatrix} 
1 &-\frac{\ii}{\tilde{u}_2}\\
0&1
\end{pmatrix}
\begin{pmatrix} 
1 &0\\
-\ii\tilde{u}_2&1
\end{pmatrix}
\begin{pmatrix} 
1 &0\\
-\ii \tilde{u}_3&1
\end{pmatrix} \, ,\\
M_{B,s_2}&=\left( \begin{array}{cc}
\tilde{x} &\  0\\
0 & \frac{1}{\tilde{x}}\end{array} \right) 
\left( \begin{array}{cc}
1 &\  0\\
\ii\tilde{u}_3\tilde{x}^2 &1 \end{array} \right) 
\left( \begin{array}{cc}
1 &\ \frac{-\ii}{\tilde{u}_2\tilde{x}^2}\\
0 &1 \end{array} \right) 
\left( \begin{array}{cc}
1 &\  \frac{-\ii}{\tilde{u}_1\tilde{u}_{12}^2\tilde{x}^2}\\
0 &1 \end{array} \right)  \,.
\end{split}
\end{align}
As before, we define $\tilde x=\frac{1}{\tilde u_{12}\nu_1}$.

\subsubsection*{The Stokes Automorphism}

Now that we have the two sets of monodromy matrices, we calculate the traces of the two sets and obtain
\begin{align}
\begin{split}
\Tr(M_{A,s_1})&= -\nu_1 \left(\frac{u_3}{u_1}+\frac{u_3}{u_2} \right)-\frac{1}{\nu_1}\left(\frac{u_1}{u_2}+\frac{u_1}{u_3}\right) \, ,\\
\Tr(M_{A,s_2})&=\frac{1}{\nu_1}(u_{31}+u_{21})+(u_{13}+u_{23})\nu_1 \, ,\\
\Tr(M_{B,s_1})&=-\nu_1\left(\frac{\tilde{u}_2}{\tilde{u}_1}+\frac{\tilde{u}_3}{\tilde{u}_2}+\frac{\tilde{u}_3}{\tilde{u}_1} \right)-\frac{1}{\nu_1}\frac{u_1}{u_2} \, ,\\
\Tr(M_{B,s_2})&=\frac{1}{\nu_1}(\tilde{u}_{21})+(\tilde{u}_{12}+\tilde{u}_{13}+\tilde{u}_{23})\nu_1 \, .
\end{split}
\end{align}
Substituting $u_{13}=u_{12}u_{23}$, and equating the expressions for
the traces in powers of $\nu_1$ (where we use the fact that the 
characteristic exponents are true invariants of the differential
equation), we can extract the Stokes automorphism formulae for the
independent contour integrals $u_{21}$ and $u_{23}$, namely:
\begin{align}
\tilde{u}_{23}&=u_{23} \, ,\\
\tilde{u}_{21}&=u_{21}(1+u_{32}) \, .
\end{align}
Since there is more than one generator of the monodromy group, one can
calculate higher-order invariants by calculating traces of products of
the matrices. Using the Stokes automorphism, one can check that the
trace of the products also coincide, thus confirming the
identification of the monodromy group. 

\subsection{Two Flavours}
\begin{figure}[ht!]
\centering
\includegraphics[width=150mm]{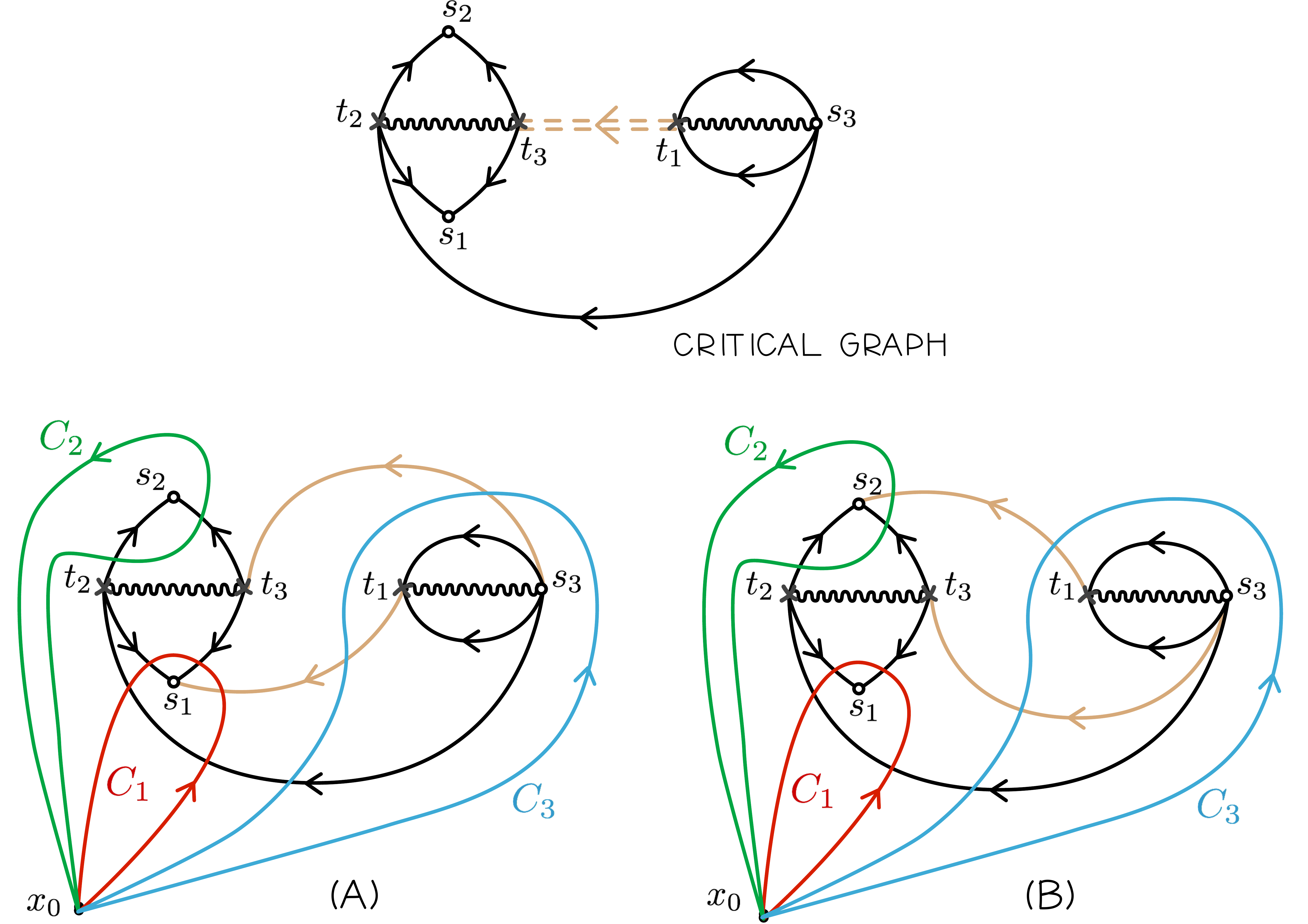}
\caption{The critical graph and the Stokes graphs for the $N_f=2$ case. While plotting the figures, we used a potential that is 
conformally equivalent to \eqref{Nf2Potential}. We set $\Lambda_2\rightarrow \ii, \widetilde{u}\rightarrow \frac{1}{2},m_3\rightarrow 0,m_4\rightarrow -2$. The 
two Stokes graphs presented were observed at $\theta=\frac{2\pi}{3}$ and $\theta=\frac{3\pi}{4}$.}
\label{Nf=2graphs}
\end{figure}

In this section, we consider the SU$(2)$ gauge theory with two flavours. 
We concentrate on the asymmetric configuration. The zeroth order potential function is given by \eqref{Nf=2 asymmetric semiclassical}
\be
\label{Nf2Potential}
Q_0(z) = \frac{(m_3+m_4)^2}{4(z-1)^2}+ \frac{m_3 m_4}{z(1-z)} +\frac{\Lambda_2^2}{z^3} +\frac{\widetilde{u}}{z^2-z^3}
\, .
\ee
In the $z$-plane, the quadratic differential has three singularities, and three turning points. One of these is an irregular singularity 
at $z=0$. 
As before, we work in a strong coupling limit, where $\widetilde{u}\ll\Lambda_2^2$. We consider the critical graph (see figure \ref{Nf=2graphs}), 
and by a flip about the $t_1-t_3$ finite WKB line, obtain the two Stokes graphs, as shown in the figure.
An important difference from the earlier cases is that we have regular singularities at $s_1$ and $s_2$.
\subsubsection*{The Monodromy Group}
In order to calculate the monodromy group, we
 first consider the Stokes graph \ref{Nf=2graphs}(A) and determine the generators
\begin{align}
\begin{split}
M_{A,s_1}& =\begin{pmatrix}
\nu_1 &0\\
0 &\frac{1}{\nu_1}
\end{pmatrix}
\begin{pmatrix}
1 &-\frac{\ii}{u_2\nu_1^2}\\
0 &1
\end{pmatrix}
\begin{pmatrix}
1 &0\\
-\ii u_2\nu_1^2 &1
\end{pmatrix}
\begin{pmatrix}
1 &0\\
-\ii u_3\nu_1^2 &1
\end{pmatrix}
\begin{pmatrix}
1 &0\\
-\ii u_1\nu_1^2 &1
\end{pmatrix}
\begin{pmatrix}
1 &\frac{\ii}{u_2}\\
0 &1
\end{pmatrix}  \, ,\\
M_{A, s_2} &= \begin{pmatrix}
\nu_2 &0\\
0 &\frac{1}{\nu_2}
\end{pmatrix}
\begin{pmatrix}
1 &0\\
-\ii u_3u_{32}^2 &1
\end{pmatrix}
\begin{pmatrix}
1 &0\\
-\ii u_2 &1
\end{pmatrix} \, , \\
M_{A, s_3} &= \begin{pmatrix}
x &0\\
0 &\frac{1}{x}
\end{pmatrix}
\begin{pmatrix}
1 &-\frac{\ii}{u_2x^2}\\
0 &1
\end{pmatrix}
\begin{pmatrix}
1 &0\\
\ii u_1\nu_1^2x^2 &1
\end{pmatrix}
\begin{pmatrix}
1 &-\frac{\ii}{u_3\nu_1^2x^2}\\
0 &1
\end{pmatrix} \, .
\end{split}
\end{align}
In the above, $x$ is the naive WKB monodromy around the branch cut connecting $t_1$ and $s_3$. This contribution satisfies the relation,
\be
u_{23}\, \nu_1\, \nu_2\, x=1 \,.
\ee
A similar calculation for Stokes graph \ref{Nf=2graphs}(B), gives us the following monodromy matrices for circling the singularities
\begin{align}
\begin{split}
M_{B, s_1} &=  \begin{pmatrix}
  \nu_1 &0\\
  0 &\frac{1}{\nu_1}
 \end{pmatrix}
\begin{pmatrix}
 1 &-\frac{\ii}{\tilde{u}_2\nu_1^2}\\
 0 &1
\end{pmatrix}
\begin{pmatrix}
1 &0\\
-\ii\tilde{u}_2\nu_1^2 &1
\end{pmatrix}
\begin{pmatrix}
1 &0\\
-\ii\tilde{u}_3\nu_1^2 &1
\end{pmatrix}
\begin{pmatrix}
1 &\frac{\ii}{\tilde{u}_2}\\
0 &1
\end{pmatrix} \, , \\
M_{B, s_2} &= \begin{pmatrix}
\nu_2 &0\\
0 &\frac{1}{\nu_2}
\end{pmatrix}
\begin{pmatrix}
1 &0\\
-\ii \tilde{u}_1\tilde{u}_{32}^2 &1
\end{pmatrix}
\begin{pmatrix}
1 &0\\
-\ii \tilde{u}_3\tilde{u}_{32}^2 &1
\end{pmatrix}
\begin{pmatrix}
1 &0\\
-\ii\tilde{u}_2 &1
\end{pmatrix} \, , \\
M_{B, s_3} &= \begin{pmatrix}
  \tilde{x} &0\\
  0 &\frac{1}{\tilde{x}}
 \end{pmatrix}
\begin{pmatrix}
  1 &-\frac{\ii}{\tilde{u}_2\tilde{x}^2}\\
  0 &1
 \end{pmatrix}
\begin{pmatrix}
  1 &-\frac{\ii}{\tilde{u}_3\nu_1^2\tilde{x}^2}\\
  0 &1
 \end{pmatrix}
\begin{pmatrix}
  1 &0\\
  \ii\tilde{u}_1\nu_1^2\tilde{x}^2 &1
 \end{pmatrix}\, .
\end{split}
\end{align}
Again we have the relation, 
\be 
u_{23} \, \nu_1\, \nu_2\, \tilde{x} = 1\,.
\ee

\subsubsection*{Stokes Automorphisms}

We now compare the traces of the generators of the monodromy group:
\begin{align}
\begin{split}
\Tr M_{A,s_1} &= \Tr M_{B,s_1} = \nu_1 + \frac{1}{\nu_1} \, ,\\
\Tr M_{A,s_2} &= \Tr M_{B,s_2} =  \nu_2 + \frac{1}{\nu_2} \, ,\\
\Tr M_{A,s_3} &=\frac{1}{\nu_1\nu_2}u_{32}+\frac{\nu_1}{\nu_2}u_{21}u_{32}+\nu_1\nu_2(u_{21}+\frac{1}{u_{32}}) \, ,\\
\Tr M_{B,s_3}
&=\frac{1}{\nu_1\nu_2}\tilde{u}_{32}(\tilde{u}_{21}\tilde{u}_{32}+1)+\frac{\nu_1}{\nu_2}\tilde{u}_{21}\tilde{u}_{32}+\nu_1\nu_2\frac{1}{\tilde{u}_{32}}
\,.
\end{split}
\end{align}
These equations illustrate  a recurring feature: the
traces of the monodromy matrices around regular singular points will always be given by the critical exponents, with no $u_{ij}$ monodromy factors entering the expression. This is
because the Stokes lines are either
all going in or coming out at such regular singular points. As a result, the relevant Stokes
matrices are all either upper triangular or lower triangular,
respectively. This leads to the trivial nature of the trace. The
irregular singularity, on the other hand, has non-trivial structure
even at the level of the simple traces.
\\
Matching the traces between the graphs \ref{Nf=2graphs}(A) and \ref{Nf=2graphs}(B) leads to the Stokes automorphism relations,
\begin{align}\label{stokesautoNf=2}
\begin{split}
u_{31} &=\tilde{u}_{31} \, ,\\
u_{32} &=\tilde{u}_{32}(1+\tilde{u}_{31}) \, ,\\
u_{21} &=\tilde{u}_{21}(1+\tilde{u}_{31})^{-1} \,.
\end{split}
\end{align} 
This is once again as expected from the general results of \cite{NakanishiandIwaki} and the intersection numbers between the various cycles. 
As a consistency check on the monodromy matrices, we have also
computed the traces of products of matrices, and a
similar analysis as above confirms the Stokes automorphisms \eqref{stokesautoNf=2}. 
%
%

\subsection{Three Flavours}
\begin{figure}[ht!]
\centering
\includegraphics[width=150mm]{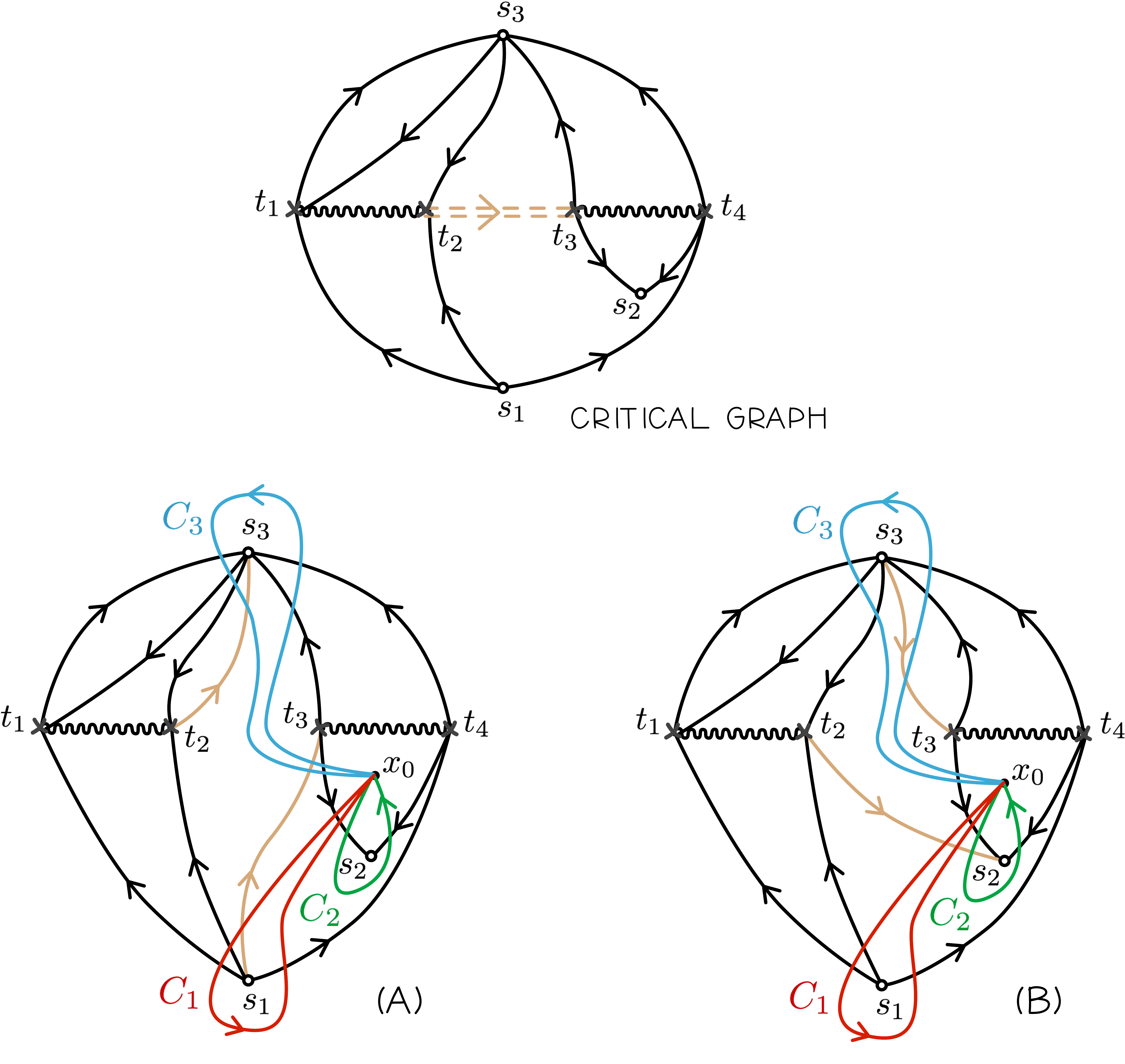}
\caption{The critical graph and the Stokes graphs for the $N_f=3$ case. While plotting the figures, we used a potential that is
conformally equivalent to \eqref{Nf3Potential}. We set $\Lambda_3\rightarrow 1, \widetilde{u}\rightarrow 2,m_1\rightarrow -1,m_3\rightarrow 0,m_4\rightarrow -2$. The 
two Stokes graphs presented were observed at $\theta=\frac{\pi}{2}$ and $\theta=\frac{7\pi}{12}$.}
\label{Nf3stokesgraphs}
\end{figure}

We move on to the SU$(2)$ theory with three flavours. The Seiberg-Witten differential is 
\begin{equation}
\label{Nf3Potential}
 Q_0(z) = \frac{(m_3+m_4)^2}{4(z-1)^2}+ \frac{m_3 m_4}{z(1-z)} +\frac{m_1\Lambda_3}{z^3}+\frac{\Lambda_3^2}{4z^4} +\frac{\widetilde{u}}{z^2-z^3}\,.
\end{equation}
There are four turning points and three singularities on the $z$-plane. The two Stokes graphs in 
figure \ref{Nf3stokesgraphs} are related by a flip about the $t_2- t_3$ finite line in the critical graph.

\subsubsection*{The Monodromy Group}
For Stokes graph \ref{Nf3stokesgraphs}(A), we find the  generators of the monodromy group:
\begin{align}
\begin{split}
M_{A, s_1} &=
\begin{pmatrix}
\nu_1& 0\\
0 &\frac{1}{\nu_1} 
\end{pmatrix}
\begin{pmatrix}
1 &0 \\
\ii u_3\nu_1^2  &1 
\end{pmatrix}
\begin{pmatrix}
1 & -\frac{\ii}{u_4\nu_1^2\nu_2^2}\\
0 &1
\end{pmatrix}
\begin{pmatrix}
1 &-\frac{\ii}{u_1} \\
0 &1 
\end{pmatrix}
\begin{pmatrix}
1 &-\frac{\ii}{u_2} \\
0 &1 
\end{pmatrix}
\begin{pmatrix}
1 &-\frac{\ii}{u_3} \\
0 &1 
\end{pmatrix}
\begin{pmatrix}
1 &0 \\
-\ii u_3 &1 
\end{pmatrix} \, ,\\
M_{A, s_2} &=
\begin{pmatrix}
\nu_2& 0\\
0 &\frac{1}{\nu_2} 
\end{pmatrix}
\begin{pmatrix}
1 &0\\
-\ii u_4 \nu_2^2 & 1\end{pmatrix}
\begin{pmatrix}
1 &0\\
-\ii u_3 & 1\end{pmatrix} \, ,
\\
M_{A, s_3} &= 
\begin{pmatrix}
\nu_3 & 0\\
0 &\frac{1}{\nu_3}
\end{pmatrix}
\begin{pmatrix}
1 &0 \\
\ii u_3\nu_3^2 &1 
\end{pmatrix}
\begin{pmatrix}
1 &\frac{\ii}{u_3\nu_3^2} \\
0 &1 
\end{pmatrix}
\begin{pmatrix}
1 &0 \\
-\ii u_2\nu_3^2 &1 
\end{pmatrix}
\begin{pmatrix}
1 &-\frac{\ii}{u_2\nu_3^2} \\
0 &1 
\end{pmatrix}
\begin{pmatrix}
1 &-\frac{\ii}{u_1\nu_3^2 u_{12}^2} \\
0 &1 
\end{pmatrix}\\
&\hspace{3cm}\times
\begin{pmatrix}
1 &0 \\
-\ii u_1\nu_3^2 u_{12}^2&1 
\end{pmatrix}
\begin{pmatrix}
1 &0 \\
-\ii u_4 u_{43}^2 &1 
\end{pmatrix}
\begin{pmatrix}
1 &0 \\
-\ii u_3 &1 
\end{pmatrix}
\begin{pmatrix}
1 &-\frac{\ii}{u_3}\\
0 &1
\end{pmatrix}
\begin{pmatrix}
1 &0 \\
-\ii u_3 &1 
\end{pmatrix} \, .
  \end{split}
\end{align}
For the Stokes graph \ref{Nf3stokesgraphs}(B), a similar calculation yields:
\begin{align}
\begin{split}
M_{B, s_1} &=
\begin{pmatrix}
\nu_1& 0\\
0 &\frac{1}{\nu_1} 
\end{pmatrix}
\begin{pmatrix}
1 &0 \\
\ii \tilde{u_3}\nu_1^2 &1 
\end{pmatrix}
\begin{pmatrix}
1 &0 \\
\ii\tilde{u}_2\nu_1^2 &1 
\end{pmatrix}
\begin{pmatrix}
1 &-\frac{\ii}{\tilde{u}_4\nu_1^2\nu_2^2} \\
0 &1 
\end{pmatrix}
\begin{pmatrix}
1 &-\frac{\ii}{\tilde{u}_1} \\
0 &1 
\end{pmatrix}
\begin{pmatrix}
1 &-\frac{\ii}{\tilde{u}_2} \\
0 &1 
\end{pmatrix}\\
&\hspace{2cm}\times 
\begin{pmatrix}
1 &0\\
-\ii\tilde{u}_2 &1
\end{pmatrix}
\begin{pmatrix}
1 &0\\
-\ii\tilde{u}_3 &1 
\end{pmatrix}\, ,\\
M_{B,s_2} &=
\begin{pmatrix}
\nu_2 &0\\
0 &\frac{1}{\nu_2}
\end{pmatrix}
\begin{pmatrix}
1 &0\\
-\ii \tilde{u}_4\nu_2^2 & 1
\end{pmatrix}
\begin{pmatrix}
1 &0 \\
-\ii \tilde{u}_2 & 1 
\end{pmatrix}
\begin{pmatrix}
1 &0 \\
-\ii \tilde{u}_3 &1
\end{pmatrix}\, ,
\\
M_{B, s_3} &=
\begin{pmatrix}
\nu_3 & 0\\
0 &\frac{1}{\nu_3}
\end{pmatrix}
\begin{pmatrix}
1 &0 \\
\ii \tilde{u}_3\nu_3^2 &1 \\
\end{pmatrix}
\begin{pmatrix}
1 &-\frac{\ii}{\tilde{u}_2\nu_3^2} \\
0 &1 \\
\end{pmatrix}
\begin{pmatrix}
1 &-\frac{\ii}{\tilde{u}_1\nu_3^2 \tilde{u}_{12}^2} \\
0 &1 
\end{pmatrix}\\
&\hspace{2cm}\times 
\begin{pmatrix}
1 &0 \\
-\ii \tilde{u}_1\nu_3^2 \tilde{u}_{12}^2&1 
\end{pmatrix}
\begin{pmatrix}
1 &0 \\
-\ii \tilde{u}_4 \tilde{u}_{43}^2 &1 
\end{pmatrix}
\begin{pmatrix}
1 &0 \\
-\ii \tilde{u}_3 &1 
\end{pmatrix}
\begin{pmatrix}
1 &-\frac{\ii}{\tilde{u}_3}\\
0 &1
\end{pmatrix}
\begin{pmatrix}
1 &0 \\
-\ii \tilde{u}_3 &1 
\end{pmatrix}\, .
\end{split}
\end{align}
\subsubsection*{The Stokes Automorphism}

As in the previous examples, the Stokes automorphism can be obtained by comparing the invariants built out of the monodromy matrices. 
%
%
The trace of the monodromy around the irregular singular point $s_3$ is non-trivial and it is important that we express it in terms of independent Stokes variables. The variables are constrained by the relation
\be
u_{12}u_{34} \nu_1\nu_2\nu_3 = 1\,,
\ee
and similarly for the $\tilde{u}$ variables. We solve for $u_{34}$ using this relation and choose the independent variables to be $u_{12}$ and $u_{23}$. In terms of these variables, we find that
\begin{align}
\Tr M_{A,s_3} &= -\nu_3 u_{12} - \nu_1\nu_2  u_{23}(1+u_{12}) - \frac{1}{\nu_3\, u_{12}} (1+u_{23}+u_{12}u_{23}) \,.
\end{align}
Similarly, from the monodromy around $s_3$ in Stokes graph \ref{Nf3stokesgraphs}(B) we find 
\be
\Tr M_{B,s_3} = -\nu_3\tilde{u}_{12}(\frac{1}{\tilde{u}_{23}}+1) - \nu_1\nu_2(\tilde{u}_{12}+\tilde{u}_{23}+\tilde{u}_{12}\tilde{u}_{23})-\frac{\tilde{u}_{23}}{\nu_3\tilde{u}_{12}}(1+\tilde{u}_{12}) \, .
\ee
Matching the traces leads to the Stokes automorphisms:
\begin{align}
u_{23} &= \tilde{u}_{23} \, , \\
u_{12} &= \tilde{u}_{12}\left(1+\frac{1}{\tilde{u}_{23}} \right) \, .
\end{align}
As a check of our monodromy matrices, we computed the traces of the products of the monodromy matrices. These imply the same Stokes automorphism as above. 



\subsection{The Conformal Theory}\label{ConfTh}

We consider  Stokes graphs in a strong coupling region of the conformal SU$(2)$ theory. The zeroth order 
potential has four regular singular points and four turning points. 

\begin{figure}[ht!]
\centering
\includegraphics[width=165mm]{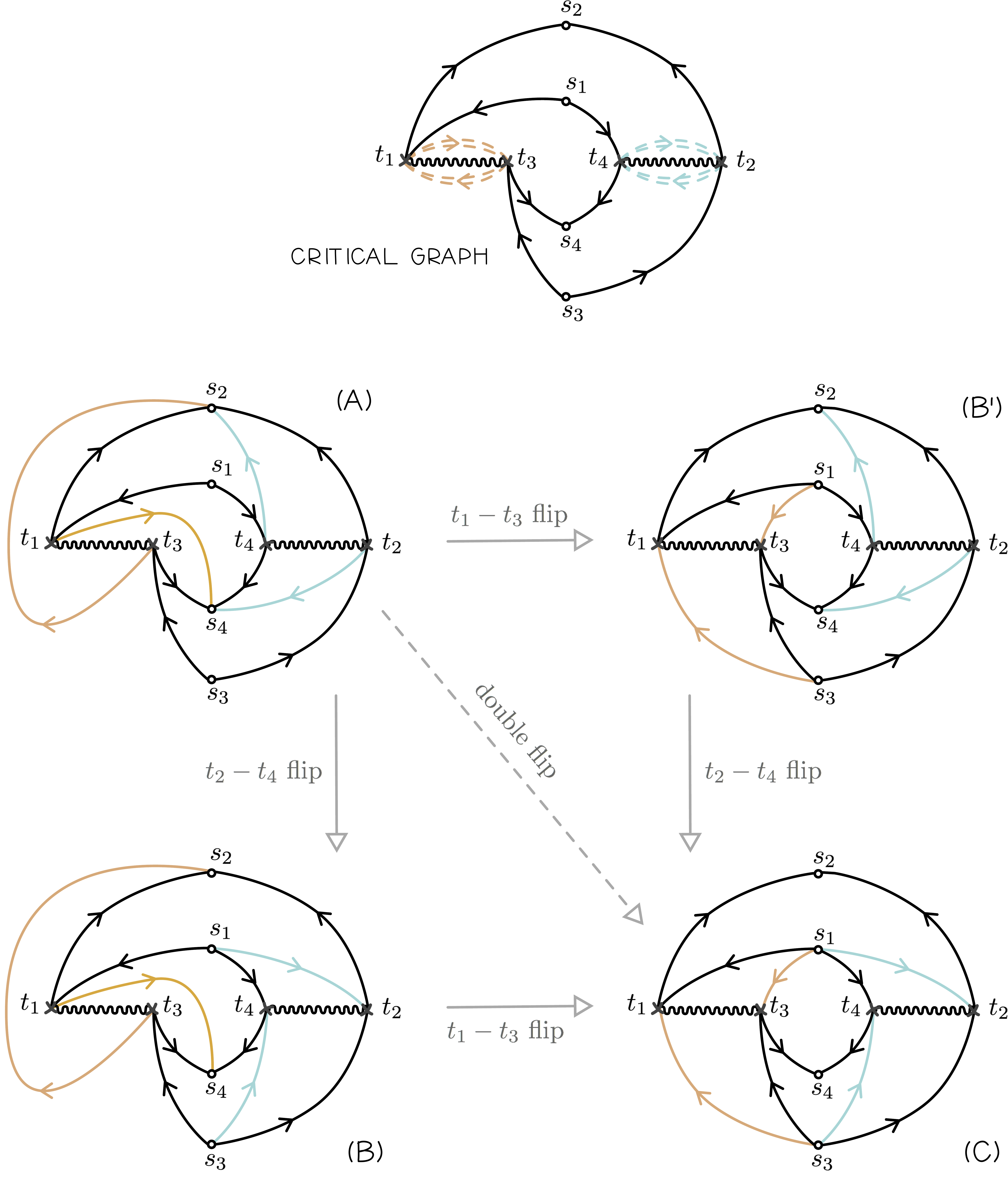}
\caption{The critical graph and a pair of Stokes graphs for the conformal SU$(2)$ theory. A double flip relates one graph to the other. We refer the reader to \ref{BDoubleFlip} for details.}
\label{Nf=4graphs}
\end{figure}

In particular, we consider the Stokes graphs corresponding to two out
of the six triangulations in figure 74 of \cite{GMN} (see figure
\ref{Nf=4graphs}(A) and (C)).\footnote{Our graphs are topologically
  equivalent to those appearing in \cite{GMN}.} It can be seen that
the two Stokes graphs are related by a double flip, a simultaneous
flip about the $t_1-t_3$ and $t_2-t_4$ finite WKB lines in the
critical graph. We realize the double flip as two alternative sequences of two single
flips. We provide the corresponding intermediate graphs after the
single flips and perform the calculation we have familiarized
ourselves with by now.

\begin{figure}[ht!]
\centering
\includegraphics[width=150mm]{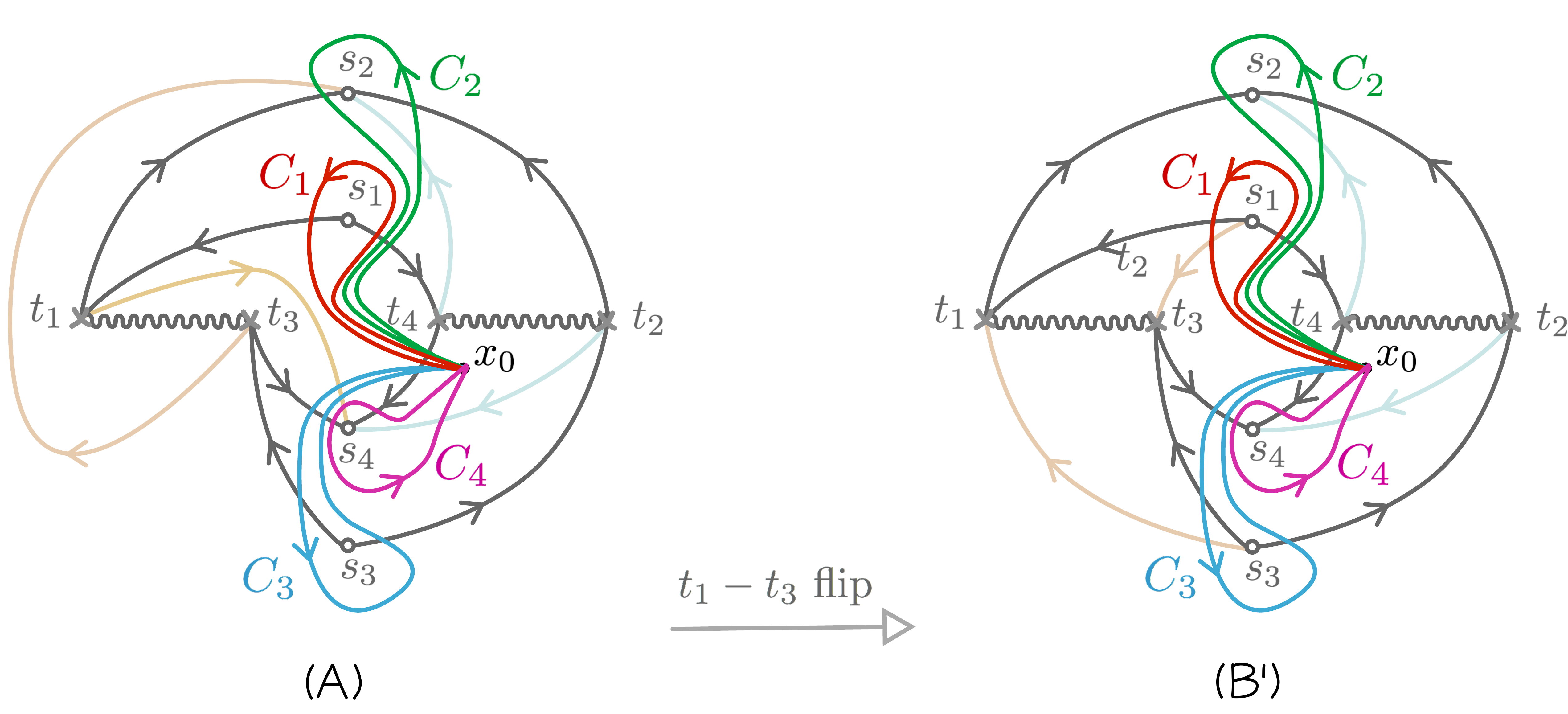}
\caption{The two Stokes graphs related by the $t_1-t_3$ flip.}
\label{t1t3flipcontours}
\end{figure}

Let us first consider the flip from Stokes graph \ref{Nf=4graphs}(A) to \ref{Nf=4graphs}(B$'$) via the $t_1-t_3$ flip. The relevant Stokes graphs and contours that generate the monodromy group are given in figure \ref{t1t3flipcontours}.

\subsubsection*{The Monodromy Group}
\begin{figure}[ht!]
\centering
\includegraphics[width=150mm]{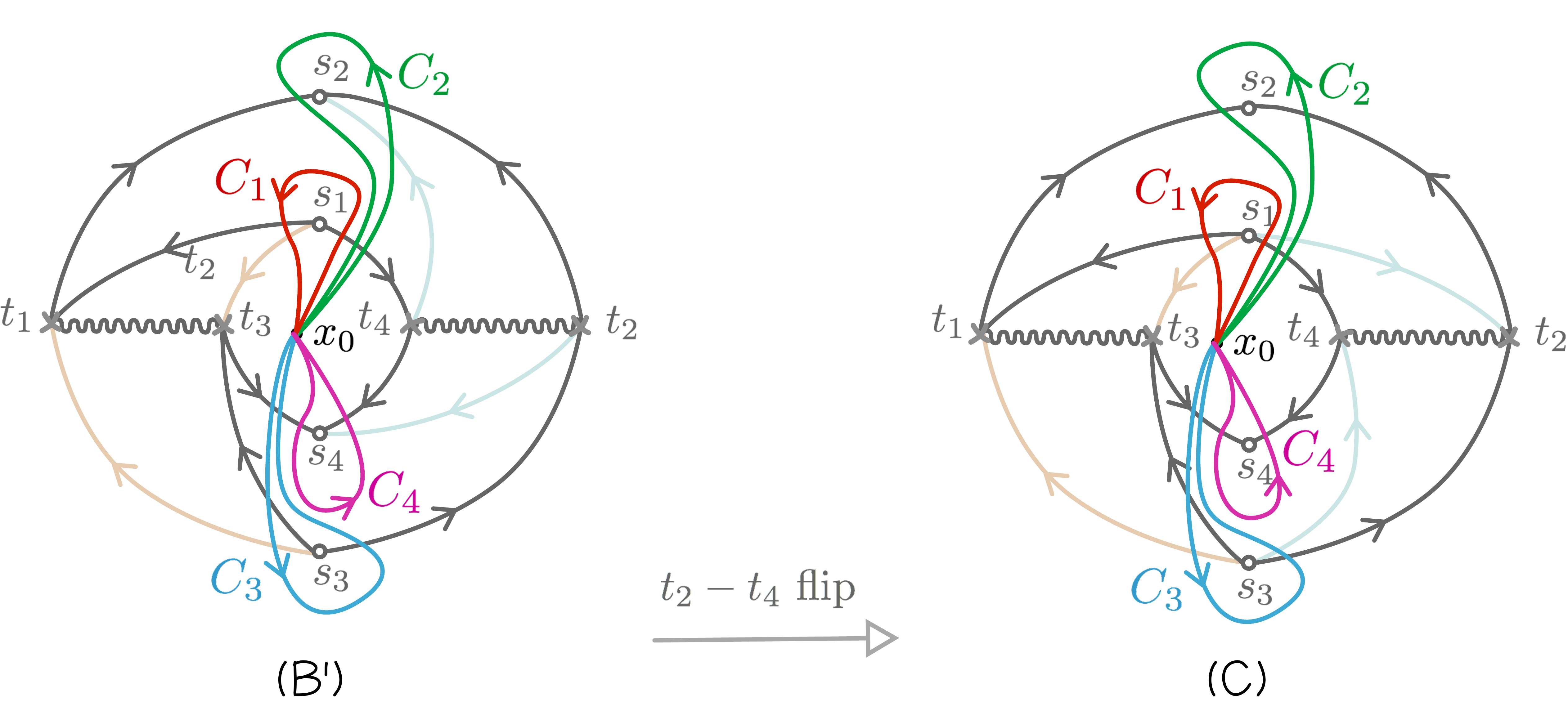}
\caption{The $t_2-t_4$ flip and the contours around the singularities.}
\label{t2t4flipcontours}
\end{figure}

For Stokes graph \ref{Nf=4graphs}(A), using the contours as shown in figure \ref{t1t3flipcontours}(A),
the monodromy matrices are given by
\begin{align}
  \begin{split}
M_{A, s_1} &=
\begin{pmatrix}
\nu_1& 0\\
0 &\frac{1}{\nu_1} 
\end{pmatrix}
\begin{pmatrix}
1& 0\\
\ii u_4\nu_1^2& 1
\end{pmatrix}
\begin{pmatrix}
1& \frac{-\ii}{u_1u_{13}^2\nu_1^2}\\
0& 1
\end{pmatrix}
\begin{pmatrix}
1&-\frac{\ii}{u_4}\\
0& 1
\end{pmatrix}
\begin{pmatrix}
1& 0\\
-\ii u_4& 1
\end{pmatrix} \, , \\
M_{A,s_2} &=
\begin{pmatrix}
\nu_2& 0\\
0& \frac{1}{\nu_2}
\end{pmatrix}
\begin{pmatrix}
1& 0\\
\ii u_4\nu_2^2& 1
\end{pmatrix}
\begin{pmatrix}
1& \frac{\ii}{u_4\nu_2^2}\\
0& 1
\end{pmatrix}
\begin{pmatrix}
1& 0\\
-\ii u_1u_{13}^2\nu_1^2\nu_2^2& 1
\end{pmatrix}
\begin{pmatrix}
1& 0\\
-\ii u_3u_{13}^2\nu_1^2\nu_2^2& 1
\end{pmatrix}
\\
&\hspace{3cm}
\times
\begin{pmatrix}
1& 0\\
-iu_2u_{24}^2& 1
\end{pmatrix}
\begin{pmatrix}
1& 0\\
-\ii u_4& 1
\end{pmatrix}
\begin{pmatrix}
1& -\frac{\ii}{u_4}\\
0& 1
\end{pmatrix}
\begin{pmatrix}
1& 0\\
-\ii u_4& 1
\end{pmatrix} \, ,\\
M_{A,s_3} &=
\begin{pmatrix}
\nu_3& 0\\
0& \frac{1}{\nu_3}
\end{pmatrix}
\begin{pmatrix}
1& 0\\
\ii u_4\nu_3^2& 1
\end{pmatrix}
\begin{pmatrix}
1& 0\\
\ii u_1u_{13}^2\nu_3^2& 1
\end{pmatrix}
\begin{pmatrix}
1& 0\\
\ii u_3\nu_3^2& 1
\end{pmatrix}
\begin{pmatrix}
1& \frac{-\ii}{u_2\nu_3^2\nu_4^2}\\
0& 1
\end{pmatrix}
\\
&
\hspace{3cm}
\times
\begin{pmatrix}
1& -\frac{\ii}{u_3}\\
0& 1
\end{pmatrix}
\begin{pmatrix}
1& 0\\
-\ii u_3& 1
\end{pmatrix}
\begin{pmatrix}
1& 0\\
-\ii u_1u_{13}^2& 1
\end{pmatrix}
\begin{pmatrix}
1& 0\\
-\ii u_4& 1
\end{pmatrix} \, ,\\
M_{A,s_4} &=
\begin{pmatrix}
\nu_4& 0\\
0& \frac{1}{\nu_4}
\end{pmatrix}
\begin{pmatrix}
1& 0\\
-\ii u_2\nu_4^2& 1
\end{pmatrix}
\begin{pmatrix}
1& 0\\
-\ii u_3& 1
\end{pmatrix}
\begin{pmatrix}
1& 0\\
-\ii u_1u_{13}^2& 1
\end{pmatrix}
\begin{pmatrix}
1& 0\\
-\ii u_4& 1
\end{pmatrix} \, . 
  \end{split}
\end{align}
Next we consider the Stokes graph \ref{Nf=4graphs}(B$'$) and move along the contours $C_k$ around the singularities as shown in figure \ref{t1t3flipcontours}(B$'$). 
We compute the monodromy matrices:
\begin{align}
\begin{split}
M_{B', s_1} &=
\begin{pmatrix}
\nu_1& 0\\
0 &\frac{1}{\nu_1} 
\end{pmatrix}
\begin{pmatrix}
1& 0\\
\ii\tilde{u}_4\nu_1^2& 1
\end{pmatrix}
\begin{pmatrix}
1& \frac{-\ii}{\tilde{u}_3\nu_1^2}\\
0& 1
\end{pmatrix}
\begin{pmatrix}
1& \frac{-\ii}{\tilde{u}_1\tilde{u}_{13}^2\nu_1^2}\\
0& 1
\end{pmatrix}
\begin{pmatrix}
1&-\frac{\ii}{\tilde{u}_4}\\
0& 1
\end{pmatrix}
\begin{pmatrix}
1& 0\\
-\ii\tilde{u}_4& 1
\end{pmatrix} \, ,\\
M_{B',s_2} &=
\begin{pmatrix}
\nu_2& 0\\
0& \frac{1}{\nu_2}
\end{pmatrix}
\begin{pmatrix}
1& 0\\
\ii \tilde{u}_4\nu_2^2& 1
\end{pmatrix}
\begin{pmatrix}
1& \frac{\ii}{\tilde{u}_4\nu_2^2}\\
0& 1
\end{pmatrix}
\begin{pmatrix}
1& 0\\
-\ii\tilde{u}_1\tilde{u}_{13}^2\nu_1^2\nu_2^2& 1
\end{pmatrix}\\
&\hspace{3cm}\times
\begin{pmatrix}
1& 0\\
-i\tilde{u}_2\tilde{u}_{24}^2& 1
\end{pmatrix}
\begin{pmatrix}
1& 0\\
-\ii\tilde{u}_4& 1
\end{pmatrix}
\begin{pmatrix}
1& -\frac{\ii}{\tilde{u}_4}\\
0& 1
\end{pmatrix}
\begin{pmatrix}
1& 0\\
-\ii\tilde{u}_4& 1
\end{pmatrix} \, ,\\
M_{B',s_3} &=
\begin{pmatrix}
\nu_3& 0\\
0& \frac{1}{\nu_3}
\end{pmatrix}
\begin{pmatrix}
1& 0\\
\ii \tilde{u}_4\nu_3^2& 1
\end{pmatrix}
\begin{pmatrix}
1& 0\\
\ii \tilde{u}_3\nu_3^2& 1
\end{pmatrix}
\begin{pmatrix}
1& \frac{-\ii}{\tilde{u}_2\nu_3^2\nu_4^2}\\
0& 1
\end{pmatrix}
\\
&\hspace{3cm}
\times
\begin{pmatrix}
1& \frac{-\ii}{\tilde{u}_1}\\
0& 1
\end{pmatrix}
\begin{pmatrix}
1& -\frac{\ii}{\tilde{u}_3}\\
0& 1
\end{pmatrix}
\begin{pmatrix}
1& 0\\
-\ii \tilde{u}_3& 1
\end{pmatrix}
\begin{pmatrix}
1& 0\\
-\ii \tilde{u}_4& 1
\end{pmatrix} \, ,\\
M_{B',s_4} &=
\begin{pmatrix}
\nu_4& 0\\
0& \frac{1}{\nu_4}
\end{pmatrix}
\begin{pmatrix}
1& 0\\
-\ii \tilde{u}_2\nu_4^2& 1
\end{pmatrix}
\begin{pmatrix}
1& 0\\
-\ii\tilde{u}_3& 1
\end{pmatrix}
\begin{pmatrix}
1& 0\\
-\ii \tilde{u}_4& 1
\end{pmatrix} \, .
\end{split}
\end{align}
We now implement the $t_2-t_4$ flip to go from Stokes graph \ref{Nf=4graphs}(B$'$) to graph \ref{Nf=4graphs}(C). The relevant Stokes graphs and 
contours are given in figure \ref{t2t4flipcontours}. Notice that the
base point of the contours in  figure \ref{t2t4flipcontours} is
different from that used in figure \ref{t1t3flipcontours}, however,
this is irrelevant because the
monodromy group is independent of the choice of a base point.
The monodromy matrices for the (B$'$) graph are given by
\begin{align}
\begin{split}
 M_{B',s_1} &=
 \begin{pmatrix}
  \nu_1& 0\\
  0& \frac{1}{\nu_1}
 \end{pmatrix}
 \begin{pmatrix}
  1& \frac{-\ii}{u_3\nu_1^2}\\
  0& 1
 \end{pmatrix}
 \begin{pmatrix}
  1& \frac{-\ii}{u_1u_{13}^2\nu_1^2}\\
  0 &1
 \end{pmatrix}
\begin{pmatrix}
 1 &-\frac{\ii}{u_4}\\
 0& 1
\end{pmatrix} \, ,\\
M_{B',s_2} &=
\begin{pmatrix}
\nu_2& 0\\
0& \frac{1}{\nu_2}
\end{pmatrix}
\begin{pmatrix}
1& \frac{\ii}{u_4\nu_2^2}\\
0& 1
\end{pmatrix}
\begin{pmatrix}
1& 0\\
-\ii u_1u_{13}^2\nu_1^2\nu_2^2& 1
\end{pmatrix}
\begin{pmatrix}
1& 0\\
-\ii u_2u_{24}^2& 1
\end{pmatrix}
\begin{pmatrix}
1& 0\\
-\ii u_4& 1
\end{pmatrix}
\begin{pmatrix}
1& \frac{-\ii}{u_4}\\
0& 1
\end{pmatrix} \, ,\\
M_{B',s_3} &=
\begin{pmatrix}
\nu_3& 0\\
0& \frac{1}{\nu_3}
\end{pmatrix}
\begin{pmatrix}
1& 0\\
\ii u_3\nu_3^2& 1
\end{pmatrix}
\begin{pmatrix}
1& \frac{-\ii}{u_2\nu_3^2\nu_4^2}\\
0& 1
\end{pmatrix}
\begin{pmatrix}
1& \frac{-\ii}{u_1}\\
0& 1
\end{pmatrix}
\begin{pmatrix}
1& \frac{-\ii}{u_3}\\
0& 1
\end{pmatrix}
\begin{pmatrix}
1& 0\\
-\ii u_3& 1
\end{pmatrix} \, ,\\
M_{B',s_4} &=
\begin{pmatrix}
\nu_4& 0\\
0& \frac{1}{\nu_4}
\end{pmatrix}
\begin{pmatrix}
1& 0\\
-\ii u_4\nu_4^2& 1
\end{pmatrix}
\begin{pmatrix}
1& 0\\
-\ii u_2\nu_4^2& 1
\end{pmatrix}
\begin{pmatrix}
1& 0\\
-\ii u_3& 1
\end{pmatrix} \, .
\end{split}
\end{align}
Finally, we consider Stokes graph \ref{Nf=4graphs}(C) and calculate the generators of the monodromy group
\begin{align}
\begin{split}
 M_{C,s_1} &=
 \begin{pmatrix}
  \nu_1& 0\\
  0& \frac{1}{\nu_1}
 \end{pmatrix}
 \begin{pmatrix}
  1& \frac{-\ii}{\tilde{u}_3\nu_1^2}\\
  0& 1
 \end{pmatrix}
 \begin{pmatrix}
  1& \frac{-\ii}{\tilde{u}_1\tilde{u}_{13}^2\nu_1^2}\\
  0 &1
 \end{pmatrix}
 \begin{pmatrix}
  1& \frac{-\ii}{\tilde{u}_2\tilde{u}_{24}^2}\\
  0& 1
 \end{pmatrix}
\begin{pmatrix}
 1 &-\frac{\ii}{\tilde{u}_4}\\
 0& 1
\end{pmatrix} \, ,\\
M_{C,s_2} &=
\begin{pmatrix}
\nu_2& 0\\
0& \frac{1}{\nu_2}
\end{pmatrix}
\begin{pmatrix}
1& \frac{\ii}{\tilde{u}_4\nu_2^2}\\
0& 1
\end{pmatrix}
\begin{pmatrix}
1& \frac{\ii}{\tilde{u}_2\tilde{u}_{24}^2\nu_2^2}\\
0 &1
\end{pmatrix}
\begin{pmatrix}
1& 0\\
-\ii\tilde{u}_1\tilde{u}_{13}^2\nu_1^2\nu_2^2& 1
\end{pmatrix}
\begin{pmatrix}
1& 0\\
-\ii\tilde{u}_2\tilde{u}_{24}^2& 1
\end{pmatrix}\\
&\hspace{3cm}\times
\begin{pmatrix}
1& \frac{-\ii}{\tilde{u}_2\tilde{u}_{24}^2} \\
0& 1
\end{pmatrix}
\begin{pmatrix}
1& \frac{-\ii}{\tilde{u}_4}\\
0& 1
\end{pmatrix} \, ,\\
M_{C,s_3} &=
\begin{pmatrix}
\nu_3& 0\\
0& \frac{1}{\nu_3}
\end{pmatrix}
\begin{pmatrix}
1& 0\\
\ii\tilde{u}_3\nu_3^2& 1
\end{pmatrix}
\begin{pmatrix}
1& \frac{-\ii}{\tilde{u}_4\nu_3^2\nu_4^2}\\
0& 1
\end{pmatrix}
\begin{pmatrix}
1& \frac{-\ii}{\tilde{u}_2\nu_3^2\nu_4^2}\\
0& 1
\end{pmatrix}
\begin{pmatrix}
1& \frac{-\ii}{\tilde{u}_1}\\
0& 1
\end{pmatrix}
\begin{pmatrix}
1& \frac{-\ii}{\tilde{u}_3}\\
0& 1
\end{pmatrix}
\begin{pmatrix}
1& 0\\
-\ii\tilde{u}_3& 1
\end{pmatrix} \, ,\\
M_{C,s_4} &=
\begin{pmatrix}
\nu_4& 0\\
0& \frac{1}{\nu_4}
\end{pmatrix}
\begin{pmatrix}
1& 0\\
-\ii\tilde{u}_4\nu_4^2& 1
\end{pmatrix}
\begin{pmatrix}
1& 0\\
-\ii\tilde{u}_3& 1
\end{pmatrix} \, .
\end{split}
\end{align}

\subsubsection*{The Stokes Automorphism for the Double Flip}
The monodromy matrices
obtained above by encircling the singularities in both the pairs of
graphs in the sequence of flips have standard traces, since all the singularities are regular.
Hence, in  order to compute the transformation of Voros symbols, we compute the traces
of products of monodromy matrices. We express the traces 
in terms of the variables $u_{13}$ and $u_{34}$ in the $t_1-t_3$ flip and in terms
of the variables $u_{42}$ and $u_{21}$ in the $t_2-t_4$ flip. 
%
%
Since the entries of the matrices are a bit cumbersome, we merely
present the results; here the variables in a given Stokes graph are denoted with the appropriate superscript:
This gives us the Stokes automorphism relations.
\begin{eqnarray}
\label{Stokes1u}
u_{13}^A=u_{13}^{B'},\ \ u_{34}^A=u_{34}^{B'}(1+u_{13}^{B'})
\end{eqnarray}
\begin{eqnarray}
 \label{Stokes2u}
u_{42}^{B'}=u_{42}^C,\ \ u_{21}^{B'}=u_{21}^C(1+u_{42}^C)
\end{eqnarray}
Upon composing the Stokes relations from the two single flips, we
obtain the desired Stokes relation for the double flip from Stokes
graph (A) to (C).
If we consider a counter-clockwise loop encircling both the branch cuts
and the four singularities in graphs
(A), (B$'$) and (C), we get the relation,
\begin{eqnarray}
\label{conditionA}
u_{13}u_{42}\nu_1\nu_2\nu_3\nu_4=1
\end{eqnarray}
There is an analogous condition that is given by 
 \begin{eqnarray}
\label{conditionB}
u_{43}u_{12}\nu_1\nu_2\nu_3\nu_4=1 \,.
\end{eqnarray}
Using these and the Stokes automorphisms for the sequence of flips, we obtain the following relations for the independent variables of the Stokes graphs (A) and (C): 
\begin{align}
u_{13}^A &=u_{13}^C \cr
u_{34}^A &=u_{34}^C\frac{1+u_{13}^C}{1+u_{42}^C} \,.
\end{align}
Because the double flip is composed of single flips, each taking place within their own arena, 
the final result for the double flip is a composition of the
result for single flips \cite{NakanishiandIwaki}. 

So far, we have implemented the double flip via a sequence of two
single flips, (A)$\rightarrow$ (B$'$)$\rightarrow$ (C) as in figure
\ref{Nf=4graphs}. The double flip can equivalently be implemented by
a different sequence of two single flips, (A)$\rightarrow$ (B)$\rightarrow$ (C) as shown in figure \ref{Nf=4graphs}. We have checked that this
results in the same Stokes automorphism relations as were arrived at
earlier. This is further confirmation of the rules for computing the monodromy
groups and of our resolution of the double flip into two
single flips.
\subsubsection*{Pops}
Finally, we consider two Stokes graphs that are
related by a pop rather than a flip, in the conformal gauge theory. 
We concentrate on the situation depicted in figure \ref{pops} (see appendix \ref{sec:StokesGraphs} for more details). 
This corresponds to
the degenerate triangulations in figure 74 of \cite{GMN}.
\begin{figure}[ht!]
\centering
\includegraphics[width=150mm]{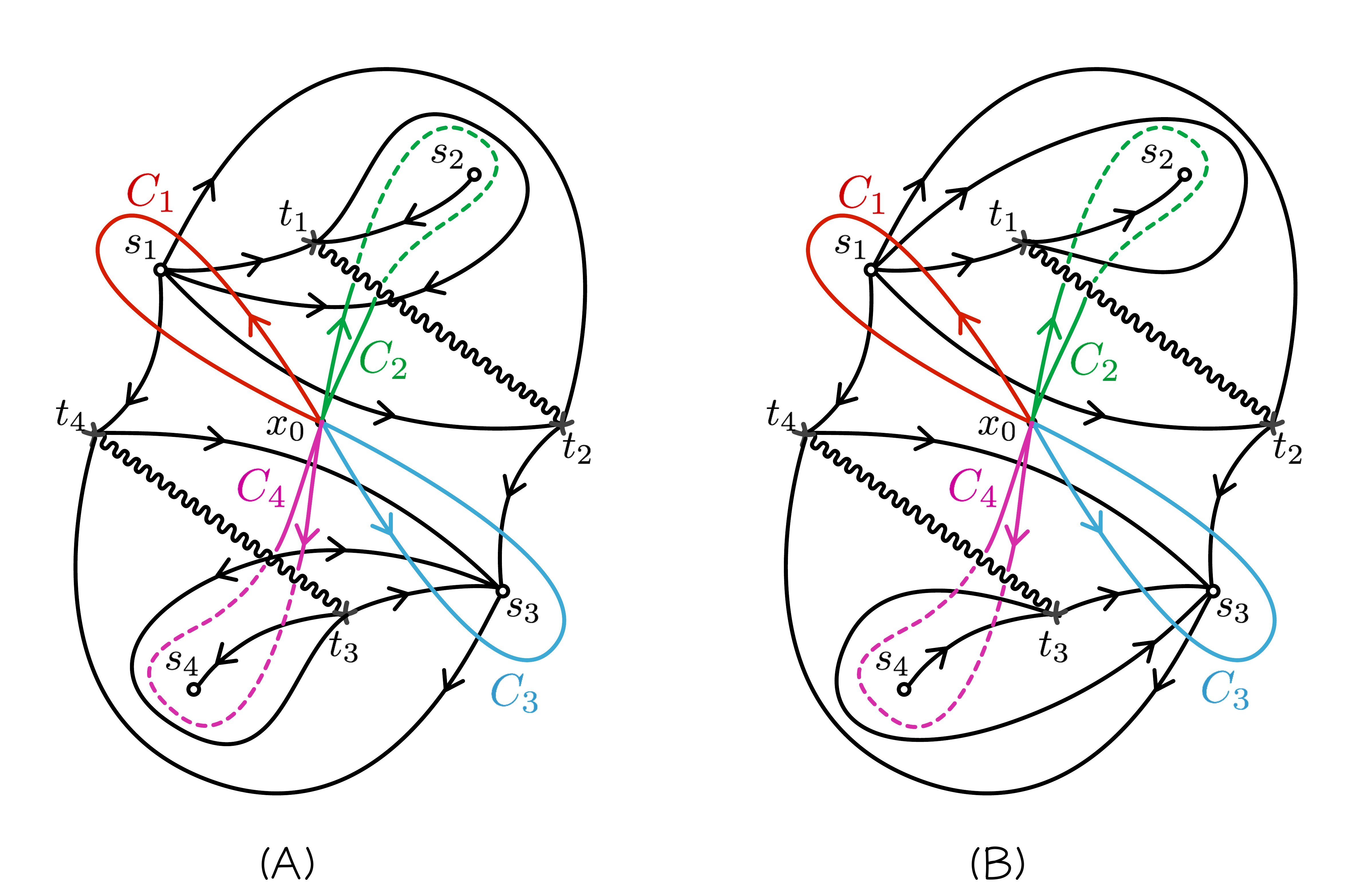}
\caption{The two Stokes graphs related by a pop. We refer the reader to \ref{BPop} for details.}
\label{pops}
\end{figure}
The pop is expected to give rise to a trivial Stokes automorphism for
our closed loop Voros symbols. We first
consider graph \ref{pops}(A) and calculate the generators of the monodromy
group. As the closed contour goes around $s_1$ counter-clockwise,
\begin{align}
M_{A,s_1} &=
\begin{pmatrix}
\nu_1& 0\\
0& \frac{1}{\nu_1}
\end{pmatrix}
\begin{pmatrix}
1& -\frac{i}{u_4\nu_1^2}\\
0& 1
\end{pmatrix}
\begin{pmatrix}
1& \frac{i\nu_2^2}{u_2u_{21}^2}\\
0& 1 
\end{pmatrix}
\begin{pmatrix}
1& \frac{-i}{u_1}\\
0& 1
\end{pmatrix}
\underbrace{
\begin{pmatrix}
0& i\\
i& 0
\end{pmatrix}
\begin{pmatrix}
1& 0\\
-\frac{i\nu_2^2}{u_1}& 1
\end{pmatrix}
\begin{pmatrix}
0& -i\\
-i& 0
\end{pmatrix}}
\begin{pmatrix}
1& -\frac{i}{u_2}\\
0& 1
\end{pmatrix} \, .
\end{align}
Next we compute the monodromy matrix as we traverse a closed contour that goes around the singularity $s_2$
\begin{align}
\begin{split}
M_{A,s_2} &=
\begin{pmatrix}
 \nu_2& 0\\
 0& \frac{1}{\nu_2}
\end{pmatrix}
\begin{pmatrix}
1 &\frac{i}{u_2\nu_2^2}\\
0 &1
\end{pmatrix}
\underbrace{
\begin{pmatrix}
0& i\\
i& 0
\end{pmatrix}
\begin{pmatrix}
1& 0\\
\frac{i}{u_1}& 1
\end{pmatrix}
\begin{pmatrix}
0& -i\\
-i& 0
\end{pmatrix}}
\begin{pmatrix}
0& i\\
i& 0
\end{pmatrix}
\begin{pmatrix}
1& iu_1\\
0& 1
\end{pmatrix}
\begin{pmatrix}
0& -i\\
-i& 0
\end{pmatrix}\\&\times
\underbrace{
\begin{pmatrix}
0& i\\
i& 0
\end{pmatrix}
\begin{pmatrix}
1& 0\\
\frac{-i\nu_2^2}{u_1}& 1
\end{pmatrix}
\begin{pmatrix}
0& -i\\
-i& 0
\end{pmatrix}}
\begin{pmatrix}
1& -\frac{i}{u_2}\\
0& 1
\end{pmatrix} \, .
\end{split}
\end{align}
Around the singularity $s_3$ and $s_4$, we find 
\begin{align}
\begin{split}
 M_{A,s_3}&=
 \begin{pmatrix}
  \nu_3& 0\\
  0& \frac{1}{\nu_3}
 \end{pmatrix}
 \begin{pmatrix}
  1& 0\\
  -iu_2\nu_3^2& 1
 \end{pmatrix}
\begin{pmatrix}
  1& 0\\
  \frac{iu_4u_{34}^2}{\nu_4^2}& 1
 \end{pmatrix}
\begin{pmatrix}
  1& 0\\
  -iu_3 &1
 \end{pmatrix}\\
&\hspace{2cm}\times
 \underbrace{
\begin{pmatrix}
  0& i\\
  i& 0
 \end{pmatrix}
\begin{pmatrix}
  1& \frac{-iu_3}{\nu_4^2}\\
  0& 1
 \end{pmatrix}
\begin{pmatrix}
  0& -i\\
  -i& 0
 \end{pmatrix}}
\begin{pmatrix}
  1& 0\\
  -iu_4& 1
 \end{pmatrix} \, ,\\
M_{A,s_4}&=
\begin{pmatrix}
\nu_4& 0\\
0& \frac{1}{\nu_4}
\end{pmatrix}
\begin{pmatrix}
1& 0\\
iu_4\nu_4^2& 1
\end{pmatrix}
\underbrace{
\begin{pmatrix}
0& i\\
i& 0
\end{pmatrix}
\begin{pmatrix}
1& iu_3\\
0& 1
\end{pmatrix}
\begin{pmatrix}
0& -i\\
-i& 0
\end{pmatrix}}
\begin{pmatrix}
0& i\\
i& 0
\end{pmatrix}
\begin{pmatrix}
1 &0\\
\frac{i}{u_3}& 1 
\end{pmatrix}
\begin{pmatrix}
0& -i\\
-i& 0
\end{pmatrix}\\&\times
\underbrace{
\begin{pmatrix}
0& i\\
i& 0
\end{pmatrix}
\begin{pmatrix}
1& \frac{-iu_3}{\nu_4^2}\\
0& 1
\end{pmatrix}
\begin{pmatrix}
0& -i\\
-i& 0
\end{pmatrix}}
\begin{pmatrix}
1& 0\\
-iu_4& 1
\end{pmatrix} \, .
\end{split}
\end{align}
In the above, the set of matrices clubbed together by an underbrace gives the rule 
for crossing a Stokes line that runs through a branch cut into the other sheet and approaches the turning point.
We repeat the above exercise for Stokes graph \ref{pops}(B) and the monodromy matrices
around the singularities in this case are given by
\begin{align}
\begin{split}
 M_{B_{s_1}} &=
 \begin{pmatrix}
 \nu_1& 0\\
 0& \frac{1}{\nu_1}
 \end{pmatrix}
\begin{pmatrix}
  1& \frac{-i}{u_4\nu_1^2}\\
  0& 1 
  \end{pmatrix}
 \begin{pmatrix}
  1& \frac{i\nu_2^2}{u_2u_{21}^2}\\
  0& 1
 \end{pmatrix}
 \begin{pmatrix}
  1 &\frac{-i\nu_2^2}{u_1}\\
  0 &1
 \end{pmatrix}
 \begin{pmatrix}
  1 &\frac{-i}{u_1}\\
  0 &1
 \end{pmatrix}
 \begin{pmatrix}
  1 &\frac{-i}{u_2}\\
  0 &1
 \end{pmatrix} \, , \\
M_{B_{s_2}}&=
\begin{pmatrix}
\nu_2& 0\\
0& \frac{1}{\nu_2}
\end{pmatrix}
\begin{pmatrix}
1& \frac{i}{u_2\nu_2^2}\\
0& 1
\end{pmatrix}
\begin{pmatrix}
0& i\\
i& 0
\end{pmatrix}
\begin{pmatrix}
1& -iu_1\nu_2^2\\
0& 1
\end{pmatrix}
\begin{pmatrix}
1& 0\\
\frac{i}{u_1}& 1
\end{pmatrix}
\begin{pmatrix}
1& 0\\
iu_1& 1
\end{pmatrix}
\begin{pmatrix}
0& -i\\
-i& 0
\end{pmatrix}
\begin{pmatrix}
 1& \frac{-i}{u_2}\\
 0& 1
\end{pmatrix} \, ,\\
 M_{B_{s_3}}&=
 \begin{pmatrix}
  \nu_3& 0\\
  0& \frac{1}{\nu_3}
 \end{pmatrix}
 \begin{pmatrix}
  1& 0\\
  -iu_2\nu_3^2& 1
 \end{pmatrix}
  \begin{pmatrix}
  1& 0\\
  \frac{iu_4u_{34}^2}{\nu_4^2}& 1
 \end{pmatrix}
  \begin{pmatrix}
  1& 0\\
  \frac{-iu_3}{\nu_4^2}& 1
 \end{pmatrix}
  \begin{pmatrix}
  1& 0\\
  -iu_3& 1
 \end{pmatrix}
  \begin{pmatrix}
  1& 0\\
  -iu_4& 1
 \end{pmatrix} \, ,\\
 M_{B_{s_4}}&=
 \begin{pmatrix}
  \nu_4& 0\\
  0& \frac{1}{\nu_4}
 \end{pmatrix}
\begin{pmatrix}
  1& 0\\
  iu_4\nu_4^2& 1
 \end{pmatrix}
\begin{pmatrix}
  0& -i\\
  -i& 0\\
 \end{pmatrix}
\begin{pmatrix}
  1& 0\\
  \frac{-i}{u_3\nu_4^2}& 1
 \end{pmatrix}
\begin{pmatrix}
  1& iu_3\\
  0& 1
 \end{pmatrix}
\begin{pmatrix}
  1& 0\\
  \frac{i}{u_3}& 1
 \end{pmatrix}
\begin{pmatrix}
  0& i\\
  i& 0
 \end{pmatrix}
\begin{pmatrix}
  1& 0\\
  -iu_4& 1
 \end{pmatrix} \, .
\end{split}
\end{align}
Again, it can be checked  that the monodromy matrices have standard 
traces in both the graphs. 
%
%
Similarly it can be checked that the product of matrices from both the graphs have the same traces.
Thus, we have a consistency check on our calculation, which is the triviality of the Stokes automorphism acting on the 
Voros symbols of graphs related by a pop \eqref{stokes pops}.




\section{The Gauge Theory Perspective}
\label{gaugetheory}
We have computed the monodromy groups associated to the differential
equations governing the instanton partition function with surface
operator insertion in terms of the exponents $\nu_i$ characterizing
the singularities, and the Borel resummed monodromies $u_{ij}$. The
characteristic exponents are readily calculated in terms of the masses
of the gauge theory using the explicit expression of the differential
equation. In this section, we further relate the exact
WKB parameters $u_{ij}$ with the Seiberg-Witten periods $a$ and $a_D$,
which are leading order approximations. As a result, we obtain the
monodromy groups in terms of the (deformed, resummed) gauge theory
data. Next, we present ideas on how to exploit this information
to obtain non-perturbative corrections to the prepotential.

\subsection{The Seiberg-Witten Variables}
We start by relating the monodromies $u_{ij}$ to more standard Seiberg-Witten
data. In the following figures, we mark only the turning points and the
singularities on the Riemann surface, and identify the $\hat{\alpha}$
and $\hat{\beta}$ cycles of the genus one Seiberg-Witten curve. For the conformal case
\cite{Ashok:2015gfa}, the cycles are identified as in figure \ref{Nf=4cycles}. 
The cycles have a smooth limit when the masses are
set to zero, i.e. when the turning points coincide with the
singularities. Further, in \cite{Ashok:2015gfa}, it was explicitly checked that the prepotential of the conformal theory, obtained by calculating the period integrals with this choice of cycles, matches the results from equivariant localization methods. Once this identification is made, it is possible to go 
down in the number of flavors sequentially, each time identifying the $\hat{\alpha}$ and
$\hat{\beta}$ cycles, until we finally reach the $N_f=0$ theory, where we obtain agreement with the results of \cite{Kashani-Poor:2015pca}.

\subsubsection*{The Conformal Theory}

\begin{figure}[ht!]
\centering
\includegraphics[width=75mm]{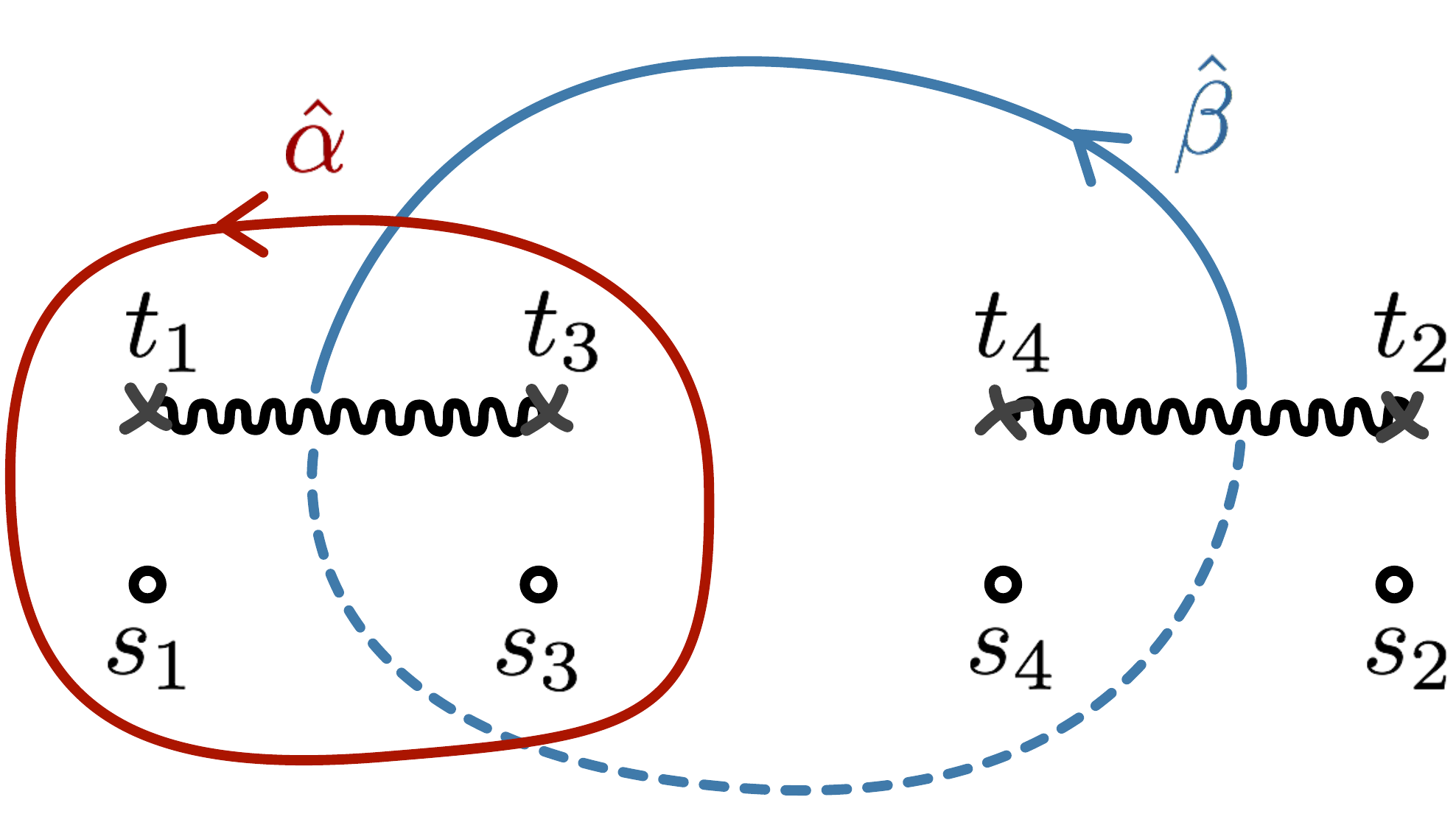}
\caption{$N_f=4$ cycles}
\label{Nf=4cycles}
\end{figure}

From figure \ref{Nf=4cycles},  we read off the identification
\begin{align}
u_{13}\nu_1\nu_3=e^{a},\ \ u_{34}\nu_3\nu_4=e^{a_D} \, . 
\end{align}

\noindent
The Stokes automorphisms we have derived for the conformal theory then imply the following relations for the gauge theory variables between the Stokes graphs \ref{Nf=4graphs}(A) and \ref{Nf=4graphs}(C):
 \begin{align}
 (e^a)_{A} &=(e^{a})_C\,\, ,\nonumber\\
(e^{a_D})_A &=
(e^{a_D})_C
\Big[
\frac{1+ \nu_1^{-1}\nu_3^{-1}(e^{a})_C}
{1+\nu_2^{-1}\nu_4^{-1}(e^{-a})_C}
\Big] \, .
\end{align}
We will comment on the meaning of these relations after we list the cycles and appropriate gauge theory variables for the asymptotically free cases. 

\subsubsection*{Three Flavours}

\begin{figure}[ht!]
\centering
\includegraphics[width=75mm]{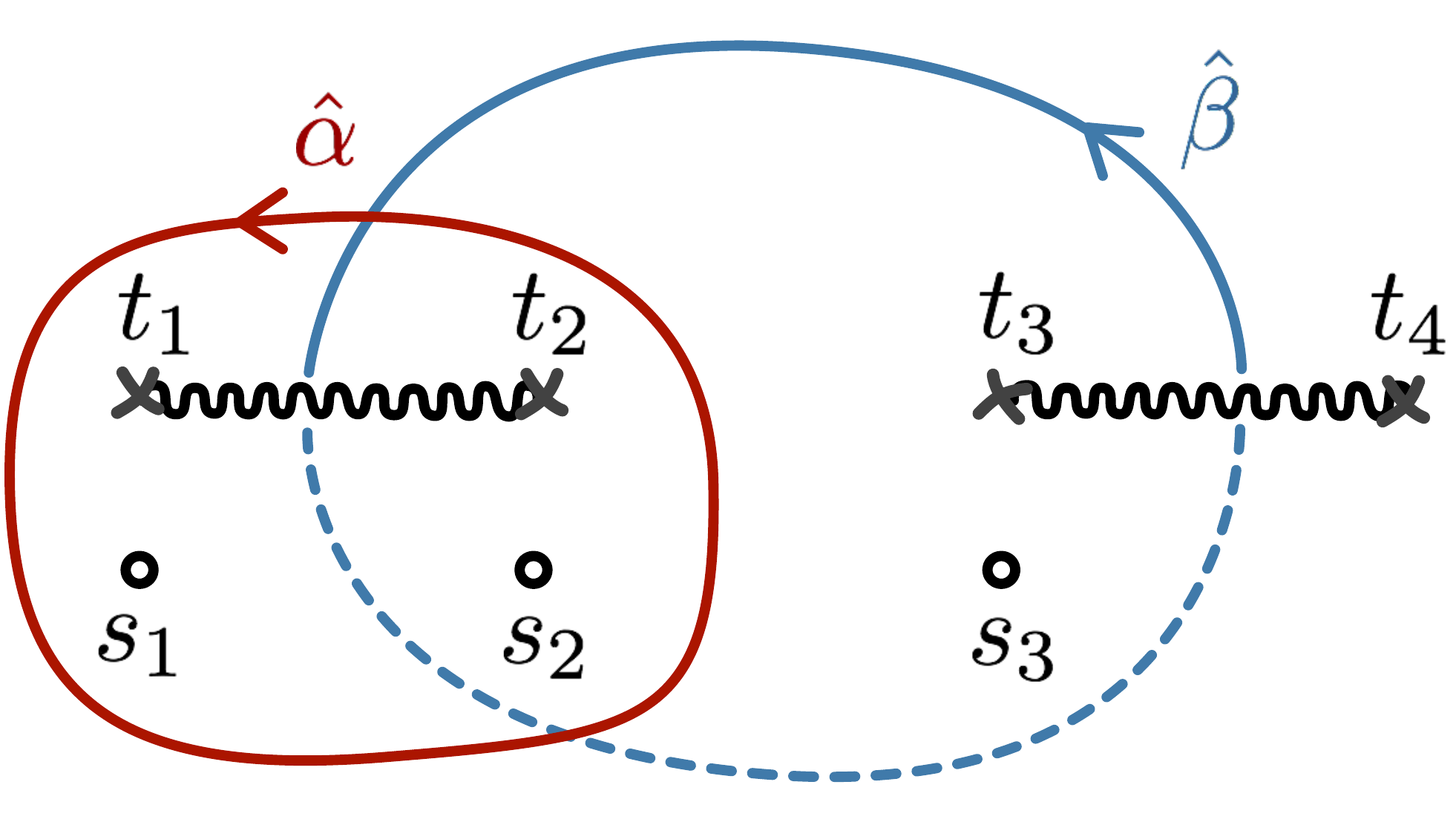}
\caption{$N_f=3$ cycles}
\label{Nf=3cycles}
\end{figure}

From figure \ref{Nf=3cycles}, for the three flavours case, we find the relation
\begin{align}
u_{12}\nu_1\nu_2=e^{a},\ \ u_{23}\nu_2\nu_3=e^{a_D}  \, .
\end{align}

\noindent
As before we can write the Stokes automorphisms in terms of the gauge
theory variables but we suppress these details and only give the
choice of cycles in our subsequent examples. 
%
\subsubsection*{Two Flavours}

\begin{figure}[ht!]
\centering
\includegraphics[width=75mm]{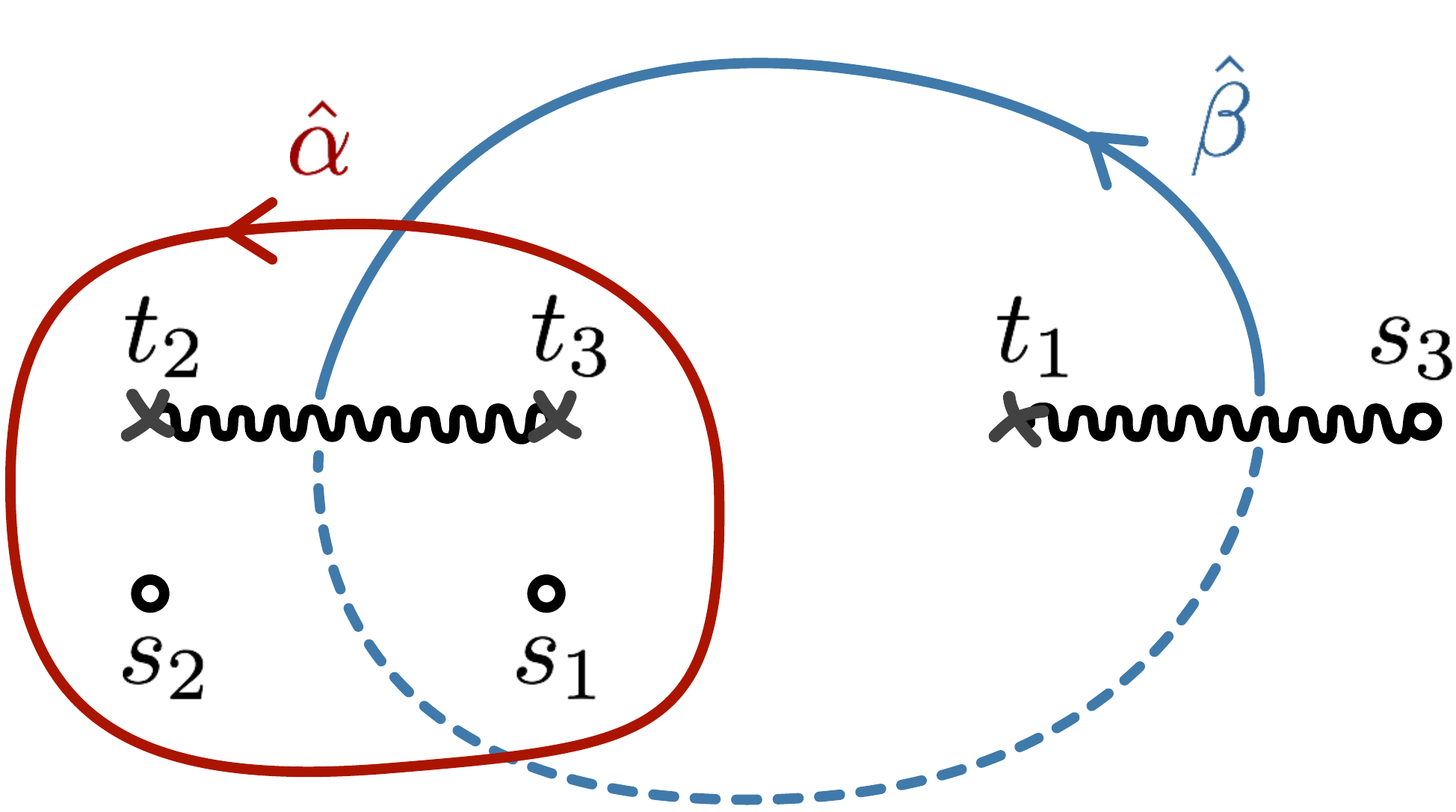}
\caption{$N_f=2$ cycles}
\label{Nf=2cycles}
\end{figure}

\noindent
For two flavours, from figure \ref{Nf=2cycles} we have,
\begin{align}
u_{23}\nu_1\nu_2=e^{a},\ \ u_{31}\nu_1=e^{a_D} \, .
\end{align}

%
%

\subsubsection*{One Flavour}

\begin{figure}[ht!]
\centering
\includegraphics[width=75mm]{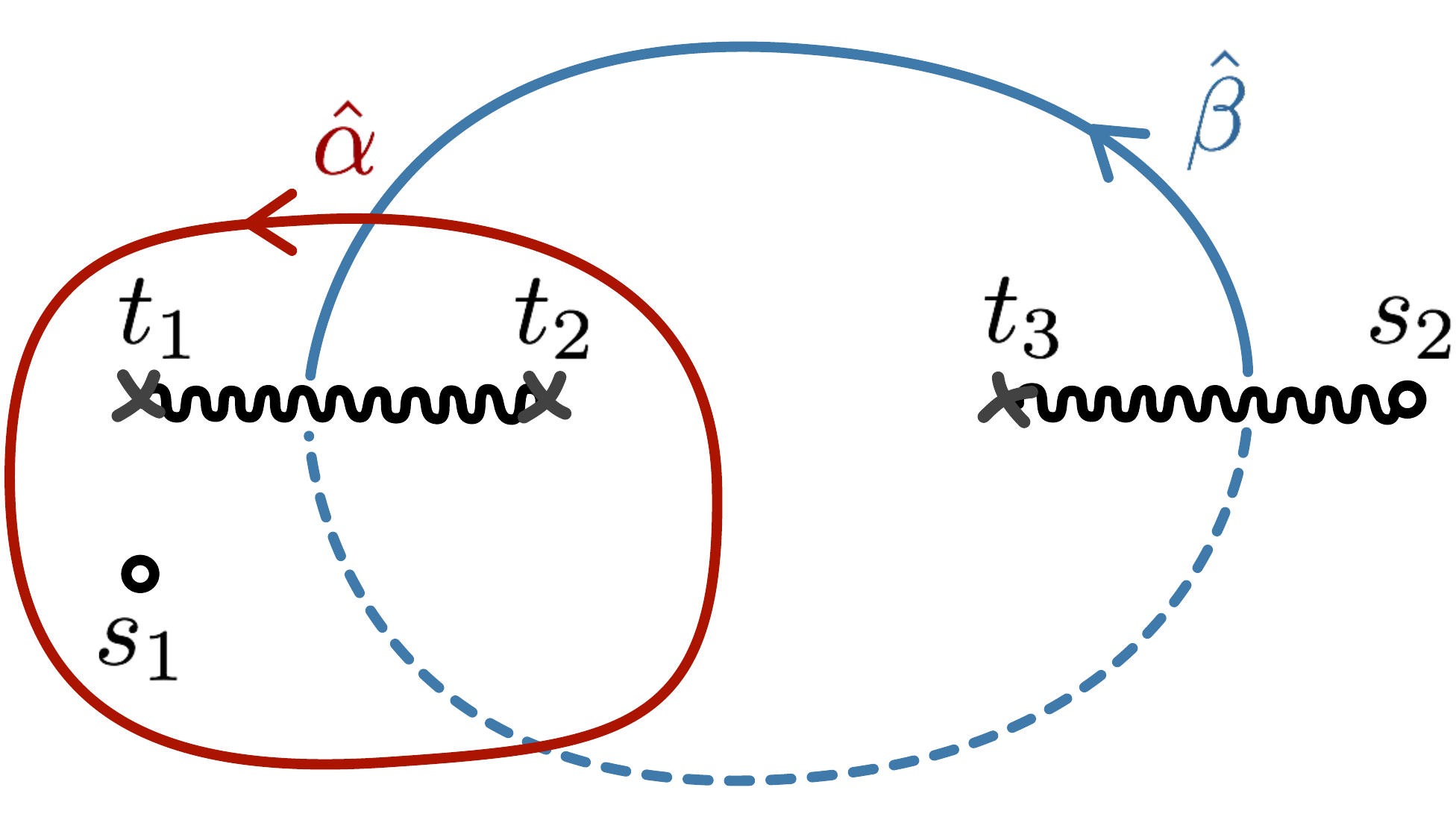}
\caption{$N_f=1$ cycles}
\label{Nf=1cycles}
\end{figure}

\noindent
When we are left with a single flavour, we find from figure \ref{Nf=1cycles}
\begin{align}
u_{12}\nu_1=e^{a},\ \ u_{23}=e^{a_D}  \, .
\end{align}
%

\subsubsection*{Pure Super Yang-Mills}

\begin{figure}[ht!]
\centering
\includegraphics[width=75mm]{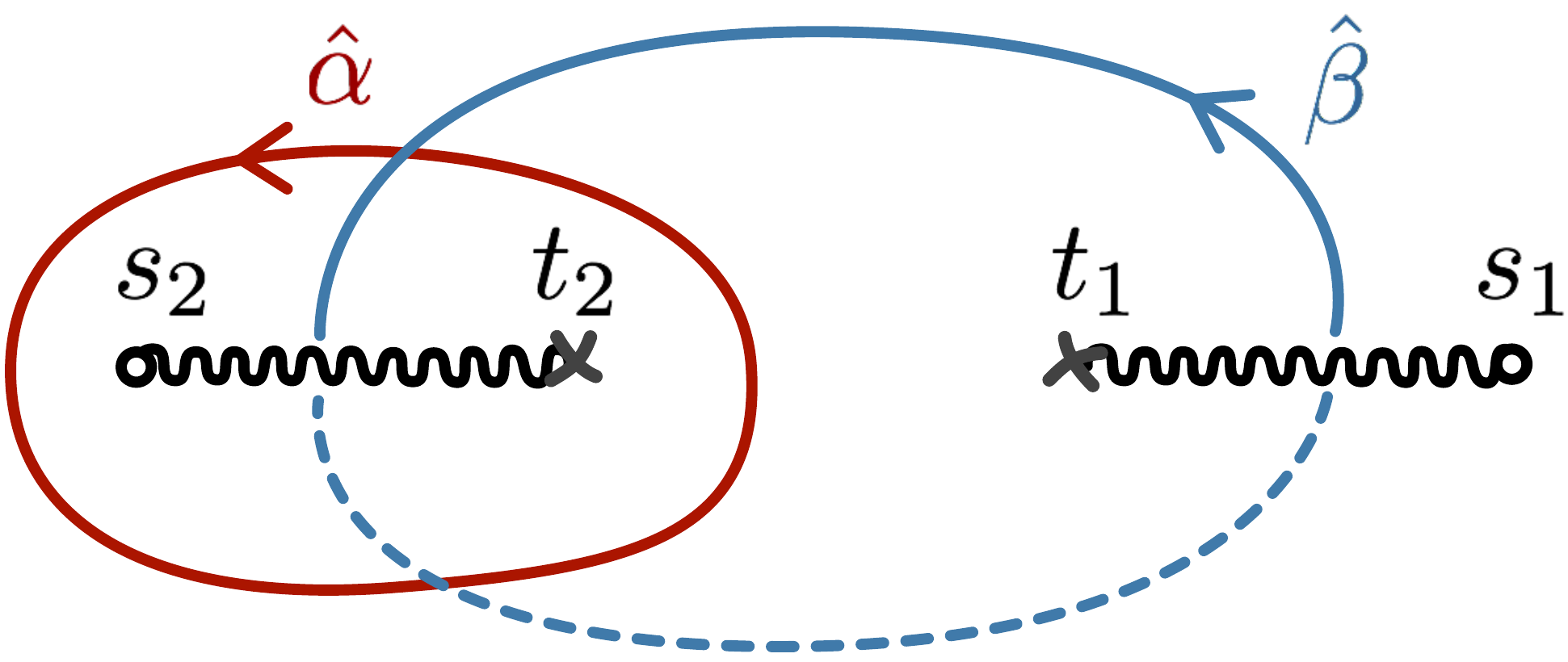}
\caption{$N_f=0$ cycles}
\label{Nf=0cycles}
\end{figure}

Finally, for pure super Yang-Mills, we have from figure \ref{Nf=0cycles}
\begin{align}
u_{21}=e^{a_D},\ \ x=e^{a}  \, .
\end{align}
The Stokes automorphisms derived in Section \ref{pureSYM} can then be
reinterpreted in terms of the gauge theory variables as a relation
between $a$ and $a_D$,
\begin{align}
e^{a_D} &=e^{\tilde{a}_D}\, ,\nonumber\\
e^{a} &=e^{\tilde{a}}(1+e^{\tilde{a}_D})\, .
\end{align}
One can check that these precisely coincide with the Stokes automorphisms obtained for the pure super Yang-Mills case in \cite{Kashani-Poor:2015pca}. 

\subsection{The Non-Perturbative Prepotential}

In this subsection, we restrict ourselves to some preliminary remarks on how to
exploit the results we have obtained on the monodromy group to find non-perturbative
corrections to the prepotential (parameterized in terms of the given exact
WKB monodromy data). We note that the expressions $e^{a(\epsilon_1)}$
and $e^{a_D(\epsilon_1)}$ used above (where we have now rendered
explicit the $\epsilon_1$ dependence) refer to the exact WKB, Borel
resummed $\epsilon_1$ perturbation series. At leading order in
$\epsilon_1$, the first term in the expansion matches the
Seiberg-Witten periods $a(0)$ and $a_D(0)$. In the $\epsilon_1$-deformed theory, these are corrected as a perturbation series in $\epsilon_1$. In
the exact WKB approach, these perturbation series are Borel resummed,
but the result depends on the region of Borel resummation, and the
resummed expressions $e^{a(\epsilon_1)}$ depend on the phase of Borel
resummation, or equivalently, on the phase of $\epsilon_1$. Thus, the
above identifications undergo the Stokes automorphisms of which we have
discussed many an example. Note that when we refer to such expressions, we
have in mind that they are valid with a given resummation angle.

The monodromy group itself, and in particular the gauge invariant
traces of products of monodromy matrices, are exact invariants of the
solution to the differential equations. In
\cite{Kashani-Poor:2015pca}, it was proposed in the context of the
theory with zero flavours to solve the equation for the Borel
resummed period $a(\epsilon_1)$ in terms of the invariant quantity
$a^{\text{exact}}$, determined by the exact periodicity (or Floquet exponent,
or monodromy) of the exact solution to the (Mathieu) differential
equation. This solution is valid at a particular resummation angle, 
and $a$ is identified with $a^{\text{exact}}$, up to non-perturbative corrections.
Near a point in moduli space where a perturbative expansion is possible, like the
weak, magnetic or dyonic point, we can then solve the relation in terms of 
a transseries \cite{Costin}.
On the other hand, since in this case $a_D(\epsilon_1)$ is independent
of the Borel resummation angle, we can posit
$a_D^{\text{exact}}=a_D(\epsilon_1)$ (independently of the specification of
the resummation angle). 
The variable $u$ determining the derivative
of the prepotential is known as a function of $a$, and therefore as a function of $a^{\text{exact}}$. Inverting this relation and calculating the dual period integrals allows one to calculate non-perturbative corrections to the prepotential ${\cal F}$.

Thus, one application of the identifications above is to attempt to integrate up the
non-perturbative relation between the period $a$ and
the parameters coding the exact monodromy group, towards
non-perturbative corrections to the prepotential ${\cal F}$. The encompassing 
case in which to execute this program is the conformal
$N_f=4$ theory.

\section{Conclusions}
\label{conclusions}
In our work, we filled in many details of the connection between
conformal field theory and four-dimensional gauge theory which was
made handily available starting from the matching of partition
functions and correlators in \cite{Alday:2009fs}. We applied the
technology developed in \cite{Awata:2010bz,AwataYamada,Gaiotto:2012sf,Gaiotto:2009ma} to study in greater detail 
SU$(2)$ super Yang-Mills theories with a varying number of flavours. We thus provided a
complete list of $\epsilon_i$-deformed differential equations
satisfied by the five-point conformal block with surface operator
insertion, in the irregular conformal block limit. Subsequently, we
took a semi-classical limit and analyzed the resulting differential equations with
exact WKB methods. We used this technology to give
a detailed parameterization of the monodromy groups in terms of 
Voros symbols and external conformal dimensions or, in the language of gauge theory, in terms of Borel resummed $\epsilon_1$-deformed Seiberg-Witten periods and masses. The Borel
resummed variables depend intrinsically on the resummation angle (and
the phase of $\epsilon_1$), and we illustrated how to bridge this
ambiguity using Stokes automorphisms, with at least one illustration for each
number of flavours.

The Stokes automorphisms we obtained are consistent with the general
theorems proved in \cite{NakanishiandIwaki}. The one subtlety arose
in the case of the conformal gauge theory: in this case, we analyzed a
pair of Stokes graphs that were related by a simultaneous flip along
two independent finite WKB lines. In such a case, we showed that the
resulting Stokes automorphisms could be obtained by treating the
double-flip as a sequence of single flips. We demonstrated consistency of
this approach by checking that the final Stokes automorphisms between
the Stokes graphs were independent of the order in which the single
flips were taken.

We believe it is instructive to develop these most elementary of
${\cal N}=2$ models in still further detail. There are many avenues to
explore. As was mentioned in the previous section, it should be
possible to calculate non-perturbative corrections in $\epsilon_1$ to
the prepotential, by suitably generalizing the analysis that was done for
the case of pure super Yang-Mills \cite{Kashani-Poor:2015pca}. Similarly, it
should be informative to match the cluster algebra
description of Stokes automorphisms of \cite{NakanishiandIwaki} to
the cluster algebra description of the spectrum of BPS states
\cite{KS}. (See also e.g. \cite{Gaiotto:2010be,Alim:2011kw,NakanishiandIwaki}.)

In the present work, we studied general properties of the Borel
resummed wave-functions of the null vector decoupling equations in the
semi-classical limit, without focusing on the specific form of the
wave-function. An interesting direction would be to study in greater
detail the dependence of the five-point conformal block on the
insertion point of the degenerate operator. This should yield
non-trivial information about the gauge theory in the presence of a
surface operator. Most interestingly, 
perturbative and non-perturbative corrections in $\epsilon_2$ can now
be studied from the differential equation point of view. 

All these projects have straightforward extensions, both to the
${\mathcal N} = 2^\star$ theory, where the Riemann surface is of genus
one \cite{Alday:2009aq}, as well as to higher rank conformal gauge
theories, the simplest of which is super Yang-Mills theory with
SU$(N)$ gauge group and $2N$ fundamental flavours. As explained in
Section \ref{gaugetheory}, one not only needs the description of the
monodromy group, but also requires a local expansion of the period
integrals in terms of the Coulomb moduli in order to carry out the
goal of calculating corrections to the prepotential that are
non-perturbative in $\epsilon_1$. For the ${\mathcal N} = 2^\star$
theory with gauge group SU$(2)$, the null vector decoupling equation
for the toroidal block has been well studied in the semi-classical
limit (see e.g.
\cite{Fateev:2009aw,Piatek:2013ifa, He:2014yka}) and the instanton series for the prepotential has been resummed
in terms of modular functions \cite{Kashani-Poor:2014mua}. One can therefore
attempt to study the monodromy group along the lines of the present
paper and hope to carry out the proposal put forward in Section
\ref{gaugetheory}. For higher rank ${\mathcal N}=2^\star$ theories,
there has been much progress on the gauge theory side in resumming the
instanton expansion for the prepotential using modular anomaly
equations in deformed gauge theories with arbitrary gauge group
\cite{Billo:2015ria, Billo:2015jta}. It remains an open problem to
reproduce these successes using conformal field theory methods and to
do the corresponding WKB analysis.

For the higher rank (and undeformed) SQCD theories, the instanton series has been resummed in a special locus with $\mathbb{Z}_N$ symmetry \cite{Ashok:2015cba, Ashok:2016oyh}. From the CFT approach to the problem, one has to work with Toda theory \cite{Wyllard:2009hg} and the corresponding null vector decoupling equations in Toda have been analyzed recently in \cite{PoghossianToda}. In the semi-classical limit, such differential equations can also be derived using a saddle point analysis of the Nekrasov integrand \cite{NPS13} and the resulting deformed Seiberg-Witten curve. It would be interesting to analyze these higher order differential equations using exact WKB methods and make contact with the general approach to these systems using spectral networks \cite{Gaiotto:2012rg, Gaiotto:2012db}.

\section*{Acknowledgments}

We would like to thank Anirban Basu, Soumyadeep Bhattacharya, Eleonora Dell'Aquila, Anshuman Maharana, Gautam Mandal, Shiraz Minwalla, and Ashoke Sen for useful discussions. We would especially like to thank Eleonora Dell'Aquila for help with the pictures. 
S.A. and D.P.J. would like to thank TIFR, Mumbai for hospitality. D.P.J. would also like
to thank IMSc, Chennai for hospitality during the course of this work. 
D.P.J. was partially supported by DAE XII-plan grant 12-R\& D-HRI-5.02-0303.
J.T. would like to acknowledge support from the grant ANR-13-BS05-0001.

\begin{appendix}

\section{The Null Vector Decoupling with Irregular Blocks}
\label{diffeqns}

In this section, we derive in detail the null vector decoupling equations that involve irregular conformal blocks. 
These equations were summarized in a different form in  \cite{Awata:2010bz}.
Our final results are listed in section \ref{CFT}.

\subsection{One Irregular Puncture}
The case of $N_f=4$ is standard and is described in section \ref{CFT} in sufficient detail.
We turn to the conformal block involving an irregular puncture. This involves
rendering one flavour very massive.
\subsubsection{$N_f=3$}
{From} the gauge theory point of view, the flavour decoupling is carried out by taking the limit:
\be
m_2 \rightarrow \infty\qquad q\rightarrow 0\quad \text{with}\quad \Lambda_3 = q\, m_2 \quad \text{finite}\,. 
\ee
The parameter $\Lambda_3$ has mass-dimension one and is the strong coupling scale of the SU$(2)$ gauge theory with $N_f=3$. 
As shown in \cite{Gaiotto:2012sf}, in the conformal field theory this involves a collision of the regular conformal block at $z=q$ and $z=0$ and leads to an irregular conformal block at $z=0$. We now consider the  five point conformal block with the insertion of the degenerate field $\Phi_{2,1}(z)$:
\be
\langle V_{\alpha_4}(\infty)\, V_{\alpha_3}(1) {\cal I}^{(4)}(0) \, \Phi_{2,1}(z) \rangle \, .
\ee
The Ward identity for this chiral conformal block can be obtained by considering the five point block relevant for the conformal $N_f=4$ and scaling the wave-function in the following way \cite{Awata:2010bz,Gaiotto:2012sf}:
\be
\Psi(z,q) = q^{-2\alpha_1\alpha_2}\, \psi_3(z, \Lambda_3) \, .
\ee
On the rescaled wave-function $\psi_3(z)$, the $q$-derivative is traded for a $\Lambda_3$-derivative. Combining these, and taking the decoupling limit, we obtain the 
 null vector decoupling equation for the $N_f=3$ theory:
\begin{align}
\Bigg[&-\epsilon_1^2 \frac{\p^2}{\p z^2} +\frac{(m_3+m_4)^2}{4(z-1)^2}+ \frac{m_3 m_4}{z(1-z)} +\frac{m_1\Lambda_3}{z^3}+\frac{\Lambda_3^2}{4z^4}-\frac{\epsilon_1^2}{4(z-1)^2} +\epsilon_2^2\frac{(3-4z)}{4z(z-1)^2}\nonumber\\ 
&+\frac{1}{z^2-z^3}\Big(-\epsilon_1\epsilon_2\Lambda_3\frac{\p}{\p\Lambda_3}+ m_1^2+m_1(\epsilon_1+\epsilon_2)\Big)+\epsilon_1\epsilon_2\Big(\frac{1-2z}{z-z^2}\frac{\p}{\p z} + \frac{1-2z}{2z(z-1)^2}\Big)\Bigg]\psi_3(z,\Lambda_3) = 0 \, .
\end{align}
The quartic pole at $z=0$ in the OPE between the stress tensor and the irregular block explains our notation: we denote such an irregular block by ${\cal I}^{(4)}(0)$.

\subsubsection{$N_f=2$: asymmetric realization}

One can take a further limit in which we decouple the mass $m_1$:
\be
m_1 \rightarrow \infty\qquad \Lambda_3\rightarrow 0\quad \text{with}\quad \Lambda_2^2 = m_1\Lambda_3 \quad \text{finite}\,. 
\ee
Simultaneously, we rescale the wave-function by 
\be
\psi_3(z,\Lambda_3) = \Lambda_3^{\frac{1}{\epsilon_1\epsilon_2}(m_1^2 + m_1(\epsilon_1+\epsilon_2))} \psi_2(z, \Lambda_2) \,.
\ee
From the conformal field theory perspective, this amounts to tuning the coefficient of the quartic pole to zero, leaving behind only a cubic pole. This corresponds to the four point conformal block
\be
\langle V_{\alpha_4}(\infty)\, V_{\alpha_3}(1) {\cal I}^{(3)}(0) \, \Phi_{2,1}(z) \rangle \, .
\ee
The null vector decoupling equation takes the  form:
\begin{multline}
\left[-\epsilon_1^2 \frac{\p^2}{\p z^2} +\frac{(m_3+m_4)^2}{4(z-1)^2}+ \frac{m_3 m_4}{z(1-z)} +\frac{\Lambda_2^2}{z^3} -\frac{\epsilon_1\epsilon_2}{2(z^2-z^3)}\Lambda_2\frac{\p}{\p\Lambda_2}
\right.\cr
\left.+\epsilon_1\epsilon_2\left(\frac{1-2z}{z-z^2}\frac{\p}{\p z} + \frac{1-2z}{2z(z-1)^2} \right)-\frac{\epsilon_1^2}{4(z-1)^2} +\epsilon_2^2\frac{(3-4z)}{4z(z-1)^2} \right]\psi_2(z,\Lambda_2) = 0 \, .
\end{multline}
The equation exhibits a cubic pole at the irregular singularity $z=0$.

\subsection{Two Irregular Punctures}

Let us begin with the five point conformal block in which we have
regular conformal primaries at $z_i$, with $i \in \{1,2,3,4\}$ and the
degenerate field at $z$. In equation \eqref{tpsi}, we have obtained
the null vector decoupling equation for this case. We now set $z_1=0$,
$z_2=q$ and $z_4=1$ and consider the simultaneous collision of
punctures such that $q\rightarrow 0$ and $z_3 \rightarrow 1$. We
rescale the wave-function as in \cite{Gaiotto:2012sf}
\be
\Psi(z_i) = q^{-2\alpha_1\alpha_2} (z_3-1)^{-2\alpha_3\alpha_4} \psi(z,\Lambda,\widetilde{\Lambda})   \, . 
\ee
In order to get finite results, we take the limit
\begin{align}
m_2 &\rightarrow \infty\quad,\quad q\rightarrow 0\quad \text{with}\quad \Lambda = q\, m_2 \quad \text{finite}\, , \cr 
m_3 &\rightarrow \infty\quad,\quad (z_3-1)\rightarrow 0\quad \text{with}\quad \widetilde{\Lambda} = (z_3-1)\, m_3 \quad \text{finite}
\, .
\end{align}
Note that this introduces two independent scales in the problem. The conformal block we are considering is a three point function, with two irregular punctures of quartic order and a degenerate field $\Phi_{2,1}(z)$:
\be
\langle {\cal I}^{(4)}(1) {\cal I}^{(4)}(0) \, \Phi_{2,1}(z) \rangle \, .
\ee
The parameters $\Lambda$ and $\widetilde{\Lambda}$ respectively represent the quartic pole coefficient at $z=0$ and $z=1$. After the  wave-function rescaling and the decoupling limits, we obtain the null vector decoupling equation for such a conformal block with two such irregular singularities and the degenerate field:
%

\begin{align}
\Bigg[-\epsilon_1^2 \frac{\p^2}{\p z^2} +\frac{\Lambda^2}{4z^4}&+\frac{\Lambda m_1}{z^3(z-1)^2}+\frac{\widetilde{\Lambda} m_4}{z(z-1)^3} +\frac{\widetilde{\Lambda}^2}{4(z-1)^4}+\epsilon_1\epsilon_2 \frac{3z-1}{z(z-1)}\frac{\p}{\p z}+\frac{(2\epsilon_1\epsilon_2+3\epsilon_2^2)(2-3z)}{4z(z-1)^2}\nonumber\\
&+\frac{1}{z^2(z-1)^2}\Big(-\epsilon_1\epsilon_2\Lambda\frac{\p}{\p\Lambda} -2\Lambda m_1+ m_1^2+m_1(\epsilon_1+\epsilon_2)\Big)\Bigg]\psi_2(z, \Lambda, \widetilde{\Lambda}) = 0\,.
\end{align}
We can tune the two scales suitably in order to obtain the null vector decoupling equations for the remaining gauge theories
of our focus.

\subsubsection{$N_f=2$: symmetric realization}

We set $\Lambda = -\widetilde{\Lambda} = \Lambda_2$, in which case we obtain the  null vector decoupling equation: 
\begin{align}
\Bigg[-\epsilon_1^2 \frac{\p^2}{\p z^2} +\frac{\Lambda_2^2}{4z^4}  &+ \frac{\Lambda_2m_1}{z^3(z-1)^2}-\frac{\Lambda_2 m_4}{z(z-1)^3} +\frac{\Lambda_2^2}{4(z-1)^4}+\epsilon_1\epsilon_2 \frac{3z-1}{z(z-1)}\frac{\p}{\p z}+\frac{(2\epsilon_1\epsilon_2+3\epsilon_2^2)(2-3z)}{4z(z-1)^2} \nonumber\\ 
&+\frac{1}{z^2(z-1)^2}\Big(-\epsilon_1\epsilon_2\Lambda_2\frac{\p}{\p\Lambda_2} -2\Lambda_2m_1+ m_1^2+m_1(\epsilon_1+\epsilon_2)\Big)\Bigg]\psi_2(z, \Lambda_2) = 0\,.
\end{align}

\subsubsection{$N_f=1$}

We set $\Lambda = \Lambda_1$ and $\widetilde{\Lambda}\rightarrow 0$ such that $\widetilde{\Lambda}m_4 = -\frac{\Lambda_1^2}{4}$ is a finite combination. In other words, the relevant conformal block of interest is 
\be
\langle {\cal I}^{(3)}(1) {\cal I}^{(4)}(0) \, \Phi_{2,1}(z) \rangle\, .
\ee
The coefficient of the cubic pole at $z=1$ is tuned and related to that of the quartic pole at $z=0$. 
Such a conformal block satisfies the  null vector decoupling equation:
\begin{align}
\Bigg[-\epsilon_1^2 \frac{\p^2}{\p z^2} &+\frac{\Lambda_1^2}{4z^4} + \frac{\Lambda_1m_1}{z^3(z-1)^2}-\frac{\Lambda_1^2}{4z(z-1)^3}+\epsilon_1\epsilon_2 \frac{3z-1}{z(z-1)}\frac{\p}{\p z}+\frac{2-3z}{4z(z-1)^2}(2\epsilon_1\epsilon_2+3\epsilon_2^2)\nonumber\\
&+\frac{1}{z^2(z-1)^2}\Big(-\epsilon_1\epsilon_2\Lambda_1\frac{\p}{\p\Lambda_1} -2\Lambda_1m_1+ m_1^2+m_1(\epsilon_1+\epsilon_2)\Big)
\Bigg]\psi_1(z, \Lambda_1) = 0\,.
\end{align}

\subsubsection{$N_f=0$}

As for the earlier case with $N_f=2$ in the asymmetric realization, in order to obtain finite results, one has to  rescale the wave-function
\be
\psi(z, \Lambda, \widetilde{\Lambda}) = \Lambda^{\frac{1}{\epsilon_1\epsilon_2}(m_1^2 + m_1(\epsilon_1+\epsilon_2))} \psi_0(z, \Lambda_0) \,.
\ee
We set $\Lambda\rightarrow 0$ and $\widetilde{\Lambda}\rightarrow 0$ such that the combinations $m_1\Lambda = \Lambda_0^2$ and $m_4\widetilde{\Lambda} = \Lambda_0^2$ are equal and finite. The relevant conformal block is given by
\be
\langle {\cal I}^{(3)}(1) {\cal I}^{(3)}(0) \, \Phi_{2,1}(z) \rangle \,.
\ee
This satisfies the null vector decoupling equation:
\begin{align}
\Bigg[-\epsilon_1^2 \frac{\p^2}{\p z^2}  + \frac{\Lambda_0^2}{z^3(z-1)^2}+ \frac{\Lambda_0^2}{z(z-1)^3}&+\epsilon_1\epsilon_2 \frac{3z-1}{z(z-1)}\frac{\p}{\p z}+\frac{2-3z}{4z(z-1)^2}(2\epsilon_1\epsilon_2+3\epsilon_2^2)\nonumber\\
&+\frac{1}{z^2(z-1)^2}\Big(-\frac{1}{2}\epsilon_1\epsilon_2\Lambda_0\frac{\p}{\p\Lambda_0 }- 2\Lambda_0^2\Big) 
\Bigg]\psi_0(z, \Lambda_0) = 0\,.
\end{align}

\section{Stokes Graphs}
\label{sec:StokesGraphs}
In this section, we plot actual machine-generated Stokes graphs for the $N_f=4$ theory as representative examples. The Stokes lines are defined by the condition
\begin{equation}
\text{Im} \left[ \int_{x_0}^{x} \text{d}x' \ \sqrt{Q_0(x')} \right] = 0 \, .
\end{equation}
The red dots indicate singularities, and the blue dots indicate
turning points. It is important to remember that what is relevant for
the monodromy calculations is the topology of the Stokes
graph. The cartoons in the body of the paper abstract away from graphs
given in this appendix.

\subsection{The Double Flip}
\label{BDoubleFlip}
\begin{figure}
    \centering
    \begin{subfigure}[b]{0.45\textwidth}
        \includegraphics[width=\textwidth]{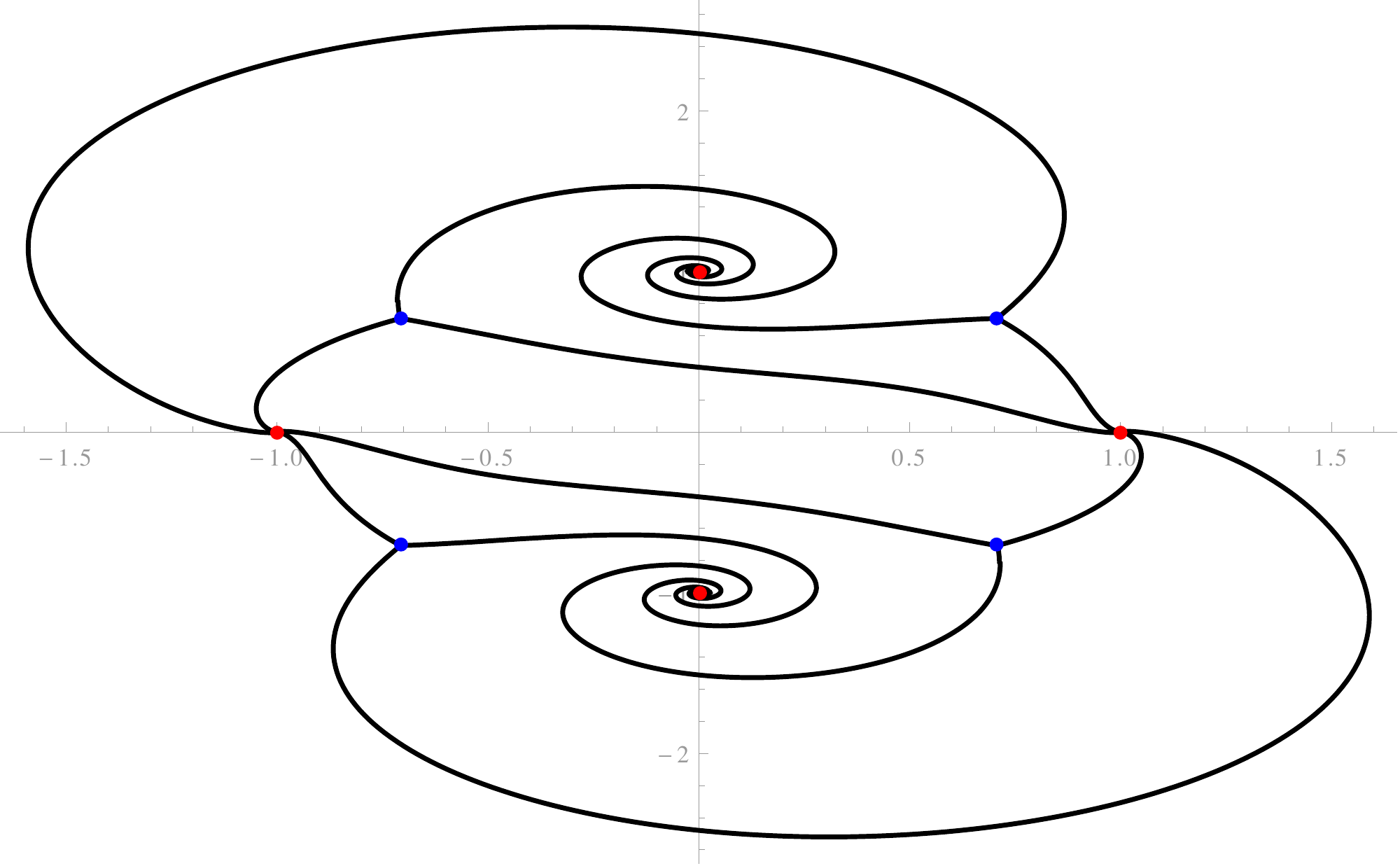}
        \caption{Before the double flip}
        \label{fig:PreFlip}
    \end{subfigure}
    ~ 
    \begin{subfigure}[b]{0.45\textwidth}
        \includegraphics[width=\linewidth]{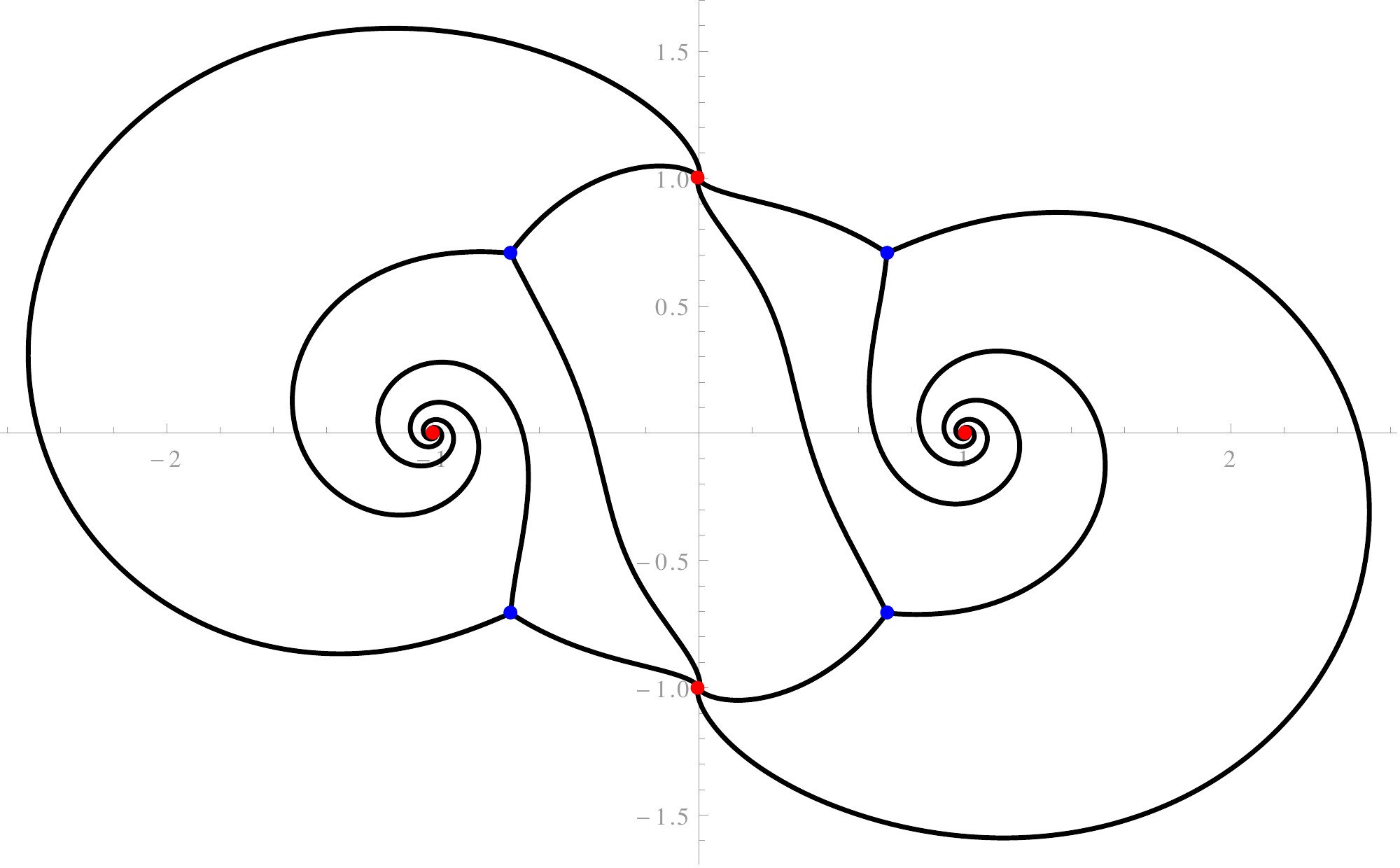}
  \caption{After the double flip.}\label{fig:PostFlip}
    \end{subfigure}
    ~ 
    \begin{subfigure}[b]{0.45\textwidth}
        \includegraphics[width=1\linewidth]{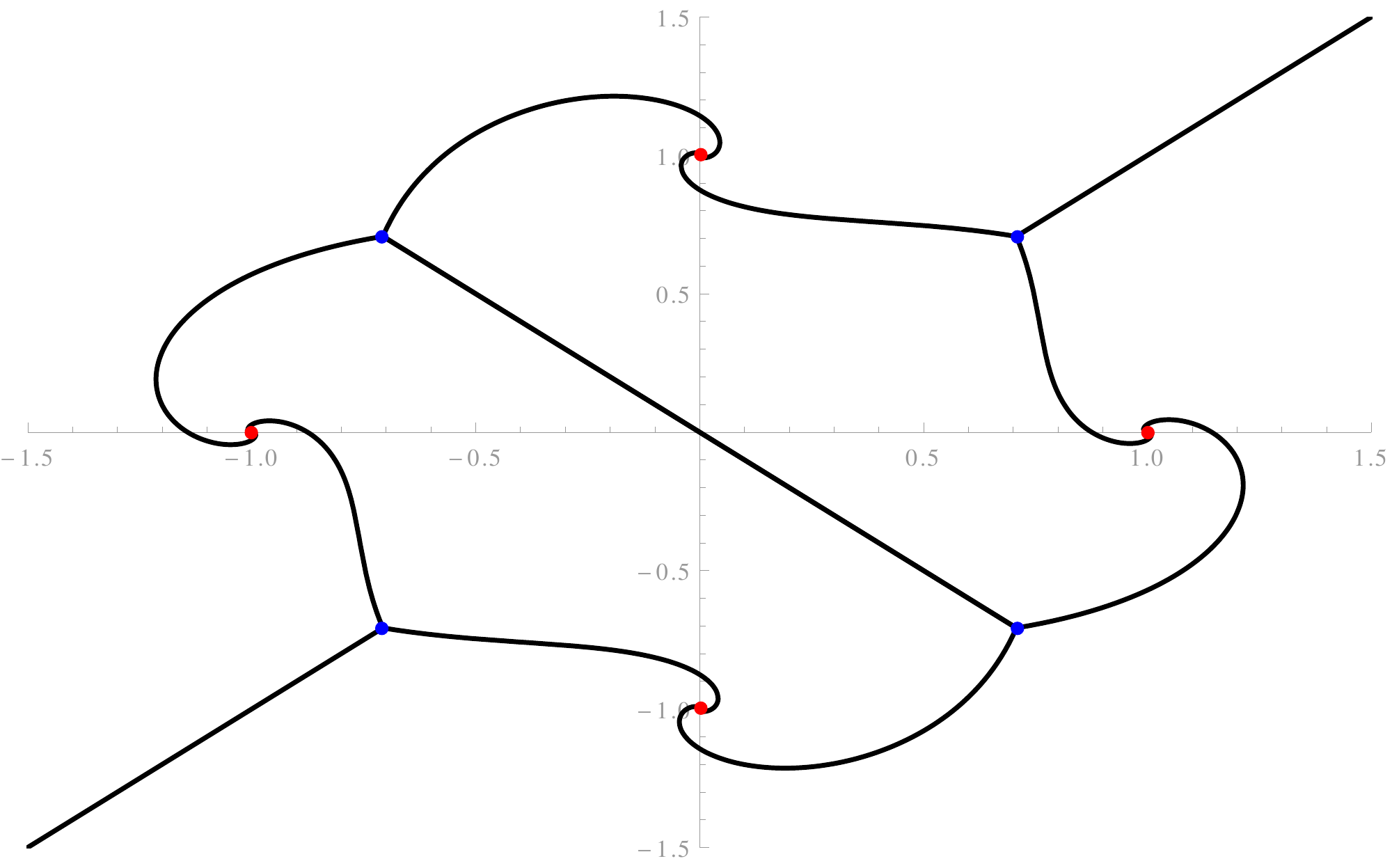}
\caption{The critical graph.}\label{fig:CritFlip}
\end{subfigure}
    \caption{Sequence of Stokes graphs related by a double flip about the critical graph}\label{fig:flips}
\end{figure}

The potential we use to plot these Stokes graphs has a convenient $\mathbb{Z}_4$ symmetric form \cite{GMN}
\begin{equation}
\label{eq:GMNZ4}
Q_0(z) = \frac{z^4 - u \left( z^4 - 1 \right)}{\left( z^4 - 1 \right)^4} \, ,
\end{equation}
where the singularities are at $z_1 = 1$, $z_2 = \text{i}$, $z_3 = -1$, and $z_4 = -\text{i}$. This potential is arrived at via an SL($2,\mathbb{C}$) transformation of \eqref{nullNf4Qfunctions} that maps all singularities to finite points for ease of plotting.
The masses of the fundamental hypermultiplets are $m_a =
\frac{1}{4} z_a$.
We have made the choice $u = \frac{1}{2}$ for plotting the double flip
Stokes graphs in figure \ref{fig:flips}.

\subsection{The Pop}
\label{BPop}
\begin{figure}
    \centering
    \begin{subfigure}[b]{0.45\textwidth}
        \includegraphics[width=\textwidth]{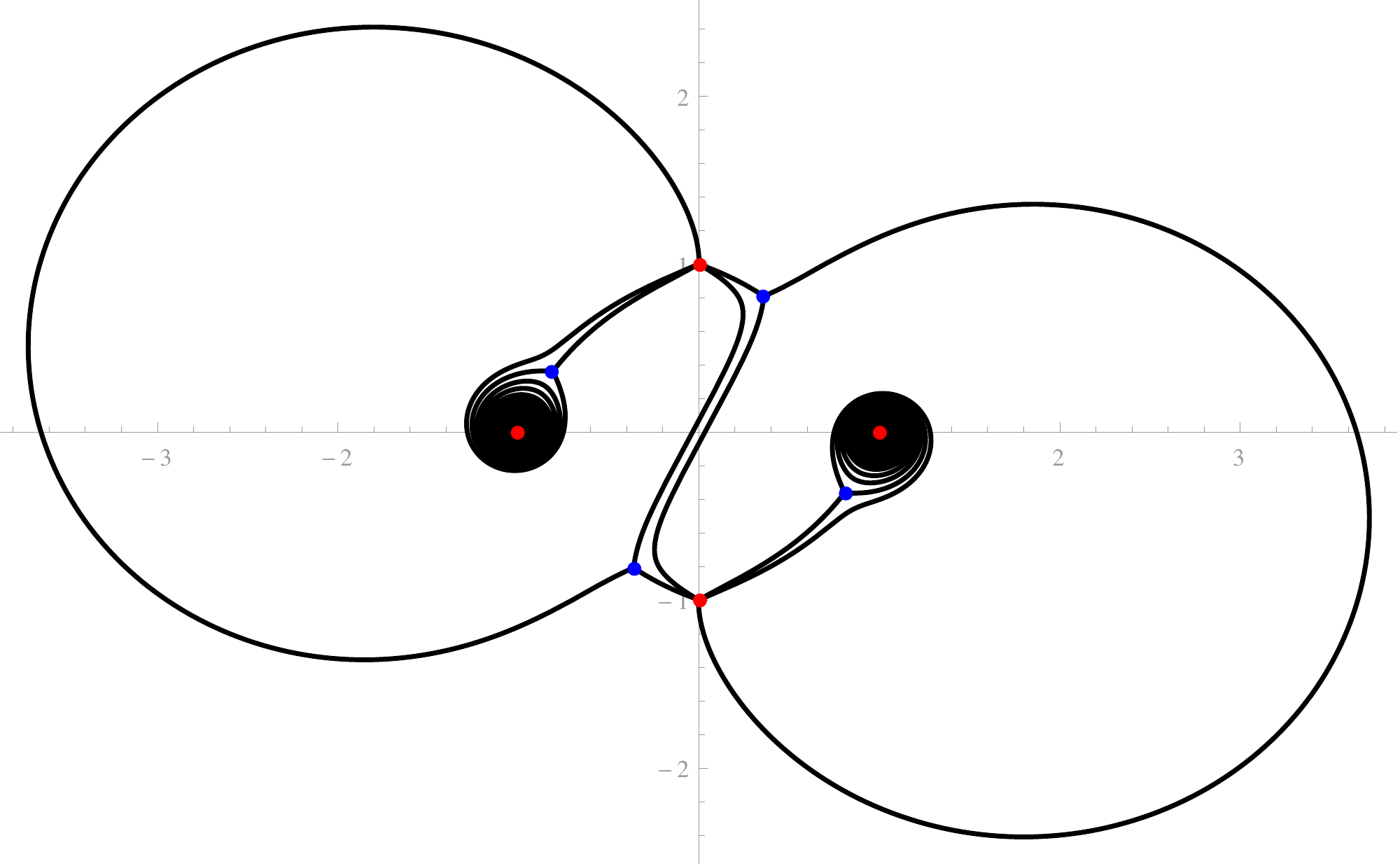}
        \caption{Before the pop.}
        \label{fig:PrePop}
    \end{subfigure}
    ~ 
    \begin{subfigure}[b]{0.45\textwidth}
        \includegraphics[width=\linewidth]{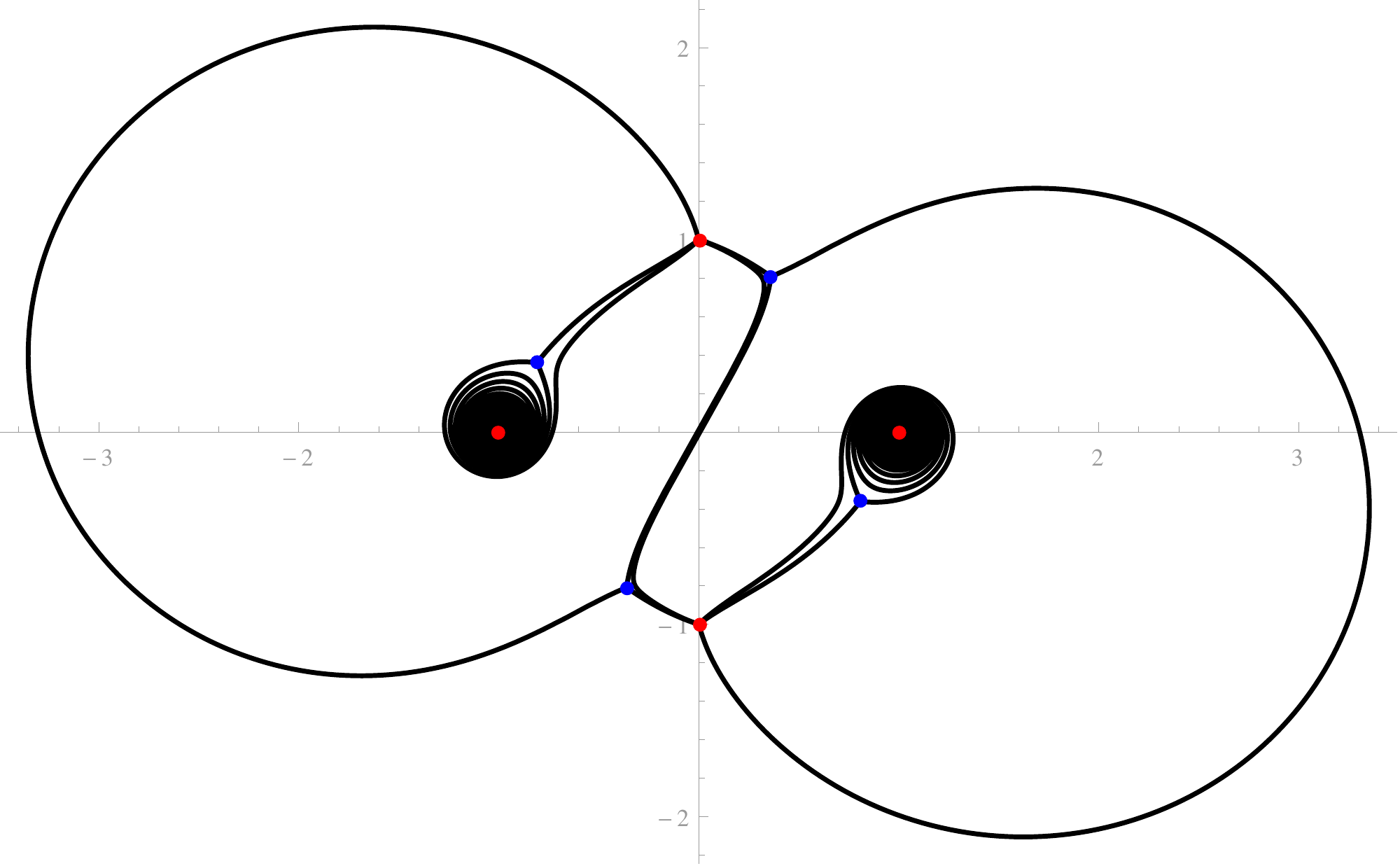}
  \caption{After the pop.}\label{fig:PostPop}
    \end{subfigure}
    ~ 
    \begin{subfigure}[b]{0.45\textwidth}
        \includegraphics[width=1\linewidth]{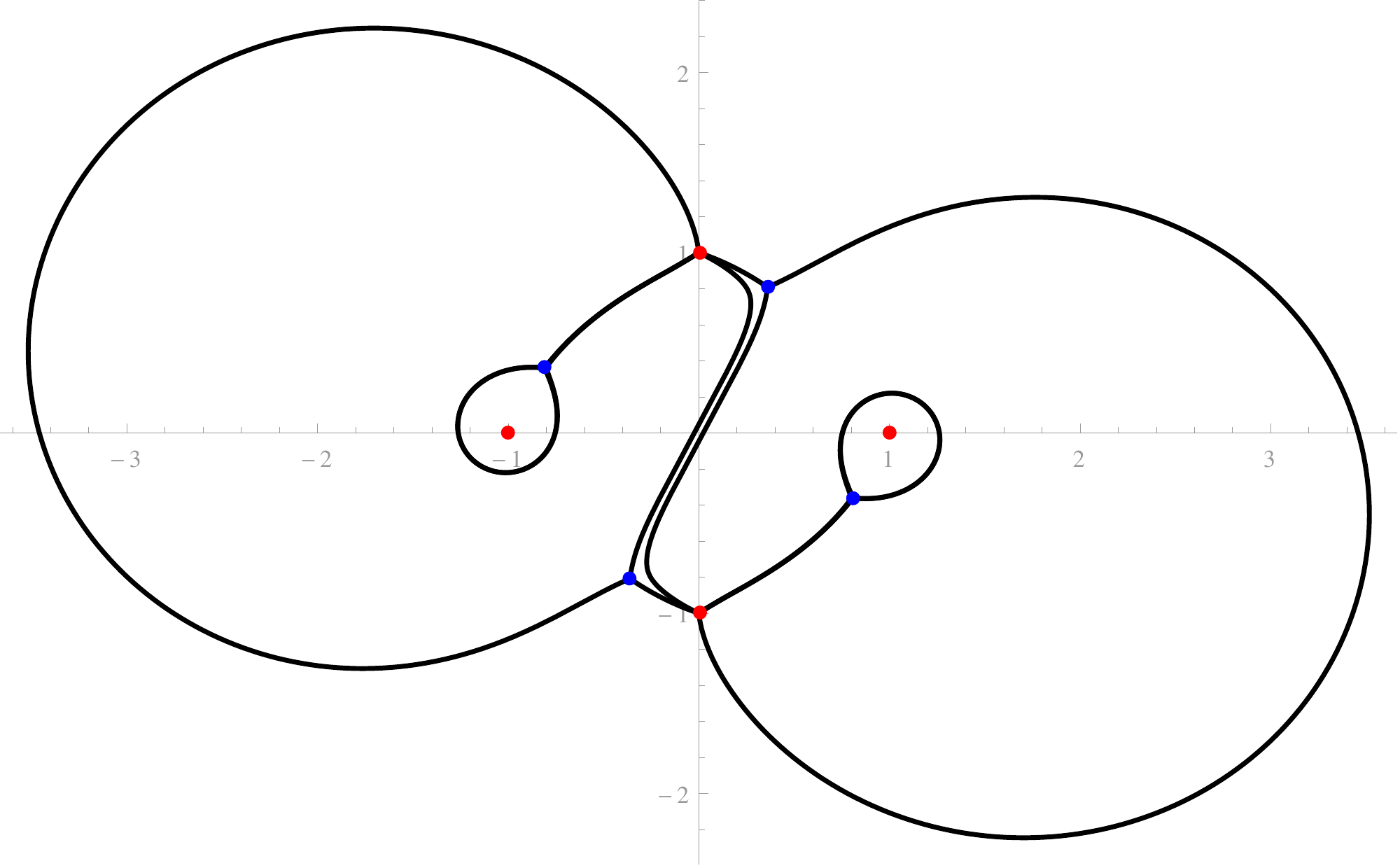}
\caption{The critical graph.}\label{fig:CritPop}
\end{subfigure}
    \caption{Sequence of Stokes graphs related by a pop about the critical graph}\label{fig:pops}
    \end{figure}

We continue to work with the potential \eqref{eq:GMNZ4} and, in order visualize the pops
in figure \ref{fig:pops}, we have made 
the choice $u = \frac{1}{2} \text{exp} \left( \frac{3}{10} \text{i} \pi \right)$. 

\section{The Saddle Point Analysis}
We discussed regular (irregular) conformal blocks associated
to two-dimensional conformal field theories that, via the 2d/4d correspondence, are dual
to  four-dimensional $\Omega$-deformed conformal (respectively,
asymptotically free) gauge theories. The null vector decoupling
equations together with the conformal Ward identities allow us to
arrive at Schr\"odinger equations that govern an integrable system
related to the $\Omega$-deformed gauge theory. While these analyses
are exact in $\epsilon_2$, it is enlightening to see how the
$\epsilon_2 \rightarrow 0$ limit of these differential equations are
derived from purely gauge-theoretic considerations. This will serve as
a consistency check of the 2d/4d correspondence, and our calculations.

In this section, we explain how to derive differential equations
starting from saddle-point difference equations, valid for any SU$(N)$
theory with $N_f=2N$ fundamental hypermultiplets. We specialize
to the case of the conformal SU$(2)$ theory ($N_f = 4$) and 
consider various decoupling limits that give
asymptotically free theories with fewer fundamental hypermultiplets.

Our analysis begins with the Nekrasov partition function, and
considers its saddle-points in the limit $\epsilon_2 \rightarrow 0$,
with $\epsilon_1$ held constant and finite
\cite{Poghossian:2010,Fucito:2011pn,Fucito:2012xc,NPS13}. The
result of this analysis is the $\epsilon_1$-deformed Seiberg-Witten
equation:
\begin{equation}
y\left( x \right) + q \frac{M\left(x-\epsilon_1\right)}{y \left( x-\epsilon_1 \right)} = \left( 1+q \right) P\left(x\right) \, .
\end{equation}
Here, $q$ is the instanton counting parameter. The gauge polynomial $P(x)$ is of degree $N$ and encodes the Coulomb moduli of the gauge theory. The flavour polynomial $M(x)$ is of degree $2N$, and we choose to factorize it into two pieces:
\begin{equation}
\label{eq:Decomp1}
M(x) = A(x) D(x) \, ,
\end{equation}
where $A(x)$ and $D(x)$ are degree $N$ polynomials. As should be evident, this decomposition is far from being unique. Different decompositions can be mapped to the different ways in which the flavour symmetry is realized in the type II construction using two NS$5$ branes and a stack of D$4$ branes. One can associate the number of semi-infinite D$4$ branes on each side of the NS$5$ branes with the degree of the polynomials $A(x)$ and $D(x)$. 
We now peel off a factor of $A(x)$ from the function $y(x)$ and express the remainder as the ratio of some rational function $\mathbf{Q}(x)$ as
\begin{equation}
\label{eq:Decomp2}
y(x) = A(x ) \frac{\mathbf{Q}(x )}{\mathbf{Q}(x -\epsilon_1)} .
\end{equation}
There are other ways to perform the split but this one will reduce, in the $\epsilon_1\rightarrow 0$ limit, to the correct M-theory curve. The result of these decompositions gives us the Baxter $T\mathbf{Q}$-relation
\begin{equation}
A(x) \mathbf{Q}(x) + q D(x-\epsilon_1) \mathbf{Q}(x-2\epsilon_1) - \left(1+q \right) P(x) \mathbf{Q}(x-\epsilon_1) = 0 .
\end{equation}
We now trade $\mathbf{Q}(x)$ for its Fourier transform \cite{NPS13}
\begin{equation}
\label{eq:FourierTransform}
\Psi(t) = \sum_{x \in \Gamma} \mathbf{Q}(x) \ e^{-x/\epsilon_1} ,
\end{equation}
and arrive at the  differential equation of order $N$:
\begin{equation}
\left[ A(x)  + q D(x-\epsilon_1) \ t^{-2} - (1+q) P(x) \ t^{-1}  \right] \Psi(t) = 0 ,
\end{equation}
where in the above equation, the Fourier transform effectively sends $x \mapsto -\epsilon_1 t \partial_t$. 

\subsection{$N_f=4$}
We  specialize to the case of the conformal SU$(2)$ gauge theory, with $N_f=4$. The matter polynomials we start with are
\begin{align}
D(x) &= \left(x-m_1+\frac{\epsilon_1}{2}\right)\left(x-m_2+\frac{\epsilon_1}{2}\right) , \\
A(x) &= \left(x-m_3+\frac{\epsilon_1}{2}\right)\left(x-m_4+\frac{\epsilon_1}{2}\right) \,. 
\end{align}
The gauge polynomial is
\begin{equation}
P(x) = x^2 - q \left( \frac{m_1 + m_2 + m_3 + m_4}{1+q} \right) x - \frac{u_S}{1+q} .
\end{equation}

Substituting this into the difference equation and taking the Fourier transform as described earlier leads to a second order differential equation. 
We would like this differential equation to be of the Schr\"odinger form, i.e.\ with no linear derivative terms. 
This may be achieved by peeling off an appropriate factor.\footnote{The terms we peel off are proportional to the products of square roots of eigenfunctions of the monodromy at $0, 1$ and $q$.}
The resulting differential equation is
\begin{equation}
\left[ -\epsilon_1^2 \frac{\partial^2}{\partial t^2} + Q(t, \epsilon_1) \right] \Psi(t) = 0 ,
\end{equation}
where the potential term $Q(t, \epsilon_1)$ has an expansion in powers
of $\epsilon_1$. In addition, we further shift of the Coulomb modulus $u_S$,
\begin{equation}
u_S = -\widetilde{u} + \frac{q-1}{2} \left( m_1^2 + m_2^2 \right) + q \left[ m_1 m_2 + m_3 m_4 + \frac{1}{2} \left(m_1 +m_2\right)\left( m_3 + m_4 \right) \right] + \frac{\epsilon_1^2}{4} \left( 1+q \right)
\end{equation}
Denoting the order $\epsilon_1^m$ coefficient in the potential by $Q_m(t)$, we obtain the  non-zero potential terms:
\begin{align}
Q_0(t) &= -\frac{\widetilde{u}}{t(t-1)(t-q)} + \frac{\left( m_1 - m_2 \right)^2}{4t^2} + \frac{\left( m_1 + m_2 \right)^2}{4(t-q)^2} + \frac{\left( m_3 + m_4 \right)^2}{4(t-1)^2} + \frac{m_1^2 + m_2^2 + 2 m_3 m_4}{2t(1-t)} \, , \\
Q_2(t) &= -\frac{1}{4t^2}-\frac{1}{4(t-1)^2}-\frac{1}{4(t-q)^2} + \frac{1}{2(t-1)(t-q)} \, .
\end{align}
This matches the potentials in the text obtained from the null vector decoupling equations (\ref{nullNf4Qfunctions}).

\subsection{$N_f=3$}
The $N_f=3$ case is obtained by decoupling one of the masses; we
choose here to send $m_2\rightarrow\infty$, while simultaneously
sending $q\rightarrow 0$ such that the combination $\Lambda_3 = q\, m_2$ remains finite: 
%
%
We identify $\Lambda_3$ to be the strong coupling scale in the SU$(2)$ gauge theory with $N_f=3$. The difference equation takes the form
\begin{equation}
A(x) \mathbf{Q}(x) -\Lambda_3 D(x-\epsilon_1) \mathbf{Q}(x-2\epsilon_1) - P(x) \mathbf{Q}(x-\epsilon_1) = 0 .
\end{equation}
The polynomial functions are given by
\begin{align}
D(x) &= \left(x-m_1+\frac{\epsilon_1}{2}\right), \\
A(x) &= \left(x-m_3+\frac{\epsilon_1}{2}\right)\left(x-m_4+\frac{\epsilon_1}{2}\right) \,,
\end{align}
while the gauge polynomial is
\begin{equation}
P(x) = x^2 - \Lambda_3 x - u_S\,.
\end{equation}
The differential equation can be obtained in a similar fashion as in the conformal case by taking the Fourier transform as in \eqref{eq:FourierTransform}. The analysis leading to the Schr\"odinger type equation can be repeated as before and we obtain the  differential equation: 
\begin{equation}\label{diffeqngeneral}
\left[ -\epsilon_1^2 \frac{\partial^2}{\partial t^2} + \sum_{m=0}^2 Q_m(t)\epsilon_1^m \right] \Psi(t) = 0 \, .
\end{equation}
The $u_S$ we use in the saddle-point analysis must be shifted in order
to make contact with the form of the potential in the text \eqref{Nf=3
  semiclassical} and the shift is given by
\begin{equation}
u_S = \widetilde{u} -\frac{\Lambda_3}{2}\left( 2 m_1 + m_3 + m_4 \right) + \frac{\epsilon_1^2}{4} \, .
\end{equation}
The shift of the Coulomb modulus is accompanied by a global rescaling $m_1 \rightarrow \frac{m_1}{2}$ and $\Lambda_3 \rightarrow \frac{\Lambda_3}{2}$. After these shifts and rescalings, the non-zero potential functions $Q_m(t)$ are given by
\begin{align}
Q_0(t) &= \frac{(m_3+m_4)^2}{4(t-1)^2}+ \frac{m_3 m_4}{t(1-t)} +\frac{m_1\Lambda_3}{t^3}+\frac{\Lambda_3^2}{4t^4} +\frac{\widetilde{u}}{t^2(1-t)} , \\
Q_2(t) &= -\frac{1}{4(t-1)^2}\,.
\end{align}
which match the potential in equation (\ref{Nf=3 semiclassical}).

\subsection{$N_f=2$: asymmetric realization}

There are two distinct cases to be considered when $N_f=2$. These correspond to the fashion in which we decouple fundamental matter. We first consider the case when 
\be
m_1\rightarrow \infty \quad \text{and} \quad \Lambda_3\rightarrow0 \quad \text{with}\quad \Lambda_2^2 = m_1\Lambda_3\quad\text{finite}\,.
\ee
The difference equation takes the form
\begin{equation}
A(x) \mathbf{Q}(x) +\Lambda_2^2 \mathbf{Q}(x-2\epsilon_1) - P(x) \mathbf{Q}(x-\epsilon_1) = 0 .
\end{equation}
The polynomials are given by 
\begin{align}
A(x) &= \left(x-m_3+\frac{\epsilon_1}{2}\right)\left(x-m_4+\frac{\epsilon_1}{2}\right) \,,
\end{align}
while the gauge polynomial is
\begin{equation}
\label{eq:GaugePolyII}
P(x) = x^2 - u_S\,.
\end{equation}
The differential equation is once again given as in \eqref{diffeqngeneral}, and after the shift
\begin{equation}
u_S = \widetilde{u} - \Lambda^2 + \frac{\epsilon_1^2}{4} \, ,
\end{equation}
We find that the non-zero potential functions $Q_m(t)$ are
\begin{align}
Q_0(t) &= \frac{(m_3+m_4)^2}{4(t-1)^2}+ \frac{m_3 m_4}{t(1-t)} +\frac{\Lambda_2^2}{t^3} +\frac{\widetilde{u}}{t^2(1-t)} , \\
Q_2(t) &= -\frac{1}{4 (t-1)^2}\,.
\end{align}
which matches the potential in \eqref{Nf=2 asymmetric semiclassical}.
\subsection{$N_f=2$: symmetric realization} 

An inequivalent way to realize the $N_f=2$ theory is to start from the $N_f=3$ differential equation and consider the limit 
\be
m_3\rightarrow \infty \quad \text{and} \quad \Lambda_3\rightarrow0 \quad \text{with}\quad \Lambda_2^2 = m_3\Lambda_3\quad\text{finite}\,.
\ee
The difference equation takes the form
\begin{equation}
\Lambda_2A(x) \mathbf{Q}(x) +\Lambda_2 D(x-\epsilon_1)\mathbf{Q}(x-2\epsilon_1) - P(x) \mathbf{Q}(x-\epsilon_1) = 0 .
\end{equation}
The polynomial $A(x)$ is given by 
\begin{align}
D(x) &= \left(x-m_1+\frac{\epsilon_1}{2}\right), \\
A(x) &= \left(x-m_4+\frac{\epsilon_1}{2}\right) \,,
\end{align}
while the gauge polynomial is given by \eqref{eq:GaugePolyII}.
The differential equation still takes the Schr\"odinger form \eqref{diffeqngeneral}. 

Finally, in order to match with the form of the differential in the text \eqref{Nf=2 symmetric semiclassical}, we need to perform a conformal transformation $\left( t \rightarrow \frac{z-1}{z} \right)$. We choose to shift away the $O(\epsilon_1^2)$ term by redefining of the Coulomb modulus:
\be
u_S = \widetilde{u} + m_1 \Lambda_2 - \frac{\Lambda^2}{2} + \frac{\epsilon_1^2}{4} \,.
\ee
After this, the differential matches the form in \eqref{Nf=2 symmetric semiclassical}, with 

\begin{equation}
Q_0(z) = \frac{\Lambda_2^2}{4z^4}  + \frac{\Lambda_2m_1}{z^3(z-1)^2}-\frac{\Lambda_2 m_4}{z(z-1)^3} +\frac{\Lambda_2^2}{4(z-1)^4} +\frac{\widetilde{u}}{z^2(z-1)^2}
\, .
\end{equation}

\subsection{$N_f=1$}

We start with the symmetric realization of the $N_f=2$ case and take the limit
\be\label{Nf=1limit}
m_4\rightarrow \infty \quad \text{and} \quad \Lambda_2\rightarrow 0 \quad\text{with}\quad \Lambda_1^3=m_4\Lambda_2^2\quad\text{finite}\,.
\ee
The difference equation in this case takes the form
\begin{equation}
\Lambda_1^2 \mathbf{Q}(x) + \Lambda_1\, D(x-\epsilon_1)\mathbf{Q}(x-2\epsilon_1) - P(x) \mathbf{Q}(x-\epsilon_1) = 0 .
\end{equation}
The gauge polynomial is once again given by \eqref{eq:GaugePolyII}, and the flavour polynomial $D(x)$ is given by
\begin{equation}
D(x) = \left( x - m_1 + \frac{\epsilon_1}{2} \right) \, .
\end{equation}
While the differential equation in this case also takes the form of a Schr\"odinger equation, with the potential (after shifting $u_S$ so as to set $Q_2(t)$ to zero for convenience)
\begin{equation}
Q_0(t) = \frac{\Lambda_1^2}{4 t^4} - \frac{m_1 \Lambda_1}{t^3} + \frac{u_S}{t^2} + \frac{\Lambda_1^2}{t} \, ,
\end{equation}
its form requires a series of conformal transformations and rescalings before it can be easily compared
with the form in \eqref{Nf=1 semiclassical}; for convenience, we reproduce the transformations below:
\begin{align}
\begin{split}
t &\longrightarrow \frac{1}{z} \\
u_S &\longrightarrow 2^{4/3} \left( \widetilde{u}+m_1 \Lambda_1 \right) \, \\
m_1 &\longrightarrow - 2^{2/3} m_1 \, \\
z &\longrightarrow 2^{2/3} w \, \\
w &\longrightarrow \frac{1-z}{z} \, .
\end{split}
\end{align}
After this sequence of transformations, we get the form of the potential as in \eqref{Nf=1 semiclassical}:
\begin{equation}
Q_0(z) = \frac{\Lambda_1^2}{4z^4}  + \frac{\Lambda_1m_1}{z^3(z-1)^2}-\frac{\Lambda_1^2}{4z(z-1)^3} 
+\frac{\widetilde{u}}{z^2(z-1)^2} \, .
\end{equation}

\subsection{$N_f=0$}

The pure super Yang-Mills case is obtained by starting with the
$N_f=1$ case and taking the limit,
\be
m_1\rightarrow \infty \quad \text{and} \quad \Lambda_1\rightarrow 0\quad\text{with}\quad \Lambda_0^4=m_1\Lambda_1^3\quad\text{finite}\,.
\ee
The difference equation takes the  form:
\be
\Lambda_0^2 \mathbf{Q}(x) +\Lambda_0^2\mathbf{Q}(x-2\epsilon_1) - P(x) \mathbf{Q}(x-\epsilon_1) = 0 .
\ee
While the differential equation in this case also takes the form of a Schr\"odinger equation, with the potential (after shifting $u_S$ so as to set $Q_2(t)$ to zero for convenience)
\begin{equation}
\phi_2 (t) =  \frac{\Lambda_0^2}{t^3} + \frac{u_S}{t^2} + \frac{\Lambda_0^2}{t} \, ,
\end{equation}
its form requires a series of conformal transformations and rescalings before it can be literally compared with the form in \eqref{PureNVD}; for convenience, we reproduce these transformations below:
\begin{align}
\begin{split}
t &\longrightarrow \text{i} w \, \\
u_S &\longrightarrow - \left( \widetilde{u} + \Lambda_0^2 \right) \, \\
\Lambda_0 &\longrightarrow \text{e}^{\text{i}\frac{\pi}{4}} \Lambda_0 \, \\
w &\longrightarrow \frac{1-z}{z} \, .
\end{split}
\end{align}
After this sequence of transformations, we get the potential as in the bulk of the paper \eqref{PureNVD}:
\begin{equation}
Q_0(z) = \frac{\Lambda_0^2}{z^3(z-1)^2} +\frac{\widetilde{u}}{z^2(z-1)^2} + \frac{\Lambda_0^2}{z(z-1)^3} \, .
\end{equation}
Thus, we completed the derivation of the null vector decoupling equations  (at $\epsilon_2=0$)
from the purely gauge theoretic instanton partition function.

\end{appendix}

\end{document}